\documentclass[fleqn,usenatbib]{mnras}

\usepackage{newtxtext,newtxmath}
\usepackage[T1]{fontenc}
\DeclareRobustCommand{\VAN}[3]{#2}
\let\VANthebibliography\thebibliography
\def\thebibliography{\DeclareRobustCommand{\VAN}[3]{##3}\VANthebibliography}

\usepackage{graphicx}	% Including figure files
\usepackage{adjustbox}  % For the orcid ids
\usepackage{amsmath}	% Advanced maths commands
\usepackage{wasysym}    % Extra astronomy symbols
\usepackage{bm}         % Bold maths
\usepackage{xspace}
\usepackage{etoolbox}
\usepackage{float}
\makeatletter % Workaround to only color the year for cite commands
  \patchcmd{\NAT@citex}
    {\@citea\NAT@hyper@{%
      \NAT@nmfmt{\NAT@nm}%
      \hyper@natlinkbreak{\NAT@aysep\NAT@spacechar}{\@citeb\@extra@b@citeb}%
      \NAT@date}}
    {\@citea\NAT@nmfmt{\NAT@nm}%
    \NAT@aysep\NAT@spacechar\NAT@hyper@{\NAT@date}}{}{}

  \patchcmd{\NAT@citex}
    {\@citea\NAT@hyper@{%
      \NAT@nmfmt{\NAT@nm}%
      \hyper@natlinkbreak{\NAT@spacechar\NAT@@open\if*#1*\else#1\NAT@spacechar\fi}%
        {\@citeb\@extra@b@citeb}%
      \NAT@date}}
    {\@citea\NAT@nmfmt{\NAT@nm}%
    \NAT@spacechar\NAT@@open\if*#1*\else#1\NAT@spacechar\fi\NAT@hyper@{\NAT@date}}
    {}{}
\makeatother

\newtoggle{referee}            %
\togglefalse{referee}          %
\iftoggle{referee}{%           %
    \usepackage{color,soul}    %
}{%                            %
}                              %
\newcommand\Msun{\text{M}_{\astrosun}} % requires the wasysym package
\newcommand\Zsun{\text{Z}_{\astrosun}} % requires the wasysym package
\newcommand\HI{\ion{H}{I}\xspace} % neutral hydrogen
\newcommand\HII{\ion{H}{II}\xspace} % ionized hydrogen
\newcommand\HeI{\ion{He}{I}\xspace} % neutral helium
\newcommand\HeII{\ion{He}{II}\xspace} % singly ionized helium
\newcommand\orcid[1]{\href{http://orcid.org/#1}{\adjustbox{trim={-.15\width} {0\height} {-.15\width} {0\height},clip}{\includegraphics[height=10pt]{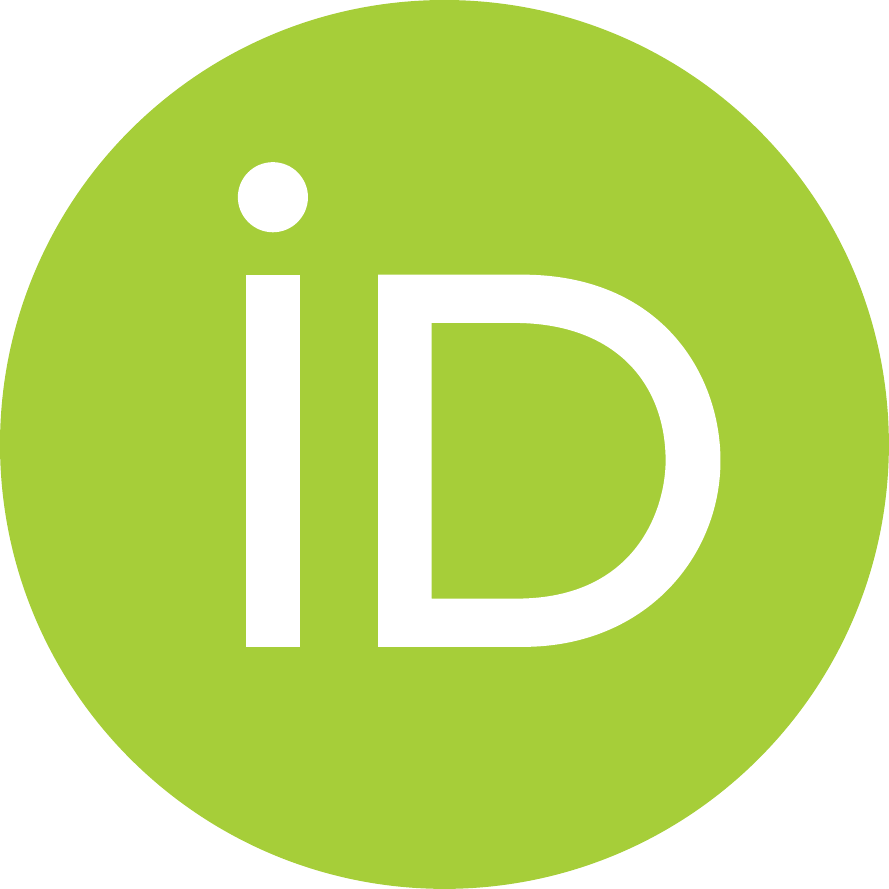}}}}
\title[Ly$\alpha$ escape from disc-like galaxies]{The physics of Lyman-$\balpha$ escape from disc-like galaxies}

\author[A.~Smith et al.]{%
Aaron~Smith\orcid{0000-0002-2838-9033},$^{1,2}$\thanks{E-mail: \href{mailto:aaron.smith@cfa.harvard.edu}{aaron.smith@cfa.harvard.edu}; NHFP Einstein Fellow.}
Rahul~Kannan\orcid{0000-0001-6092-2187},$^{1}$\thanks{E-mail: \href{mailto:rahul.kannan@cfa.harvard.edu}{rahul.kannan@cfa.harvard.edu}}
Sandro~Tacchella\orcid{0000-0002-8224-4505},$^{3,4}$\thanks{E-mail: \href{mailto:st578@cam.ac.uk}{st578@cam.ac.uk}}
Mark~Vogelsberger\orcid{0000-0001-8593-7692},$^{2}$
Lars~Hernquist,$^{1}$
\newauthor
Federico~Marinacci\orcid{0000-0003-3816-7028},$^{5}$
Laura~V.~Sales\orcid{0000-0002-3790-720X},$^{6}$
Paul~Torrey\orcid{0000-0002-5653-0786},$^{7}$
Hui~Li\orcid{0000-0002-1253-2763},$^{8}$\thanks{NHFP Hubble Fellow.}
Jessica~Y.-C.~Yeh\orcid{0000-0002-5721-7679}$^{2}$
and
Jia~Qi$^{7}$ \\
  $^{1}$Center for Astrophysics $\vert$ Harvard \& Smithsonian, 60 Garden St, Cambridge, MA 02138, USA \\
  $^{2}$Department of Physics,
  Massachusetts Institute of Technology, Cambridge, MA 02139, USA \\
  $^{3}$Kavli Institute for Cosmology, University of Cambridge, Madingley Road, Cambridge, CB3 0HA, UK \\
  $^{4}$Cavendish Laboratory, University of Cambridge, 19 JJ Thomson Avenue, Cambridge, CB3 0HE, UK \\
  $^{5}$Department of Physics and Astronomy ``Augusto Righi'', University of Bologna, via Gobetti 93/2, 40129 Bologna, Italy \\
  $^{6}$Department of Physics and Astronomy, University of California, Riverside, 900 University Avenue, Riverside, CA 92521, USA \\
  $^{7}$Department of Astronomy, University of Florida, 211 Bryant Space Sciences Center, Gainesville, FL 32611, USA \\
  $^{8}$Department of Astronomy, Columbia University, New York, NY 10027, USA
}

\date{Accepted 2022 September 12. Received 2022 July 01; in original form 2021 December 03}

\pubyear{2022}

\begin{document}
\label{firstpage}
\pagerange{\pageref{firstpage}--\pageref{lastpage}}
\maketitle

\begin{abstract}
  Hydrogen emission lines can provide extensive information about star-forming galaxies in both the local and high-redshift Universe. We present a detailed Lyman continuum (LyC), Lyman-$\alpha$ (Ly$\alpha$), and Balmer line (H$\alpha$ and H$\beta$) radiative transfer study of a high-resolution isolated Milky Way simulation using the state-of-the-art \textsc{arepo-rt} radiation hydrodynamics code with the SMUGGLE galaxy formation model. The realistic framework includes stellar feedback, non-equilibrium thermochemistry accounting for molecular hydrogen, and dust grain evolution in the interstellar medium (ISM). We extend our publicly available Cosmic Ly$\alpha$ Transfer (\textsc{colt}) code with photoionization equilibrium Monte Carlo radiative transfer and various methodology improvements for self-consistent end-to-end (non-)resonant line predictions. Accurate LyC reprocessing to recombination emission requires modelling pre-absorption by dust ($f_\text{abs} \approx 27.5\%$), helium ionization ($f_\text{He} \approx 8.7\%$), and anisotropic escape fractions ($f_\text{esc} \approx 7.9\%$), as these reduce the available budget for hydrogen line emission ($f_\text{H} \approx 55.9\%$). We investigate the role of the multiphase dusty ISM, disc geometry, gas kinematics, and star formation activity in governing the physics of emission and escape, focusing on the time variability, gas phase structure, and spatial, spectral, and viewing angle dependence of the emergent photons. Isolated disc simulations are well-suited for comprehensive observational comparisons with local H$\alpha$ surveys, but would require a proper cosmological circumgalactic medium (CGM) environment as well as less dust absorption and rotational broadening to serve as analogs for high-redshift Ly$\alpha$ emitting galaxies. Future applications of our framework to next-generation cosmological simulations of galaxy formation including radiation-hydrodynamics that resolve $\lesssim 10$\,pc multiphase ISM and $\lesssim 1$\,kpc CGM structures will provide crucial insights and predictions for current and upcoming Ly$\alpha$ observations.
\end{abstract}


%%%%%%%%%%%%%%%%%%%%%%%%%%%%%%%%%%%%%%%%%%%%%%%%%%

%%%%%%%%%%%%%%%%% BODY OF PAPER %%%%%%%%%%%%%%%%%%

\section{Introduction}
\label{sec:intro}
Star-forming galaxies produce copious amounts of Lyman continuum (LyC) photons, primarily from young massive stars, that efficiently ionize their surroundings but have low galactic escape fractions. The subsequent recombination of hydrogen atoms generates strong line emission, including the Lyman-$\alpha$ (Ly$\alpha$) and Balmer (H$\alpha$ and H$\beta$) channels, which provide rich information about these galaxies throughout cosmic history \citep{Partridge1967,KennicuttEvans2012,Kewley2019,ForsterSchreiberWuyts2020}. In practice, the observable line luminosities can vary enormously as a result of dust extinction, star-formation and feedback duty cycles, galaxy and stellar population properties, and complex radiation transport effects imposed by the intervening interstellar medium (ISM), circumgalactic medium (CGM), and intergalactic medium (IGM) en route to our telescopes. Understanding Ly$\alpha$ emitting galaxies (LAEs) is particularly challenging due to resonant scattering with neutral hydrogen atoms, which changes the influence of dust, couples transport to gas flows and geometries, and overall reduces the surface brightness compared to non-resonant lines and continuum radiation \citep{Dijkstra2014,Dijkstra2019}. Still, observational approaches and technologies that reach greater sensitivities, cover larger areas, or mitigate atmospheric and foreground contamination are affording greater access to the Ly$\alpha$ Universe, both locally and at high redshifts \citep[see][and references therein]{Hayes2015,Hayes2019,Ouchi2019,Ouchi2020}. In particular, sensitive instruments such as the Multi Unit Spectroscopic Explorer (MUSE) Integral Field Spectrograph on the Very Large Telescope (VLT) have further revolutionized our knowledge of LAEs at intermediate redshifts $z = 3$--$6$ \citep[e.g.][]{Leclercq2017,Wisotzki2018}.

Cosmological simulations of galaxy formation have seen significant improvements over the past decades \citep{Vogelsberger2020}. However, the majority of such simulations still lack essential ingredients for fully self-consistent Ly$\alpha$ radiative transfer modelling. This is mainly due to the extreme numerical challenges encountered as well as various trade-offs between simulation accuracy and efficiency \citep{SpringelSPH2010,Teyssier2015}. Beyond this, we seek a consensus understanding of the physics regulating the structure of the ISM, star formation, and the driving of galactic outflows \citep{NaabOstriker2017}. Given the current state-of-the-art, the ideal Ly$\alpha$ study would be based on fully-coupled radiation-magneto-hydrodynamics simulations with a proper cosmological environment, reliable galaxy formation model, realistic feedback connecting subresolution and galactic scales, dust prescriptions tracking growth and destruction processes, possibly other important phenomena such as black holes and cosmic rays, high-enough spatial resolution to capture ISM, CGM, and IGM effects, and sufficient redshift, volume, and galaxy mass coverage to ensure representative LAE statistics. While this is certainly ambitious, progress in the community on any of these aspects individually or collectively is encouraging.

Furthermore, as realistic galaxy models continue to include additional physical ingredients, it is important to evaluate their reliability with accurate and stringent tests. This can be achieved with self-consistent radiative transfer calculations that clearly illustrate and quantify the role of various processes governing the physics of emission and escape. Moreover, such exploratory post-processing simulations bring to light non-trivial trends that are sometimes easiest to understand in hindsight. For example, we now recognize that anisotropic escape fractions are naturally induced by galaxy geometry and orientation with respect to the line of sight (LOS), temporal variability of average luminosities is connected to the burstiness and history of star formation, spatial morphology, and phase space structure is related to feedback patterns and recycling of high-density gas, and emergent spectral line profiles are heavily shaped by the gas kinematics on both small and large scales \citep[e.g.][]{Smith2019}. Likewise, observed (non)correlations among galaxy populations can arise from different host properties, for example from a range of dust attenuation and covering fraction scenarios \citep{Hayes2013,Huang2021,Reddy2022}. Finally, when applied to realistic cosmological environments, it is also beneficial to assess the relative impacts from ISM, CGM, and IGM scale physics on Ly$\alpha$ radiation transport and observational signatures \citep{Byrohl2021,Garel2021}.

In recent years there has been an increasing number of Ly$\alpha$ radiative transfer studies in different galaxy formation contexts. Indeed, analytical studies have been beneficial in elucidating fundamental escape properties and mechanisms \citep{Harrington1973,Neufeld1990,LoebRybicki1999,HansenOh2006,LaoSmith2020}, but are also complemented by tests in slab and spherical geometries \citep{Ahn2002,Zheng2002,Dijkstra2006,Verhamme2006,Smith2017}. Beyond this there are non-trivial qualitative and quantitative insights revealed by simulations of expanding shell environments \citep{Verhamme2008,GronkeBullDijkstra2015,Yang2017,Orlitova2018,Gurung-Lopez2022}, clumpy multiphase media \citep{DijkstraKramer2012,Laursen2013,Duval2014,LiSteidel2021}, and other idealized set-ups \citep{Behrens2014,Zheng2014,Smith2015,Gronke2017,Remolina-Gutierrez2019,Seon2020,Song2020}. However, although the intuition and physics is generally applicable, the path towards reliable Ly$\alpha$ predictions based on hydrodynamical simulations is inherently tied to the success of galaxy modelling \citep{Dayal2018}. This is by no means straightforward and is further complicated by the unique dynamic range, resolution, and physics requirements that are not always prioritized in simulation design strategies. Still, Ly$\alpha$ forward modelling is proving to be a worthwhile endeavor to reveal physical (in)consistencies in hydrodynamical simulations \citep{Tasitsiomi2006,Laursen2009,Yajima2012,Behrens2019,Smith2019,Laursen2019,Kimock2021}.

Understanding the physics of Ly$\alpha$ escape has important implications for other related radiation signatures as well. In fact, self-consistently incorporating (non-)ionizing continuum and other hydrogen, nebular, and metal absorption lines into Ly$\alpha$ studies helps to disentangle certain radiative transfer effects that are either amplified or suppressed by resonant scattering. For example, different types of line and continuum radiation may have unique spatial, spectral, and angular escape or absorption features that help to determine the relative importance of turbulent ISM porosity compared to smoother spherical and disc-like density gradients. This is also significant because ionizing radiation and stellar feedback act to clear low-column density channels that facilitate the escape of Ly$\alpha$ photons as suggested by several theoretical works \citep[e.g.][]{Yajima2014,Dijkstra2016,Kimm2019,Kakiichi2021,Mauerhofer2021} and observational studies \citep[e.g.][]{NakajimaOuchi2014,Henry2015,Chisholm2018,Gazagnes2018,Gazagnes2020,Jaskot2019}. Of course, due to the implications for cosmic reionization and other epochs, dedicated simulation-based efforts to quantify LyC escape is an active field on its own even without drawing specific connections to Ly$\alpha$ radiative transfer \citep{WiseCen2009,Wise2014,Ma2015,Ma2020,Paardekooper2015,Kimm2017,Barrow2020,Yeh2022}.

Similarly, there are numerous theoretical investigations including the H$\alpha$ line \citep[and similar emission lines, e.g.][]{Katz2019,Shen2020,Wilkins2020,KannanLIM2022}, which is a powerful star-formation rate (SFR) indicator observationally accessible in current surveys out to $z \sim 2.5$, both on global \citep[e.g.][]{Kennicutt1983,Lee2009,Koyama2015,Shivaei2015} and spatially-resolved scales \citep[e.g.][]{Tacchella2015,nelson2016,Belfiore2018,Ellison2018,Belfiore2022}. Upcoming programs with the \textit{James Webb Space Telescope}~(\textit{JWST}) will extend this window to $z \sim 7$, providing a critical perspective on cosmic star formation. However, the complex ionization states of galaxies arising from both compact and diffuse \HII regions makes it crucial to model LyC-reprocessed H$\alpha$ emission with similar care and accuracy as Ly$\alpha$ radiative transfer studies, especially in the context of resolved ISM simulations with radiation and dust physics. Different methodologies have been developed in a handful of studies with various advantages and limitations. We highlight applications focusing on the scattered light contribution of the diffuse galactic background \citep{Wood1999,Barnes2015} star formation relations (without dust) from simulations of isolated dwarfs and high-redshift galaxies \citep{Kim2013,Kim2019}, subresolution population synthesis for a Milky Way like galaxy \citep{PellegriniEMP2020,Pellegrini2020} and periodic tall box simulations capable of exploring subparsec scale feedback and emission \citep{Peters2017,Kado-Fong2020}. Our work further adds to this class of detailed emission line modelling by achieving resolved emission and dust extinction throughout entire galaxies to better understand the physics and assist with observational interpretations.

From a numerical standpoint, there has recently been a strong movement towards more robust subgrid modelling ($\lesssim 10$\,pc) in galaxy simulations capable of low-temperature gas cooling and stellar feedback that produces a spatially-resolved multiphase ISM \citep[e.g.][and references therein]{Hopkins2014,Hopkins2018,SmithSijackiShen2018,Marinacci2019, Gutcke2021}. In this paper we further explore the high-resolution Milky Way simulation from \cite{Kannan2020}, which combines the state-of-the-art \textsc{arepo-rt} \citep{Kannan2019} radiation hydrodynamics solver with explicit photoheating and radiation pressure feedback from young stars, a non-equilibrium thermochemistry module that accounts for molecular hydrogen (H$_2$), coupled to explicit dust formation and destruction, all of which are integrated into a novel stellar feedback framework, the Stars and MUltiphase Gas in GaLaxiEs (SMUGGLE) feedback model \citep{Marinacci2019}. We employ our COsmic Ly$\alpha$ Transfer code \citep[\textsc{colt};][for public code access and documentation see \href{https://colt.readthedocs.io}{\texttt{colt.readthedocs.io}}]{Smith2015,Smith2019} to perform post-processing Monte Carlo radiative transfer (MCRT) calculations for ionizing radiation and hydrogen Ly$\alpha$, H$\alpha$, and H$\beta$ emission lines. In a companion study \citet{Tacchella2022}, we investigate H$\alpha$ emission as a SFR tracer on spatially resolved scales including the time-variability and diffuse ionized gas from these simulations. We anticipate other applications of our data sets and methodology, including cosmological zoom-in simulations based on the same framework in conjunction with the \textsc{thesan} project \citep{KannanThesan2022,GaraldiThesan2022,SmithThesan2022}.

In light of the simulated physics, this paper also represents a follow-up study of previous Ly$\alpha$ investigations of isolated disc-like galaxies. In particular, \citet{Verhamme2012} showed that the ISM model can significantly affect the Ly$\alpha$ escape fraction and spectral features, as metal line cooling down to $\sim10$\,K leads to more clumpy, clustered star-forming regions. The authors argue that discrepancies in Ly$\alpha$ properties from warmer $\sim10^4$\,K ISM models demonstrate that radiation transfer calculations can only lead to realistic properties in simulations where galaxies are resolved into giant molecular clouds. This requirement has been achieved by several groups, although the included physics varies dramatically. Another relevant work is from \citet{BehrensBraun2014} who study the inclination dependence from a turbulent disc simulation with broad agreement to \cite{Verhamme2012} with additional discussion about time variability. However, both of these investigations set the ionization states according to local collisional ionization equilibrium calculations, which results in unrealistic Ly$\alpha$ sourcing and transport. Furthermore, our radiation hydrodynamics simulations achieve over an order of magnitude higher resolution ($\lesssim 1$\,pc minimum cell sizes compared to $18$~and~$30$\,pc, respectively), allowing explorations of smaller-scale effects influencing the scattering and absorption. 

The remainder of the paper is as follows: in Section~\ref{sec:methods}, we describe the simulation and methods with an emphasis on the most relevant or updated radiative transfer schemes for ionizing and line photons; in Section~\ref{sec:results}, we present various post-processing results including insights about time, spatial, spectral, angular, and inclination dependence for the emission and escape of radiation; and finally in Section~\ref{sec:summary}, we provide a summary of the conclusions and a discussion on the outlook for future work.

\begin{figure*}
  \centering
  \includegraphics[width=\textwidth]{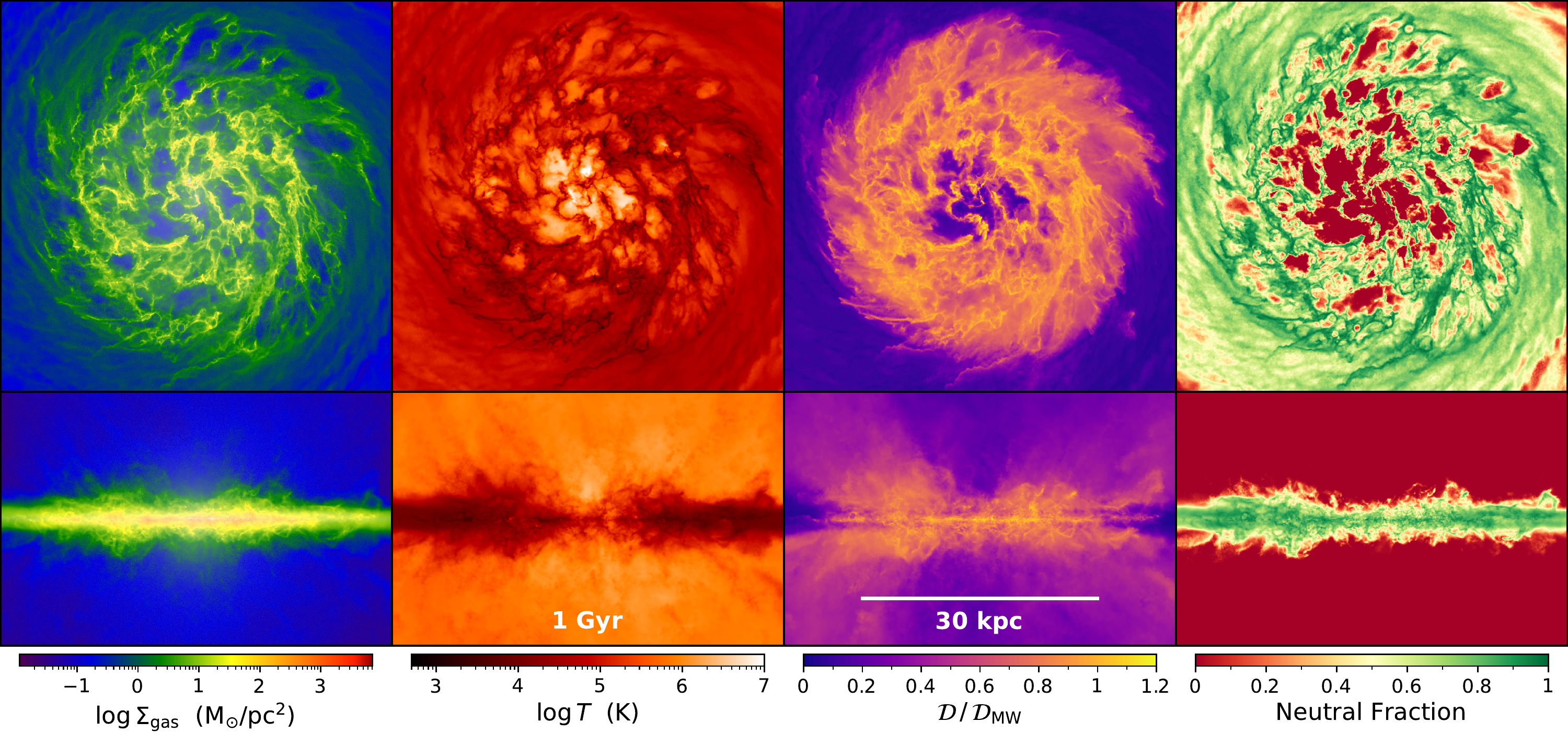}
  \caption{Projected images of the MW galaxy at 1\,Gyr illustrating face-on and edge-on views of the gas surface density $\Sigma_\text{gas}$, gas temperature $T$, dust-to-gas ratio $\mathcal{D}$ relative to the canonical MW value (0.01), and neutral hydrogen fraction $x_\text{\HI} \equiv n_\text{\HI} / n_\text{H}$. In the leftmost panel, stars are shown as whitened pixels with the transparency scaled by the logarithmic intrinsic ionizing photon rate $\dot{N}_\text{ion}$. The ISM structure consists of a patchwork of large hot ionized bubbles surrounded by dense clumps, filaments, and walls providing escape channels and absorption barriers for radiative transfer calculations.}
  \label{fig:rho_T_D_HI}
\end{figure*}

\section{Methods}
\label{sec:methods}
We briefly describe the simulations in Section~\ref{subsec:simulation}, which are based on a novel framework to self-consistently model the effects of radiation fields, dust physics, and molecular chemistry (H$_2$) in the ISM of galaxies. In Sections~\ref{subsec:rt-ion}~and~\ref{subsec:rt}, we explain in detail the MCRT calculations employed in order to predict the line emission from idealized disc-like galaxies.

\begin{figure*}
  \centering
  \includegraphics[width=\textwidth]{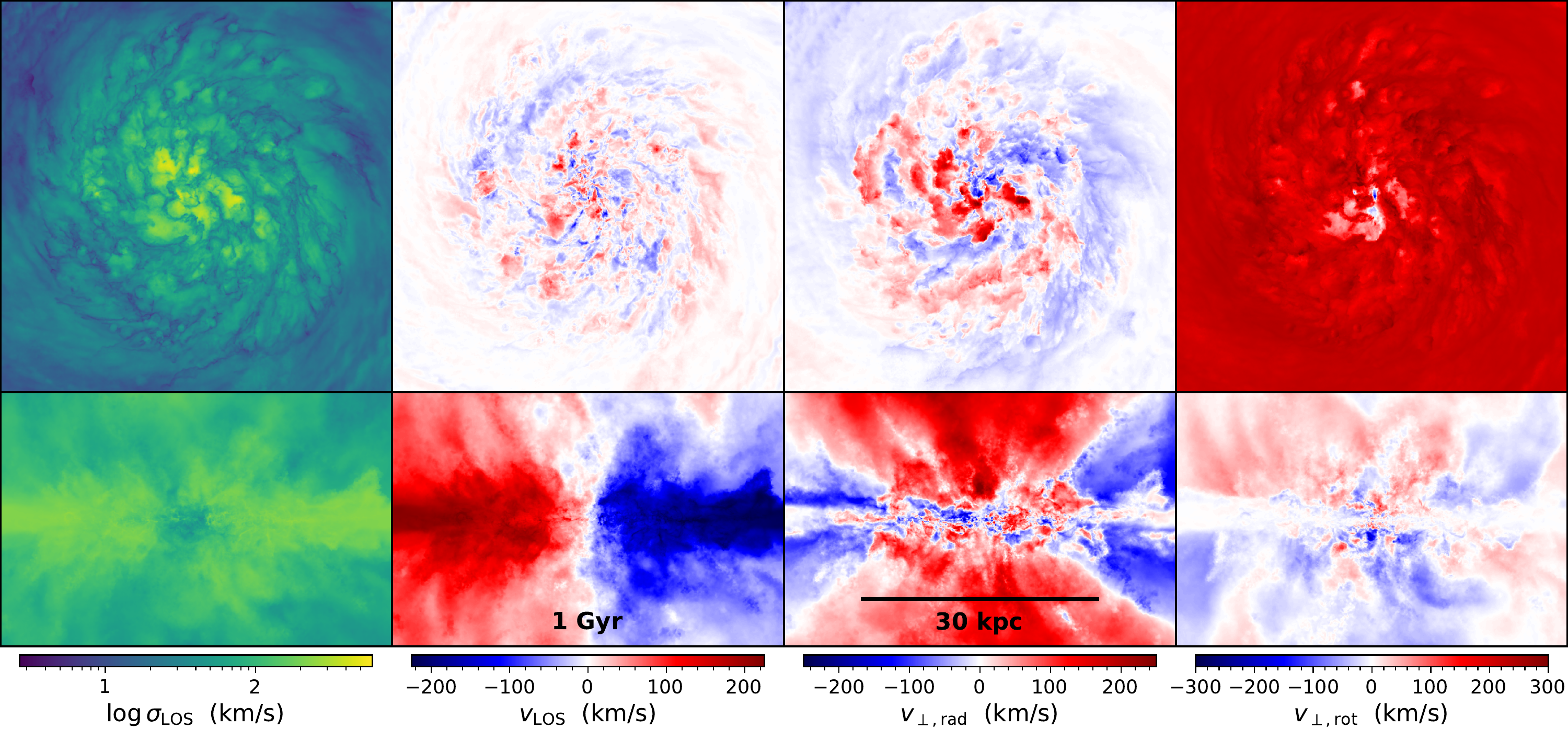}
  \caption{Projected images of the MW galaxy at 1\,Gyr illustrating face-on and edge-on views of the gas mass-weighted LOS velocity dispersion $\sigma_\text{LOS}$, LOS velocity $v_\text{LOS}$, perpendicular (to the image plane) radial velocity $v_{\perp,\text{rad}}$, and perpendicular rotational velocity $v_{\perp,\text{rot}}$. The LOS velocity statistics are particularly relevant for shaping intrinsic line profiles, although there is a density bias when converting to line emission. Likewise, the perpendicular velocity components are important for Ly$\alpha$ transport due to resonant scattering in the comoving frame of the gas flows.}
  \label{fig:v_4}
\end{figure*}

\subsection{Isolated disc Milky Way simulation}
\label{subsec:simulation}
We make use of the simulation presented in \citet{Kannan2020, Kannan2021}, which is a high-resolution isolated disc simulation of a Milky Way like galaxy (MW; $M_\text{halo} = 1.5 \times 10^{12}\,\Msun$). The simulations were performed with \mbox{\textsc{arepo-rt}} \citep{Kannan2019}, a novel radiation hydrodynamic extension of the moving mesh hydrodynamic code \textsc{arepo} \citep[][\href{https://arepo-code.org}{\texttt{arepo-code.org}}]{Springel2010,Weinberger2020}. The adopted SMUGGLE sub-grid models for star formation and feedback are described in \citet{Marinacci2019} and \citet{Kannan2020}. Briefly, gas is allowed to cool down to $10$\,K with the cooling function divided into primordial cooling from hydrogen (both atomic and molecular) and helium, metal cooling scaled linearly with the metallicity of the gas and cooling through gas-dust and radiation field interactions, in addition to photoelectric and photoheating from far ultraviolet (FUV) and Lyman continuum photons, respectively. Star particles are probabilistically formed from cold gas above a density threshold of $n = 10^{3}\,\text{cm}^{-3}$. Additionally, gas clouds must be self-gravitating in order to form stars. There are three feedback mechanisms implemented related to stars: radiative feedback, stellar winds from young O, B, and asymptotic giant branch (AGB) stars, and supernova feedback. Photoheating, radiation pressure, and photoelectric heating are modelled self-consistently through the radiative transfer scheme. Furthermore, the simulations employ a novel self-consistent dust formation and destruction model \citep{McKinnon2017}, which accounts for three distinct dust production channels: SNII, SNIa, and AGB stars \citep{Dwek1998}. The dust is assumed to be dynamically coupled to the gas. The dust mass in the ISM increases due to the gas-phase elements colliding with existing grains \citep{Dwek1998} and decreases due to shocks from SN remnants \citep{McKee1989} and sputtering in high-temperature gas \citep{Tsai1995}. The SMUGGLE framework has been shown to produce a realistic multiphase ISM \citep{Marinacci2019}, reasonable dense ISM and star cluster properties in simulated late-type and merging galaxies \citep{Li2020GMC,Li2022}, and constant-density cores in idealized dwarf galaxies \citep{Jahn2021}.

The MW simulation consists of a dark matter halo, a bulge, and a stellar and gaseous disc set up following the techniques described in \citet{Hernquist1993} and \citet{Springel2005}. The full set-up parameters are listed in Table~2 of \citet{Kannan2020} but we outline the most relevant details here. The dark matter halo is modelled as a static background gravitational field that is not impacted by the baryonic physics. The dark matter halo and the bulge are modelled as Hernquist profiles \citep{Hernquist1990}. The simulation box size is 600\,kpc and the initial radial profiles of the gas and stellar discs are exponential with effective radii of 6\,kpc and 3\,kpc, respectively. The vertical profile of the stellar disc follows a sech$^2$ functional form with a scale height of 300\,pc and initial stellar ages are taken to be 5\,Gyr to minimize spurious photoionization. The initial gas fraction is 10 per cent and the gas particle distributions are computed self-consistently assuming hydrostatic equilibrium. The initial gas temperature is set to $10^4\,\text{K}$ and the initial metallicity to $1\,\Zsun$. The production of new metals is turned off in order to suppress unrealistic gas metallicities, caused by the lack of cosmological gas inflow into the disc. The simulation is run with a stellar mass resolution of $2.8 \times 10^3\,\Msun$ and a gas mass resolution of $1.4 \times 10^3\,\Msun$. The corresponding gravitational softening length is $\varepsilon_{\star} = 7.1\,\text{pc}$ and the simulation is run for approximately 1\,Gyr.

To illustrate the typical properties of the isolated disc simulation, in Fig.~\ref{fig:rho_T_D_HI}, we show projected images of the MW galaxy at 1\,Gyr for both face-on and edge-on views, obtained using an adaptive quadrature ray-tracing integration scheme implemented in \textsc{colt} to guarantee convergence for small-scale structures. Specifically, we include the gas surface density $\Sigma_\text{gas} \equiv \int \rho\,\text{d}\ell$ where $\ell$ denotes the line-of-sight (LOS) projection path, with stars shown as whitened pixels with the transparency scaled by the logarithmic intrinsic ionizing photon rate $\dot{N}_\text{ion}$. We also give mass-weighted projections defined as $\langle f \rangle_m \equiv \int f \rho\,\text{d}\ell / \Sigma_\text{gas}$ for the gas temperature $T$, dust-to-gas ratio $\mathcal{D}$ relative to the canonical MW value (0.01), and neutral hydrogen fraction $x_\text{\HI} \equiv n_\text{\HI} / n_\text{H}$. These quantities highlight the ISM structure consisting of a patchwork of large hot ionized bubbles surrounded by dense clumps, filaments, and walls providing escape channels and absorption barriers for radiative transfer calculations. The kinematic structure is also important in setting, shifting, and broadening hydrogen emission line profiles. Therefore, we also calculate the mass-weighted projected gas velocity components for each camera unit vector, $\{v_\text{LOS}, v_{\perp,x}, v_{\perp,y}\} \equiv \langle \bmath{v} \bmath{\cdot} \{\bmath{n}_\text{LOS}, \bmath{n}_{\perp,x}, \bmath{n}_{\perp,y}\} \rangle_m$, corresponding to directions out of the image plane towards the viewer as well as horizontal and vertical image coordinate axes, respectively. In Fig.~\ref{fig:v_4}, we show for the same face-on and edge-on views the LOS velocity $v_\text{LOS}$ and the LOS velocity dispersion $\sigma_\text{LOS} \equiv (\langle (\bmath{v} \bmath{\cdot} \bmath{n}_\text{LOS})^2 \rangle_m - v_\text{LOS}^2)^{1/2}$. To better visualize the perpendicular velocity components with respect to the image plane, we utilize the image coordinates $\{x,y\}$ and define the cylindrical radius $\varrho \equiv (x^2 + y^2)^{1/2}$ to conveniently express the radial velocity $v_{\perp,\text{rad}} \equiv (x v_{\perp,x} + y v_{\perp,y}) / \varrho$ and perpendicular counter-clockwise rotational velocity $v_{\perp,\text{rot}} \equiv (x v_{\perp,y} - y v_{\perp,x}) / \varrho$, both of which are shown in Fig.~\ref{fig:v_4}. We note that there will be an additional bias when converting to an actual line profile due to $\propto \rho^2$ emission, but the LOS velocity statistics provide relevant information to guide our expectations. Likewise, the perpendicular velocity components are important for Ly$\alpha$ transport due to resonant scattering in the comoving frame of the gas flows. Additional properties of the simulation are discussed in later sections along with the radiative transfer connections.

\subsection{Radiative transfer of ionizing photons}
\label{subsec:rt-ion}

The dominant production mechanism for hydrogen line emission is via cascade recombination of recently ionized hydrogen atoms. While pursuing this study, we found that the on-the-fly ionization states overestimate the recombination emission due to not fully resolving the temperature and density substructure of a fraction of the young \HII regions,\footnote{It is more difficult to resolve the Str{\"o}mgren radius at higher densities given that $r_\text{s} \propto n_\text{H}^{-2/3}$ \citep[see the discussion corresponding to equation~7 of][]{Kannan2020}. Thus, for a fixed star formation density criterion (in our case $n = 10^3\,\text{cm}^{-3}$) this issue would become less apparent by improving the gas particle resolution. Otherwise, explicit correction factors are needed for more accurate modelling of unresolved \HII regions \citep[e.g. see][]{Jaura2020}.} which is a challenging problem for radiation hydrodynamics simulations in general (see Appendix~\ref{appendix:ion_accuracy} for further discussion). We emphasize that this transient numerical phenomenon does not significantly affect the overall dynamics of the simulation, which is largely determined by sub-grid feedback mechanisms. The role of photoheating is relatively minor in low-gas surface density galaxies like the Milky Way but becomes important in high-redshift analogues, so it is essential to rectify this for cosmological runs. In the meantime, these simulations capture the multiphase density--temperature states more accurately and self-consistently than equivalent simulations without such sophisticated radiation hydrodynamics and ISM physics modelling. Still, it is non-trivial to robustly identify and reconcile the small fraction of unreliable ionization states for our detailed spatially-resolved study of reprocessed line emission. Therefore, we performed post-processing ionization equilibrium calculations with \textsc{colt}, which was implemented for this study as an MCRT module mirroring the physics of galaxy formation simulations as detailed in \citet{Rosdahl2013} and \citet{Kannan2019}. The MCRT approach is ideal for our application as it employs subresolution ray-tracing, preserves photon directionality, incorporates scattering physics, which generates accurate ionization states for line emission modelling.

Specifically, we retain the gas internal energy as this is already faithfully modelled\footnote{We note that the conversion between gas internal energy and temperature depends on the electron abundance as this affects the mean molecular weight.}, but we iteratively recalculate the ionizing radiation field in three bands (\HI, \HeI, and \HeII, or 13.6, 24.59, and 54.42 eV, respectively) and update the ionization states assuming photoionization equilibrium \citep{Katz1996}. As we require accurate line luminosities, we stop this process of alternating between updating the ionization states and the radiation field when the global recombination emission is converged within a specified relative difference ($<0.1\%$). For consistency with the existing simulations we assume Case B recombination rates but the MCRT framework allows us to include continuous dust absorption and anisotropic scattering. As with the original simulations the ionization solver also includes collisional ionization and a meta-galactic UV background \citep{Faucher-Giguere2009} with a self-shielding prescription \citep{Rahmati2013}. Due to the high resolution of these simulations we first run preconditioning MCRT iterations with $10^7$ photon packets until reaching $1$ per cent convergence, followed by final iterations with $10^8$ packets stopping at $0.1$ per cent convergence. We have tested that this is adequate to represent the ionizing radiation field based on sampling from the age and metallicity dependent stellar SEDs in terms of position, direction, and frequency. All calculations in this study are based on the updated (MCRT-based) ionization states, which explicitly conserve photons for the generation of recombination line emission.

As this is the first paper describing the \textsc{colt} ionization implementation we now provide additional technical details. The emission from each star particle is calculated from age--metallicity tabulated SEDs, which in this case is taken from the high-resolution \citet{BruzualCharlot2003} model for consistency with the original simulations although we note that other SEDs are implemented as well. If the total rate of ionizing photons from all stars is $\dot{N}_\text{ion}^\text{tot}$ then the initial weight of each packet is $w_{0,j} \equiv \dot{N}_{\text{ion},j} / \dot{N}_\text{ion}^\text{tot}$, so by construction the initial weights sum to unity $\sum w_{0,j} = 1$. In practice, we construct the cumulative distribution function to assign photons to stellar sources. The photon packets are inserted employing a power-law luminosity boosting technique to better sample emission from dim stars according to $\propto L^\gamma$ normalized to conserve energy. We use an exponent of $1/2$ for the ionization calculations. The emission direction is isotropic and the initial position corresponds to the location of the star. We employ native ray-tracing through the
% three-dimensional
Voronoi tesellation \citep[implemented in \textsc{colt} for][]{SmithDCBH2017}. We use temperature dependent collisional ionization and recombination coefficients from \citet{Cen1992} and \citet{HuiGnedin1997}. Frequency dependent photoionization cross-sections are taken from \citet{Verner1996}, treating the transport of radiation in the grey approximation. For notational compactness we introduce the photon rate integral defined by
\begin{equation}
  \int_i f(\nu) \equiv \int_{\nu_{\text{min},i}}^{\nu_{\text{max},i}} \text{d}\nu \frac{4\pi J_\nu}{h\nu} f(\nu) \, ,
\end{equation}
where the subscript $i$ denotes the particular band covering the unique frequency range $\nu \in [\nu_{\text{min},i}, \nu_{\text{max},i}]$. The rates for ionizing photons, photoionization, and photoheating are respectively given by
\begin{equation}
  \left\{\dot{N}_{\text{ion},i}, \; \Gamma_{x,i}, \; \mathcal{E}_{x,i}\right\} \equiv \int_i \left\{ 1, \; \sigma_x, \; \sigma_x (h\nu - h\nu_x) \right\} \, ,
\end{equation}
where the subscript $x \in \{\HI, \HeI, \HeII\}$ denotes the species. The total emission rate for each star is simply $\dot{N}_\text{ion} = \sum_i \dot{N}_{\text{ion},i}$. For reference the average photoionization cross-section and average energy transferred to the gas per ionizing photon is
\begin{equation}
  \sigma_{x,i} \equiv \frac{\Gamma_{x,i}}{\dot{N}_{\text{ion},i}} \qquad \text{and} \qquad \varepsilon_{x,i} \equiv \frac{\mathcal{E}_{x,i}}{\Gamma_{x,i}} \, ,
\end{equation}
and the effective absorption coefficient for transport is given by
\begin{equation}
  k_{\text{ion},i} = \sum_x n_x \sigma_{x,i} \, .
\end{equation}
Likewise, the dust distribution is self-consistently taken from the simulation such that the local dust absorption coefficient is
\begin{equation}
  k_{\text{d},i} = \kappa_{\text{d},i} \mathcal{D} \rho \, ,
\end{equation}
where the dust opacity is $\kappa_{\text{d},i} \equiv \int_i \kappa_{\text{d},\nu} / \dot{N}_\text{ion}$ in $\text{cm}^2/\text{g}$ of dust and the dust-to-gas ratio is $\mathcal{D}$.

Both photoionization and dust absorption are treated as continuous processes, which is a standard MCRT variance reduction technique. This is achieved by multiplying the photon weight $w_j$ by the transmission fraction $e^{-\tau_\text{a}}$ each time the photon moves a distance $\Delta \ell$, where for notational simplicity the traversed absorption optical depth is $\tau_\text{a} \equiv k_\text{a} \Delta \ell$. The dust absorption is separated according to the scattering albedo $A_i \equiv \int_i A(\nu) / \dot{N}_\text{ion}$, which allows us to write the combined absorption coefficient for each path segment as $k_\text{a} = k_{\text{ion},i} + k_\text{a,d}$ with $k_\text{a,d} = (1 - A_i) k_{\text{d},i}$. We terminate photon packets when the weight falls below $10^{-14}$ times the global intrinsic emission rate, and warn that significantly less stringent thresholds can lead to artificial absorption. Dust scattering distances are determined by drawing the scattering optical depth from an exponential distribution with the scattering coefficient for each path segment given by $k_\text{s} = A_i k_{\text{d},i}$. Anisotropic dust scattering is included based on the Henyey--Greenstein phase function with an asymmetry parameter of $g_i = \langle \cos\theta \rangle_i \equiv \int_i g(\nu) / \dot{N}_\text{ion}$. All of these frequency-dependent properties are based on the fiducial Milky Way dust model from \citet{Weingartner2001}.

We employ path-based estimators for calculating photon interaction rates, which are determined by the local properties of each gas cell and photon source SED. The absolute photoionization and photoheating rates for each species are
\begin{equation} \label{eq:photoheating}
  \left\{\mathcal{I}_x, \; \mathcal{H}_x\right\} = \sum_{\text{paths},i} w_j \left\{\Gamma_{x,i}, \; \mathcal{E}_{x,i}\right\} \frac{n_x}{k_\text{a}} \left( 1 - e^{-\tau_\text{a}} \right) \, ,
\end{equation}
where $w_j$ is the weight at the beginning of the segment and the sum is over all frequency bins and photon paths within the given cell. We note that the photoionization equilibrium calculation considers the rate per species so the actual quantity calculated by \textsc{colt} internally is divided by the number of interacting species within the cell volume, i.e. $n_x V$. Also, we only track the total photoheating rate over all species; i.e. $\mathcal{H} = \sum_x \mathcal{H}_x$. Finally, for each photon we also track the amount of the initial weight that is removed by dust absorption as
\begin{equation}
  w_{\text{abs},j} = \sum_\text{paths} w_j \frac{k_\text{a,d}}{k_\text{a}} \left( 1 - e^{-\tau_\text{a}} \right) \, ,
\end{equation}
which can then be used to compute global statistics for the pre-absorption of ionizing photons by dust as $f_\text{abs}^\text{ion} \equiv \sum_j w_{\text{abs},j}$. To account for every absorption channel we also implement analogous photon trackers for the weight removed by \HeI and \HeII ionization. Furthermore, the average distance to absorption provides a powerful metric to understand the photon transport physics. Thus, we aim to calculate the following path integral over the trajectory $\ell$:
\begin{equation} \label{eq:mean_dist}
  \langle \ell \rangle \equiv \ell^{-1} \int_0^\ell \ell' e^{-k_\text{a} \ell'} \text{d}\ell' \, .
\end{equation}
In a constant density medium we get $\langle \ell \rangle = [1 - (1 + \tau_\text{a}) e^{-\tau_\text{a}}] \ell /\tau_\text{a}^2$, which in the optically thin limit of $\tau_\text{a} \ll 1$ is slightly below the mid-point $\langle \ell \rangle \approx \ell (1/2 - \tau_\text{a}/3)$, but reveals strong local biasing when absorption is significant. In the case of MCRT discretization, the weight $w_j$ is continually reduced via factors of $e^{-\tau_\text{a}}$. Therefore we split up the integral, such that any given path segment defined in terms of the starting $\ell_j$ and ending $\ell_{j+1} = \ell_j + \Delta \ell_j$ lengths will be
\begin{align}
  \langle \ell \rangle_j
  &= (w_{0,j} \ell)^{-1} \sum_\text{paths} w_j \int_0^{\Delta\ell_j} (\ell_j + \ell') e^{-k_\text{a} \ell'} \text{d}\ell' \notag \\
  &= (w_{0,j} \ell)^{-1} \sum_\text{paths} w_j \left[ 1 + \tau_j - (1 + \tau_{j+1}) e^{-\Delta\tau_j} \right] / k_j^2 \notag \\
  &\approx (w_{0,j} \ell)^{-1} \sum_\text{paths} w_j \Delta\ell_j \left[ \ell_j + \frac{\Delta\ell_j}{2} - \Delta\tau_j \left( \frac{\ell_j}{2} + \frac{\Delta\ell_j}{3} \right) \right] \, ,
\end{align}
where $\tau_j = k_j \ell_j$ and $\tau_{j+1} = k_j \ell_{j+1}$, and the final expression is only valid for segments with $\Delta\tau_j \ll 1$. Finally, the global mean distance averaged over all photons is $\langle \ell \rangle = \sum_j w_{0,j} \langle \ell \rangle_j$.

We emphasize that the grey approximation is not strictly necessary in our implementation, but helps reduce noise in sampling photon frequencies. Future applications may adopt a continuous frequency approach if beneficial for including additional physics \citep[e.g. see][]{Vandenbroucke2018}. An advantage of the MCRT approach is that each photon packet has unique ionization and dust properties inherited from the SED of the stellar source. In comparison, moment-based transport methods typically employ global cross-sections and heating rates. For reference we provide a summary of the global ionization statistics in Table~\ref{tab:time_avg_ion}. Throughout this paper time-averaged calculations utilize all simulation snapshots starting from a time of $400$\,Myr, which is well after the system has relaxed from the initial violent episodes of star formation.

\subsection{Radiative transfer of hydrogen emission lines}
\label{subsec:rt}

We also employ \textsc{colt} for our post-processing MCRT emission line radiative transfer calculations. We calculate the resolved luminosity caused by radiative recombination as
\begin{equation} \label{eq:L_rec}
  L_X^\text{rec} = h \nu_X \int P_{\text{B},X}(T,n_e) \alpha_\text{B}(T)\,n_e n_p\,\text{d}V \, ,
\end{equation}
where $X \in \{\text{Ly}\alpha, \text{H}\alpha, \text{H}\beta\}$ denotes the line, the energy at line centre is $h \nu_X = \{10.2, 1.89, 2.55\}$\,eV, and the number densities $n_e$ and $n_p$ are for free electrons and protons, respectively. For consistency with the ionization equilibrium calculations we use the same case B recombination coefficient $\alpha_\text{B}$ from \citet{HuiGnedin1997}. Departing from previous versions of \textsc{colt} we now take the conversion probability per recombination event from \citet{StoreyHummer1995}, for reference $P_\text{B}(10^4\,\text{K}) \approx \{0.68, 0.45, 0.12\}$, who provide accurate recombination rate tables as a function of temperature and electron number density. To avoid spuriously high Balmer decrements in artificially cold \HII regions we impose a temperature floor for $P_\text{B}$ in these simulations of $7000$\,K as a reasonable lower limit for ionized gas with metal line cooling \citep{Pequignot2001}.

We also include the contribution of radiative de-excitation of collisional excitation of neutral hydrogen by free electrons. As recommended by previous studies \citep[e.g.][]{Osterbrock2006,Hansen2018} we take into account all excitations for the $n \leq 5$ levels using Maxwellian-averaged effective collision strengths $\Upsilon_{ij}$ taken from \citet{Anderson2000,Anderson2002}. Including cascades from higher levels is arguably more consistent and accurate than the common approach of adopting $q_{1s2p}$ for Ly$\alpha$ \citep[e.g.\ from][]{Scholz1991} and $q_{13}$ for H$\alpha$ \citep[e.g.\ from][]{Aggarwal1983}, specifically we find Ly$\alpha$ collisional excitation rates are about $10$--$30\%$ too low. Additionally, our detailed treatment of collisional excitation allows us to calculate rates for other emission lines such as H$\beta$. In Appendix~\ref{appendix:col_rates}, we provide fitting formulae for the temperature dependent rate coefficients $q_{\text{col},X}$, which then allow us to calculate the resolved luminosity caused by collisional excitation as
\begin{equation} \label{eq:L_col}
  L_X^\text{col} = h \nu_X \int q_{\text{col},X}(T)\,n_e n_\text{\HI}\,\text{d}V \, ,
\end{equation}
with $n_\text{\HI}$ denoting the number density of neutral hydrogen. While pursuing this study, we found that the temperature of some dense photoheated gas can be slightly too high, e.g. increasing by a factor of two to an order-of-magnitude excess in collisional excitation emission due to the strong temperature dependence of the rate coefficient \citep[for a discussion of related issues see][]{Osterbrock2006,Faucher-Giguere2010}. Such issues are commonly encountered when combining complex thermochemistry and radiation hydrodynamics at high densities \citep[e.g.][]{Jaura2020}. Improvements to resolving such non-linear cooling physics without resorting to prohibitively expensive time stepping criteria for the explicit \textsc{arepo-rt} solver are being investigated. For our current study, we devised a scheme to eliminate unphysical line cooling emission while retaining the contributions from cells requiring no correction. We first calculate the standard collisional excitation luminosity from equation~(\ref{eq:L_col}) and limit this value by the maximal luminosity emitted through the line channel:
\begin{equation} \label{eq:L_col_max}
  L_{X,\text{max}}^\text{col} = f_\text{lim} \mathcal{H}_\text{\HI} \frac{h\nu_X q_{\text{col},X}}{\psi_\text{\HI}} \, .
\end{equation}
Here we assume that photoheating is the dominant mechanism bringing dense gas to $\gtrsim 10^4\,\text{K}$. Thus, we include a limiting factor $f_\text{lim}$, which we set to $10$ to allow for other heating sources and out of equilibrium effects. The photoheating rate $\mathcal{H}_\text{\HI}$ is from equation~(\ref{eq:photoheating}) and is scaled by the fraction of cooling occurring through the line relative to the total cooling rate $\psi_\text{\HI}$ from equation~(\ref{eq:cooling}). We note that some previous studies facing similar issues often take more conservative approaches by limiting the temperature for the rate coefficient \citep{Kim2013} or ignoring emission from cells with unresolved cooling \citep{Mitchell2021,Garel2021}. We tested several alternative correction methods and found ours to be the most robust and physically motivated for the present simulations.

\begin{table}
  \centering
  \caption{Time-averaged ionization properties for the simulation defined as $\mu_f = \int f(t)\,\text{d}t / \int \text{d}t$, also including the 1$\sigma$ standard deviation for select quantities, defined as $\sigma_f^2 = \int (f(t)-\mu_f)^2\,\text{d}t / \int \text{d}t$. In cases with LOS radiative transfer fluctuations we provide the median and asymmetric 1$\sigma$ (68.27\%) time-weighted confidence levels, based on collecting the escaped photon packets into $3072$ healpix directions of equal solid angle. Specifically, we include the total ($> 13.6\,\text{eV}$) and band ($i \in \{\HI, \HeI, \HeII\}$) values for the intrinsic photon rate ($\dot{N}_{\text{ion},i}$), emitted luminosity ($L_{\text{ion},i}$), fractional band photon rate ($\dot{N}_{\text{ion},i} / \dot{N}_\text{ion}$), mean energy per photon ($e_i$), photoionization cross-sections ($\sigma_{x,i}$), average energy transferred to the gas per photon ($\varepsilon_{x,i}$), and dust opacity ($\kappa_{\text{d},i}$), scattering albedo ($A_i$), and scattering cosine ($\langle\cos\theta\rangle_i$). In the lower portion we report statistics based on the radiative transfer outcomes, including the fractions of photons ionizing hydrogen and helium ($f_{\text{\HI},i},f_{\text{\HeI},i},f_{\text{\HeII},i}$), fraction of photons absorbed by dust ($f_{\text{abs},i}$), ionizing escape fraction ($f_{\text{esc},i}$), degree of isotropy ($F_{\text{LOS},i} / F_{\Omega,i}$), and fractional band escaped photon rate ($\dot{N}_{\text{ion},i}^\text{esc} / \dot{N}_\text{ion}^\text{esc}$). We note that the sum of photon outcome fractions falls slightly below unity because the sightline median escape fraction is lower than the mean value of $f_\text{esc} \approx 7.9\pm3.7\%$.}
  \label{tab:time_avg_ion}
  \addtolength{\tabcolsep}{-3pt}
  \renewcommand{\arraystretch}{1.1}
  \begin{tabular}{@{} l cccc @{}}
    \hline
    Quantity & $> 13.6\,\text{eV}$ & \HI & \HeI & \HeII \\
    \hline
    $\log \dot{N}_{\text{ion},i}\ [\text{s}^{-1}]$ & $53.6\pm0.2$ & $53.5\pm0.2$ & $52.9\pm0.2$ & $51.6\pm0.2$ \\
    $\log L_{\text{ion},i}\,[\text{erg\,s}^{-1\!}]\!\!\!\!\!\!$ & $43.1\pm0.2$ & $42.9\pm0.2$ & $42.7\pm0.2$ & $41.6\pm0.2$ \\
    $\dot{N}_{\text{ion},i} / \dot{N}_\text{ion}\ [\%]$ & $100$ & $77.3\pm0.4$ & $21.7\pm0.4$ & $0.96\pm0.19$ \\
    $e_i\ [\text{eV}]$ & $21.4$ & $17.8$ & $32.1$ & $73.3$ \\
    $\sigma_{\HI,i}\ [\text{cm}^2]$ & $2.74 \times 10^{-18}$ & $3.48 \times 10^{-18}$ & $2.23 \times 10^{-19}$ & $4.08 \times 10^{-21}$ \\
    $\sigma_{\HeI,i}\ [\text{cm}^2]$ & $4.92 \times 10^{-18}$ & 0 & $5.13 \times 10^{-18}$ & $1.82 \times 10^{-19}$ \\
    $\sigma_{\HeII,i}\ [\text{cm}^2]$ & $8.51 \times 10^{-19}$ & 0 & 0 & $8.51 \times 10^{-19}$ \\
    $\varepsilon_{\HI,i}\ [\text{eV}]$ & $3.20$ & $2.97$ & $15.9$ & $51.8$ \\
    $\varepsilon_{\HeI,i}\ [\text{eV}]$ & $5.63$ & 0 & $5.57$ & $42.2$ \\
    $\varepsilon_{\HeII,i}\ [\text{eV}]$ & $11.7$ & 0 & 0 & $11.7$ \\
    $\kappa_{\text{d},i}\ [\text{cm}^2\text{g}^{-1}]\!$ & $9.79 \times 10^4$ & $1.13 \times 10^5$ & $4.77 \times 10^4$ & $2.69 \times 10^4$ \\
    $A_i$ & $0.241$ & $0.216$ & $0.325$ & $0.422$ \\
    $\langle\cos\theta\rangle_i$ & $0.738$ & $0.699$ & $0.866$ & $0.969$ \\
    \hline
    $f_{\text{\HI},i}\ [\%]$ & $55.9\pm6.5$ & $63.1\pm7.6$ & $32.5\pm3.1$ & $0.8\pm0.3$ \\
    $f_{\text{\HeI},i}\ [\%]$ & $7.9\pm1.1$ & $-$ & $36.4\pm4.7$ & $1.7\pm0.6$ \\
    $f_{\text{\HeII},i}\ [\%]$ & $0.8\pm0.2$ & $-$ & $-$ & $83.6\pm4.4$ \\
    $f_{\text{abs},i}\ [\%]$ & $27.5\pm6.0$ & $29.5\pm6.1$ & $21.3\pm6.0$ & $4.4\pm1.0$ \vspace{.05cm} \\
    $f_{\text{esc},i}\ [\%]$ & $5.6^{+9.6}_{-5.3}$ & $4.9^{+9.4}_{-4.7}$ & $7.8^{+10.8}_{-7.1}$ & $4.0^{+16.1}_{-3.7}$ \vspace{.1cm} \\
    $F_{\text{LOS},i} / F_{\Omega,i}$ & $1^{+0.84}_{-0.95}$ & $1^{+0.88}_{-0.97}$ & $1^{+0.74}_{-0.92}$ & $1^{+1.16}_{-0.95}$ \vspace{.05cm} \\
    $\dot{N}_{\text{ion},i}^\text{esc} / \dot{N}_\text{ion}^\text{esc}\ [\%]$ & $100$ & $70.9\pm2.7$ & $27.9\pm2.6$ & $1.17\pm0.35$ \\
    \hline
  \end{tabular}
  \addtolength{\tabcolsep}{3pt}
  \renewcommand{\arraystretch}{0.9090909090909090909}
\end{table}

The line photon packets are inserted according to the cell recombination and collisional luminosities. We employ $5 \times 10^7$ ($10^8$) photon packets including power-law luminosity boosting with an exponent of $3/4$ ($1/2$) for Ly$\alpha$ (H$\alpha$, H$\beta$) to better sample emission from low-surface brightness regions. The photon locations are then drawn uniformly within the selected Voronoi cell, the direction is isotropic, and the frequency is taken from the thermal velocity distribution (also accounting for natural broadening). To derive line equivalent widths we also include $10^8$ stellar continuum photon packets with the same sampling strategy but with the frequency sampled uniformly in $\Delta v \in [-3000, 3000]\,\text{km\,s}^{-1}$ with respect to the comoving gas frame. The spectral luminosities $L_{\lambda,\text{cont}}$ for each line are tabulated by age and metallicity with values calculated as the logarithmic average of the SEDs over windows of $50$\,\AA\ centred on the reference wavelengths. The dust density is once again self-consistently taken from the simulations such that the local dust absorption coefficient is $k_{\text{d},X} = \kappa_{\text{d},X} \mathcal{D} \rho$, where the dust opacity is $\kappa_{\text{d},X} = \{58020, 6627, 10220\}\,\text{cm}^2/\text{g}$ of dust for Ly$\alpha$, H$\alpha$, and H$\beta$, respectively. Notably, we do not incorporate an ad-hoc dust survival factor in ionized gas as dust destruction mechanisms are included in the simulations, however see Appendix~\ref{appendix:fion_test} for a discussion of the impact of also including this factor for Ly$\alpha$ radiative transfer observables. Dust absorption is treated continuously by tracking the cumulative absorption optical depth. The dust scattering albedos are $A = \{0.3251, 0.6741, 0.6650\}$ and the asymmetry parameters are $g = \langle \cos\theta \rangle = \{0.6761, 0.4967, 0.5561\}$, based on the same fiducial Milky Way dust model from \citet{Weingartner2001}. For the transport physics we also include microturbulence with $v_\text{turb} = 10\,\text{km\,s}^{-1}$, such that the effective thermal velocity becomes $b = (v_\text{th}^2 + v_\text{turb}^2)^{1/2}$ where $v_\text{th}^2 = 2 k_\text{B} T / m_\text{H}$.

The \textsc{colt} output files capture the escaping photon properties and next-event estimation (peel-off) calculations for high-signal-to-noise images at very high-(10\,pc) pixel resolution oriented in face-on and edge-on directions, including first and second frequency moment maps. We also calculate ray-tracing based images of the intrinsic and dust attenuated line emission based on an adaptive convergence algorithm similar to the one described in Appendix~A of \citet{Yang2020}. This method eliminates the noise due to Monte Carlo sampling at the expense of treating dust as purely absorbing; i.e. assuming an albedo of $A = 0$. For each of these lines and cameras we retain information about the collisional-to-total emission, which allows us to calculate statistics and images isolating the collisional and recombination contributions without running separate simulations. The remaining implementation details, including for Ly$\alpha$ resonant scattering, are described in \citet{Smith2015} or may be found in the \textsc{colt} source code and documentation.

\begin{figure}
  \centering
  \includegraphics[width=\columnwidth]{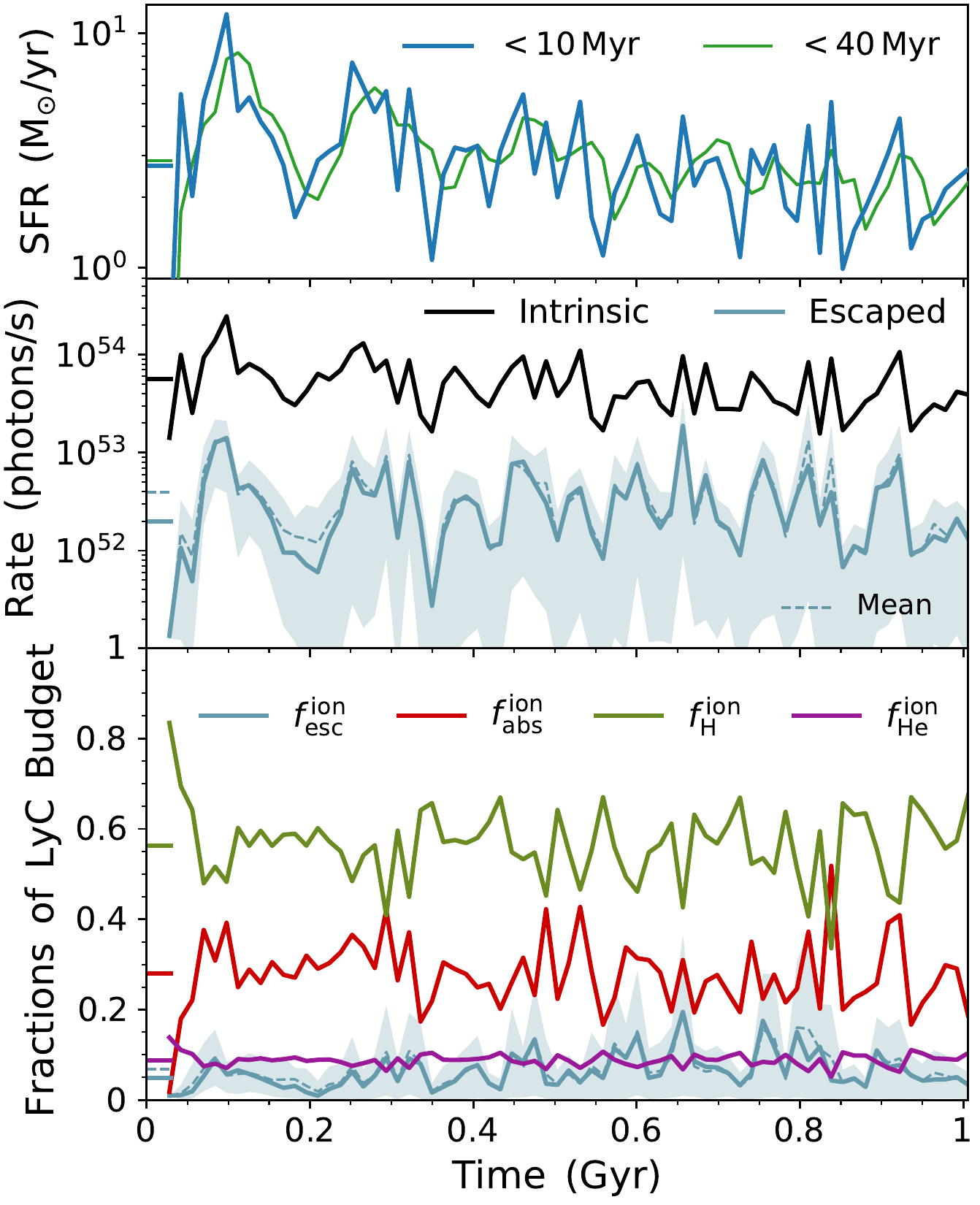}
  \caption{\textit{Top:} The evolution of the star-formation rate as probed by stars with ages less than 10\,Myr (blue) and 40\,Myr (green). \textit{Middle:} The corresponding intrinsic and escaped ionizing ($> 13.6\,\text{eV}$) photon emission rates. The shaded regions show the $1\sigma$ confidence levels considering the angular dependent escape properties across 3072 different viewing angles. \emph{Bottom:} The escape fractions (blue), dust pre-absorption fractions (red), helium ionization fractions (purple), and available budget for conversion to hydrogen recombination line emission (green) representing the majority outcome. Throughout this paper the coloured markers on the axes denote the median values.}
  \label{fig:f_esc_ion}
\end{figure}

\begin{figure}
  \centering
  \includegraphics[width=\columnwidth]{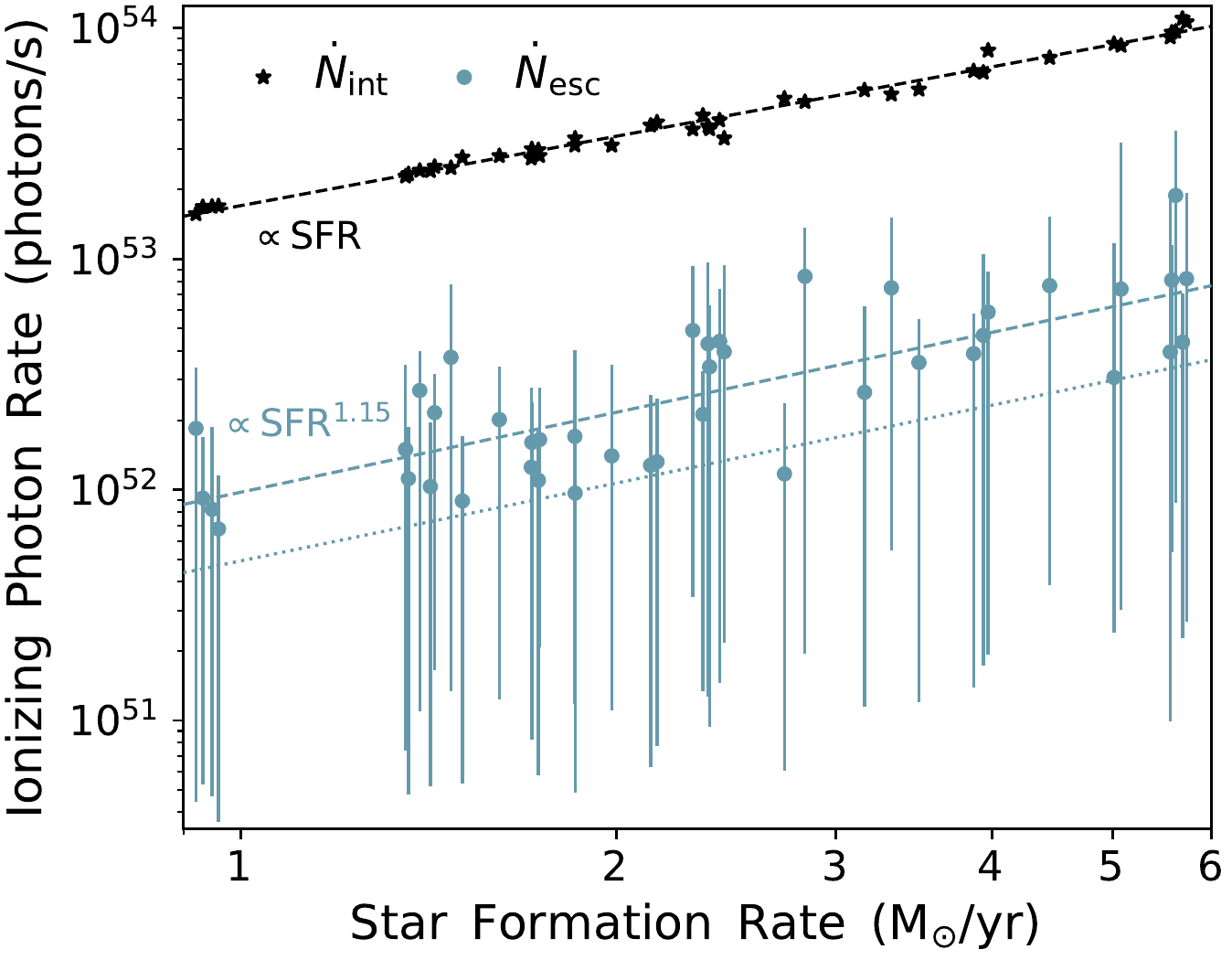}
  \caption{Correlation across different snapshots between the star-formation rate as probed by stars with ages less than 5\,Myr and ionizing photon rate ($\dot{N}$). The intrinsic emission has very little scatter around the expected scaling with SFR, while escaped rates vary significantly and show a super-linear relationship due to a slight enhancement of escape fractions at higher SFR. The dashed curves are power-law fits to the median values, and the dotted curve includes all sightlines as represented by the vertical $1\sigma$ confidence levels.}
  \label{fig:SFR_Ndot}
\end{figure}

\section{Radiative transfer results}
\label{sec:results}
We now explore the results of the radiative transfer calculations, which provide a picture of the time and LOS variability as well as the escape and absorption properties of ionizing and line photons. For reference in Tables~\ref{tab:time_avg_ion} and \ref{tab:time_avg_lines}, we provide the main intrinsic and emergent statistics for ionizing and line photons, respectively.

\subsection{Ionizing radiation}
In Fig.~\ref{fig:f_esc_ion}, we show the evolution of the SFR, intrinsic and escaping ionizing photon emission rates, and LyC escape fractions, including asymmetric $1\sigma$ confidence levels considering different viewing angles. The bursty nature of the emission is comparable in cadence and amplitude to the star formation activity, noting that $\langle \text{SFR} \rangle \approx 3\,\Msun\,\text{yr}^{-1}$. Given the assumed initial mass function and simulated distribution of stellar ages we calculate a time-averaged intrinsic ionizing ($> 13.6\,\text{eV}$) photon emission rate of $\langle \dot{N}_\text{ion} \rangle \approx 4.66 \times 10^{53}\,\text{s}^{-1}$ and a luminosity of radiation from stars of $\langle L_\text{ion} \rangle \approx 1.60 \times 10^{43}\,\text{erg\,s}^{-1}$. Furthermore, from the radiative transfer outputs we calculate the time-averaged ionizing escape fraction as $\langle f_\text{esc}^\text{ion} \rangle \approx 7.9$ per cent (in non-cosmological isolated simulation environments we define this as the fraction of photons escaping the domain in terms of $\dot{N}_\text{ion}$) with significant fluctuations across sightlines biased by LyC leakage such that the median value reduces to $5.6^{+9.6}_{-5.3}$ per cent. Interestingly, we find the fraction of photons absorbed by dust is $\langle f_\text{abs}^\text{ion} \rangle \approx 27.5$ per cent and the fraction of photons ionizing helium is $\langle f_\text{He}^\text{ion} \rangle \approx 8.7$ per cent, such that only $\langle 1 - f_\text{esc}^\text{ion} - f_\text{abs}^\text{ion} - f_\text{He}^\text{ion} \rangle \approx 55.9$ per cent of the photons are available for conversion to hydrogen recombination line emission.

We emphasize that the consumption of ionizing photons by dust, helium, and global escape is a non-negligible fraction of the intrinsic budget, and these losses should be kept in mind in the remaining discussions throughout this paper. In particular, spectral energy distributions and cross-sections complicate intuition, e.g. the cross-section for \HeI is 23 times higher than \HI for the second band, and about 23\% of the ionizing radiation is capable of ionizing helium. A helium-to-hydrogen absorption fraction of $\approx 15\%$ is understandable, given that the dust absorption is higher for the \HI band (30\%) compared to \HeI band (21\%). Our values are in line with previous observational inferences \citep[e.g.][]{Inoue2001} and high-resolution periodic simulations exploring subparsec scale feedback \citep[e.g.][]{Kado-Fong2020}. Overall, the remarkable consistency of these values and predictable pattern of variations is a consequence of the idealized disc set-up and the behaviour of the feedback model.

\begin{figure}
  \centering
  \includegraphics[width=\columnwidth]{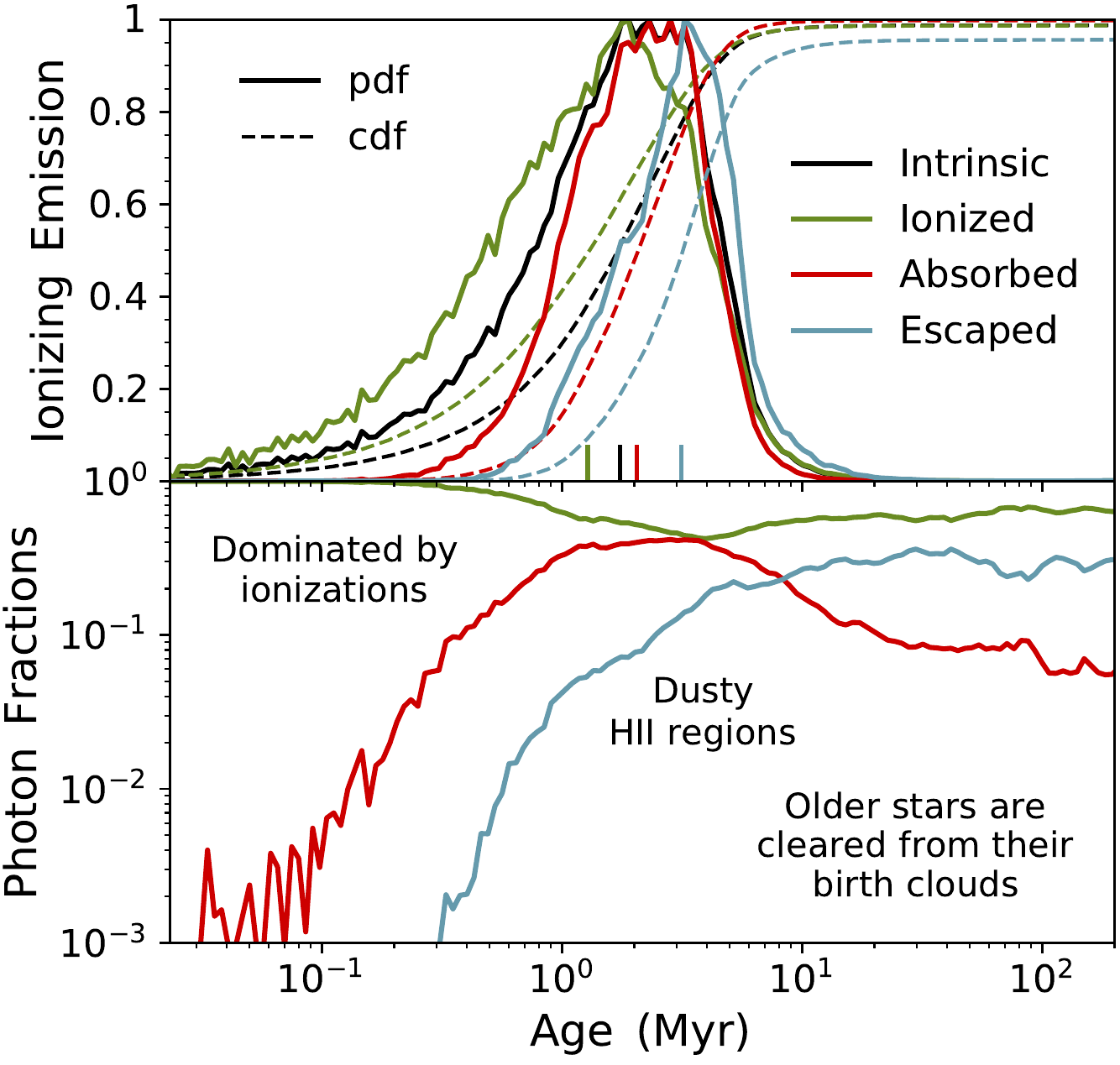}
  \caption{\textit{Top:} The relative distribution of stellar ages contributing to the ionizing photon rates, with (reverse) cumulative distribution functions shown as (dotted) dashed curves. The different coloured curves denote various radiative transfer outcomes, including intrinsic emission (black), direct ionization (green), dust absorption (red), and successful escape (blue). Ionizing radiation traces the youngest stellar populations with median ages of approximately $2$\,Myr (vertical markers). \textit{Bottom:} The corresponding ionization, absorption, and escape fractions as a function of stellar age, revealing clear trends for both younger and older stars as a consequence of the environmental evolution.}
  \label{fig:age_hist}
\end{figure}

\begin{figure}
  \centering
  \includegraphics[width=\columnwidth]{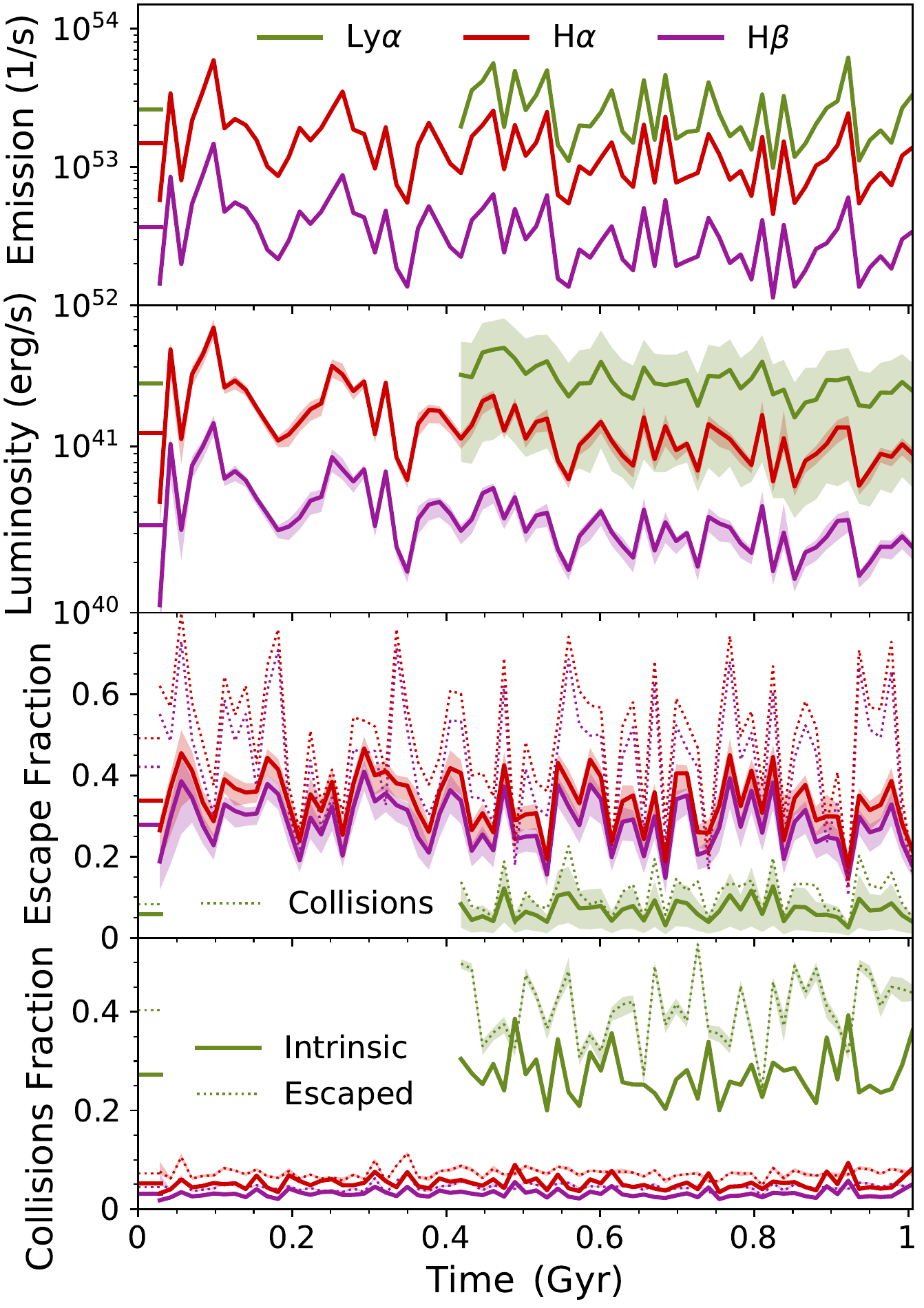}
  \caption{\emph{Top:} The evolution of the intrinsic photon emission rate $\dot{N}$ for the Ly$\alpha$ (green), H$\alpha$ (red), and H$\beta$ (purple) hydrogen emission lines. \emph{Second:} The escaping luminosity where shaded regions show the $1\sigma$ confidence levels considering different viewing angles. \emph{Third:} The corresponding escape fractions for each line, including for photons originating from collisional excitation emission. \emph{Bottom:} The fraction of intrinsic and escaped emission due to collisional excitation.}
  \label{fig:f_esc_lines}
\end{figure}

\begin{figure}
  \centering
  \includegraphics[width=\columnwidth]{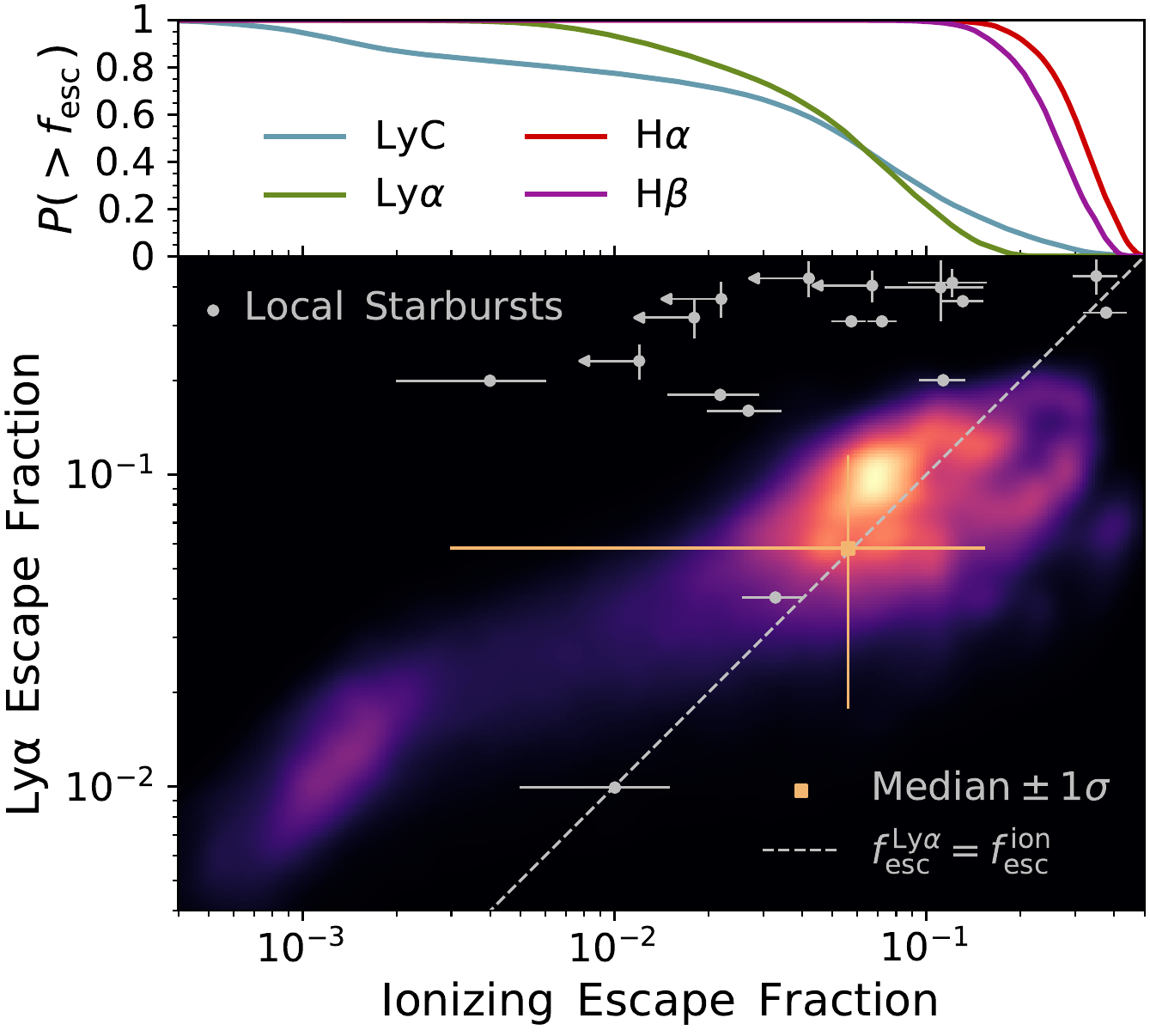}
  \caption{\textit{Top:} The probability in time and viewing angle that a given sightline has an escape fraction larger than a certain value $P(>f_\text{esc})$ for ionizing radiation (blue) and the Ly$\alpha$ (green), H$\alpha$ (red), and H$\beta$ (purple) hydrogen emission lines. Escape fractions for LyC and Ly$\alpha$ exhibit significant variation, while the H$\alpha$ and H$\beta$ lines are more predictable. \textit{Bottom:} Probability distribution of the directional LyC and Ly$\alpha$ escape fractions over all snapshots. Most sightlines have $f_\text{esc}^{\text{Ly}\alpha} > f_\text{esc}^\text{ion}$, consistent with theory and observation. For reference we show observational data from low-mass local starburst galaxies \citep[as reported by][]{Izotov2021}, although these LyC leakers typically have much higher specific SFRs than the Milky Way.}
  \label{fig:f_esc_cdf}
\end{figure}

\begin{figure*}
  \centering
  \includegraphics[width=\textwidth]{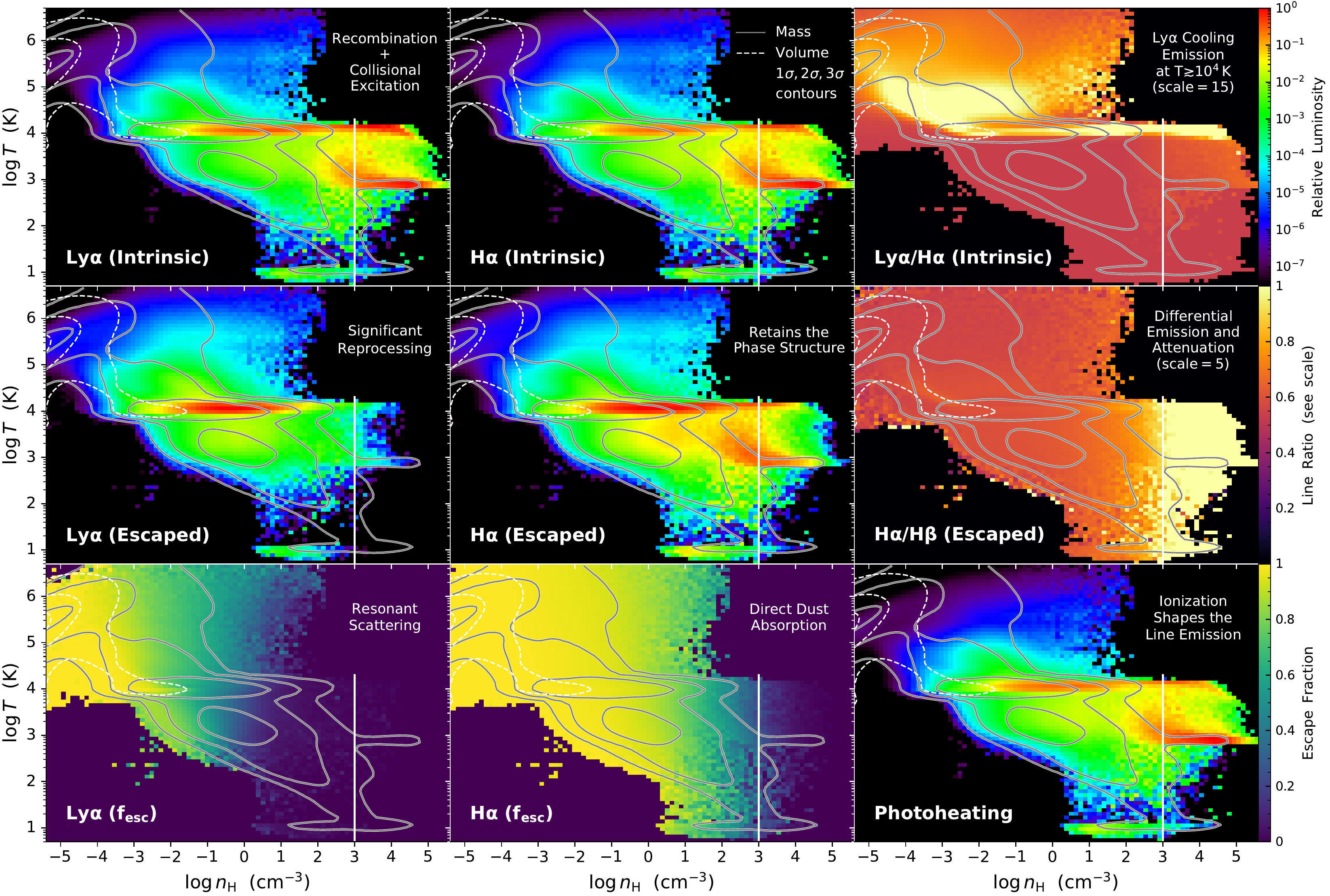}
  \caption{The density--temperature phase space dependence of numerous quantities related to the sources and physics of line radiative transfer. In the first two columns from top to bottom we illustrate the relative intrinsic luminosities, relative escaped luminosities, and escape fractions for the Ly$\alpha$ and H$\alpha$ lines, respectively. In the third column we show the $\text{Ly}\alpha / \text{H}\alpha$ intrinsic and $\text{H}\alpha / \text{H}\beta$ escaped line ratios and the relative photoheating rates. Most of the emission comes from dense \HII regions with a characteristic temperature of $T \sim 10^4$\,K. The impact of dust absorption can also be seen, including the impact of resonant scattering for Ly$\alpha$ photons. See the text for additional details but for reference we also provide mass and volume probability contours, vertical lines indicating the star formation threshold, and brief summaries in each panel to guide the narrative.}
  \label{fig:phase_nT}
\end{figure*}

The quasi-equilibrium set-up leads to only a factor of $\approx 6$ variation in SFR, so it is difficult to discern evolutionary trends. To make the time series data more transparent in Fig.~\ref{fig:SFR_Ndot}, we illustrate the strong correlation across different snapshots between the SFR as probed by stars with ages less than 5\,Myr and ionizing photon rate. We find the intrinsic emission has very little scatter around the expected scaling with SFR, while escaped rates vary significantly and reveal a super-linear relationship ($\text{SFR}^{1.15}$) due to a slight enhancement of escape fractions at higher SFR. We expect steeper slopes from lower-mass starburst galaxies with larger escape fractions and SFR variations.

To explore the sources of these photons in Fig.~\ref{fig:age_hist}, we show the relative distribution of stellar ages contributing to the ionizing photon rates, which is possible due to the source and weight tracking of Monte Carlo photon packets. For comparison we also include (reverse) cumulative distribution functions shown as (dotted) dashed curves. Various radiative transfer outcomes are shown, including intrinsic emission (black), direct ionization (green), dust absorption (red), and successful escape (blue). Ionizing radiation traces the youngest stellar populations with median and $1\sigma$ ages of $\{1.74^{+2.00}_{-1.19}, 1.28^{+1.92}_{-0.93}, 2.05^{+1.65}_{-0.99}, 3.13^{+2.32}_{-1.52}\}$\,Myr for each process, respectively. In the lower panel we show the corresponding ionization (green), absorption (red), and escape (blue) fractions as a function of stellar age. We can understand the trends in terms of the environmental evolution. The youngest stars are heavily obscured by neutral hydrogen gas such that the radiation is efficiently depleted by photoionizations. Eventually the \HII regions become well resolved and almost half of the photons are absorbed by dust before interacting with the gas. Finally, older stars have relatively higher escape fractions and lower dust absorption as they are cleared from their birth clouds. We emphasize that neutral hydrogen has a much higher cross-section than dust, specifically taking the values for \HI from Table~\ref{tab:time_avg_ion} and ignoring scattering gives
\begin{equation} \label{eq:k_ion_dust_ratio}
  \frac{k_{\text{ion},\HI}}{k_\text{d}} \approx \frac{x_\text{\HI} X \sigma_\text{\HI}}{\kappa_\text{d,\HI} \mathcal{D} m_\text{H}} \approx 1.17 \left( \frac{x_\text{\HI}}{10^{-3}} \right) \left( \frac{\mathcal{D}}{0.012} \right)^{-1} \, .
\end{equation}
Here we note that the dust-to-gas ratio is a strong function of gas density and we have chosen to normalize to the median value found within the highest density environments \citep{Kannan2020}. Thus, hydrogen must generally be highly ionized before dust becomes important. More practically, \HII regions must be well resolved in the sense that star particles fully ionize the cells in their immediate vicinity, which is facilitated by photoheating and other forms of feedback that over time reduce the gas density around young stars \citep{Kannan2020b}. From Fig.~\ref{fig:age_hist} this appears to be the case for stars with ages $\gtrsim 0.5\,\text{Myr}$. Furthermore, the absorption ratios we find are consistent with equation~(\ref{eq:k_ion_dust_ratio}) and more careful theoretical investigations of a multiphase ISM comprised of dusty \HII regions \citep{Draine2011}. Indeed we find many bright sources in dust optically thick regions ($\tau_\text{d} \gtrsim 1$), as quantified by the column densities for neutral hydrogen and surface densities for gas and dust. These are calculated by ray-tracing along 192 healpix sightlines from every star and weighting by the intrinsic photon rate, i.e. $N_\text{\HI} \equiv \sum \dot{N}_{\text{int},i} N_\text{\HI,i} / \sum \dot{N}_{\text{int},i}$. We obtain time-averaged values of $\langle \log N_\text{\HI}[\text{cm}^{-2}] \rangle \approx 22.3 \pm 0.3$ for neutral hydrogen, $\langle \log \Sigma_\text{gas}[\text{g\,cm}^{-2}] \rangle \approx -1.3 \pm 0.3$ for gas, and $\langle \log \Sigma_\text{dust}[\text{g\,cm}^{-2}] \rangle \approx -3.2 \pm 0.3$, which decrease by several orders of magnitude when weighted by the escaped photon rate. We note that our dust absorption is higher than previous studies such as \citet{Gnedin2008} and \citet{Mauerhofer2021} from cosmological simulations run down to $z \sim 3$, achieving minimum cell resolutions of $65\,\text{pc}$ and $14\,\text{pc}$, respectively. We attribute this to a combination of lower gas mass, lower metallicity, and lower resolution, all of which impact dust distributions in high density star-forming environments. Furthermore, in simulations without an on-the-fly dust model, the dust density is typically assumed to scale with metallicity adopting a small survival fraction in ionized regions (see Appendix~\ref{appendix:fion_test}). We emphasize that underestimating the importance of dust pre-absorption for any of these reasons still affects line emission even if subsequent photoionization implies the escape fraction remains low.

\begin{table}
  \centering
  \caption{Time-averaged line properties before radiative transfer (\textsc{int}) and after accounting for ISM scattering (\textsc{ism}) with a similar layout as Table~\ref{tab:time_avg_ion}. The quantities in the upper portion are the intrinsic photon rate $\dot{N}$, luminosity $L$, fraction of emission originating from collisional excitation $f_\text{col}$, and equivalent width (EW) of each line. The quantities in the lower portion are median and asymmetric $1\sigma$ confidence levels from all snapshots and sightlines for the escaping luminosity $L_\text{esc}$, fraction of observed photons arising from collisional excitation $f_\text{col}^\text{esc}$, line escape fraction $f_\text{esc}$, escape fraction of photons due to collisional excitation $f_\text{esc}^\text{col}$, escape fraction of the stellar continuum around the line $f_\text{esc}^\text{cont}$, degree of isotropy $F_\text{LOS} / F_\Omega$, half-light radii for line emission $R_{1/2}$ and stellar continuum $R_{1/2}^\text{cont}$, radial exponential scale lengths for line emission $R_\text{h}$ and stellar continuum $R_\text{h}^\text{cont}$, red-to-blue flux ratio $F_\text{red} / F_\text{blue}$, flux-weighted frequency centroid $\langle \Delta v \rangle$ and standard deviation $\sigma_{\Delta v}$, characteristic peak velocity offset $\Delta v_\text{peak}$, full-width at half-maximum (FWHM), and observed EW of each line.}
  \label{tab:time_avg_lines}
  \addtolength{\tabcolsep}{1pt}
  \renewcommand{\arraystretch}{1.1}
  \begin{tabular}{@{} lc ccc @{}}
    \hline
    Quantity & Type & Ly$\alpha$ & H$\alpha$ & H$\beta$ \\
    \hline
    $\log \dot{N}\ [\text{s}^{-1}]$ & \textsc{int} & $53.4\pm0.2$ & $53.0\pm0.2$ & $52.4\pm0.2$ \\
    $\log L\ [\text{erg\,s}^{-1}]$ & \textsc{int} & $42.6\pm0.2$ & $41.5\pm0.2$ & $41.0\pm0.2$ \\
    $f_\text{col}\ [\%]$ & \textsc{int} & $27.2\pm3.7$ & $5.3\pm1.1$ & $3.1\pm0.7$ \\
    $\text{EW}\ [\text{\AA}]$ & \textsc{int} & $49.4\pm12.9$ & $21.4\pm8.1$ & $5.3\pm2.0$ \\
    \hline
    $\log L_\text{esc}\ [\text{erg\,s}^{-1}]$ & \textsc{ism} & $41.4^{+0.22}_{-0.52}$ & $41.0^{+0.15}_{-0.16}$ & $40.5^{+0.15}_{-0.17}$ \vspace{.1cm} \\
    $f_\text{col}^\text{esc}\ [\%]$ & \textsc{ism} & $40.3^{+7.9}_{-7.4}$ & $7.3^{+0.9}_{-0.9}$ & $4.5^{+0.6}_{-0.6}$ \vspace{.1cm} \\
    $f_\text{esc}\ [\%]$ & \textsc{ism} & $5.8^{+5.6}_{-4.0}$ & $31.7^{+8.7}_{-8.3}$ & $26.1^{+8.5}_{-7.5}$ \vspace{.1cm} \\
    $f_\text{esc}^\text{col}\ [\%]$ & \textsc{ism} & $8.3^{+10.6}_{-5.8}$ & $48.3^{+15.3}_{-20.1}$ & $41.7^{+14.8}_{-19.0}$ \vspace{.1cm} \\
    $f_\text{esc}^\text{cont}\ [\%]$ & \textsc{ism} & $53.5^{+16.1}_{-27.3}$ & $98.7^{+2.3}_{-11.3}$ & $97.3^{+3.2}_{-15.2}$ \vspace{.1cm} \\
    $F_\text{LOS} / F_\Omega$ & \textsc{ism} & $1^{+0.62}_{-0.70}$ & $1^{+0.11}_{-0.13}$ & $1^{+0.13}_{-0.15}$ \vspace{.1cm} \\
    $R_{1/2}\ [\text{kpc}]$ & \textsc{ism} & $5.2^{+1.2}_{-1.0}$ & $3.8^{+1.5}_{-1.4}$ & $3.8^{+1.5}_{-1.4}$ \vspace{.1cm} \\
    $R_{1/2}^\text{cont}\ [\text{kpc}]$ & \textsc{ism} & $2.1^{+1.0}_{-1.0}$ & $3.3^{+0.7}_{-0.5}$ & $3.3^{+0.6}_{-0.5}$ \vspace{.1cm} \\
    $R_\text{h}\ [\text{kpc}]$ & \textsc{ism} & $5.6^{+2.1}_{-1.4}$ & $11.1^{+1.6}_{-1.8}$ & $11.2^{+1.6}_{-1.8}$ \vspace{.1cm} \\
    $R_\text{h}^\text{cont}\ [\text{kpc}]$ & \textsc{ism} & $12.9^{+1.9}_{-1.7}$ & $17.2^{+1.4}_{-1.3}$ & $17.1^{+1.4}_{-1.3}$ \vspace{.1cm} \\
    $F_\text{red} / F_\text{blue}$ & \textsc{ism} & $1.2^{+0.3}_{-0.2}$ & $1.0^{+0.5}_{-0.4}$ &  $1.0^{+0.5}_{-0.4}$ \vspace{.1cm} \\
    $\langle \Delta v \rangle\ [\text{km\,s}^{-1}]$ & \textsc{ism} & $19.3^{+27.0}_{-19.4}$ & $2.0^{+27.9}_{-28.7}$ & $2.1^{+27.2}_{-28.0}$ \vspace{.1cm} \\
    $\sigma_{\Delta v}\ [\text{km\,s}^{-1}]$ & \textsc{ism} & $199.4^{+22.9}_{-31.2}$ & $131.6^{+22.4}_{-39.6}$ & $132.4^{+21.7}_{-39.2}$ \vspace{.1cm} \\
    $\Delta v_\text{peak}\ [\text{km\,s}^{-1}]$ & \textsc{ism} & $52.0^{+91.9}_{-87.8}$ & $-0.4^{+133.6}_{-136.0}$ & $-2.1^{+133.8}_{-133.9}$ \vspace{.1cm} \\
    $\text{FWHM}\ [\text{km\,s}^{-1}]$ & \textsc{ism} & $369.8^{+102.2}_{-141.1}$ & $106.9^{+71.5}_{-47.1}$ & $123.5^{+81.7}_{-55.4}$ \vspace{.1cm} \\
    $\text{EW}\ [\text{\AA}]$ & \textsc{ism} & $5.5^{+2.1}_{-2.1}$ & $6.3^{+2.4}_{-1.7}$ & $1.3^{+0.5}_{-0.4}$ \\
    \hline
  \end{tabular}
  \addtolength{\tabcolsep}{-1pt}
  \renewcommand{\arraystretch}{0.9090909090909090909}
\end{table}

\begin{figure}
  \centering
  \includegraphics[width=\columnwidth]{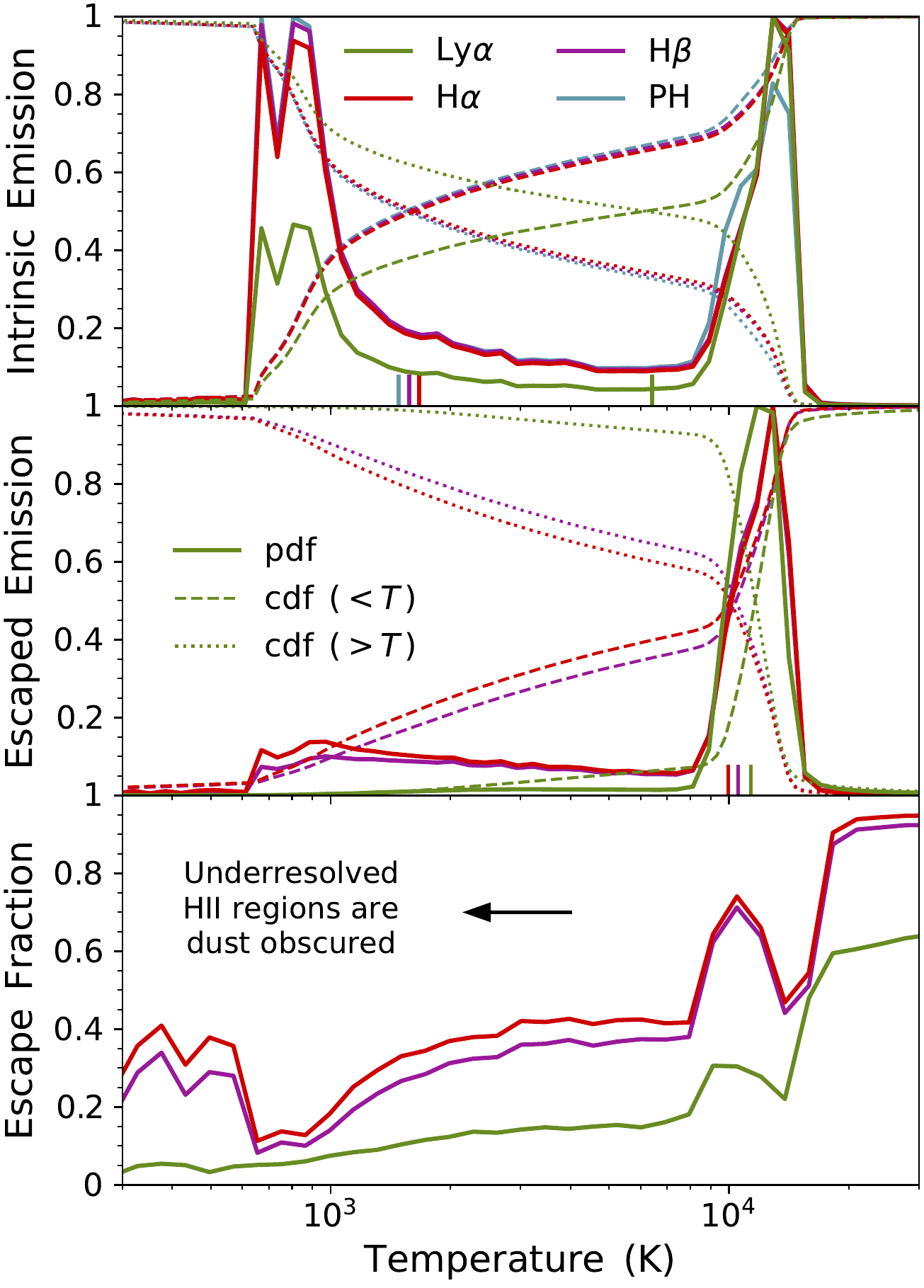}
  \caption{\textit{Top and middle:} The relative distribution of gas temperature contributing to the intrinsic and escaped Ly$\alpha$ (green), H$\alpha$ (red), and H$\beta$ (purple) line emission and photoheating rates (blue) with (reverse) cumulative distribution functions shown as (dotted) dashed curves. The emitted radiation is split between underheated ($\sim 10^3\,\text{K}$) and resolved ($\sim 10^4\,\text{K}$) \HII regions, while the medians shift to $\sim 10^4\,\text{K}$ for escaped photons (vertical markers). \textit{Bottom:} The corresponding escape fractions as a function of temperature, showing a clear difference in the populations as well as Ly$\alpha$ behaviour.}
  \label{fig:T_hist}
\end{figure}

\begin{figure}
  \centering
  \includegraphics[width=\columnwidth]{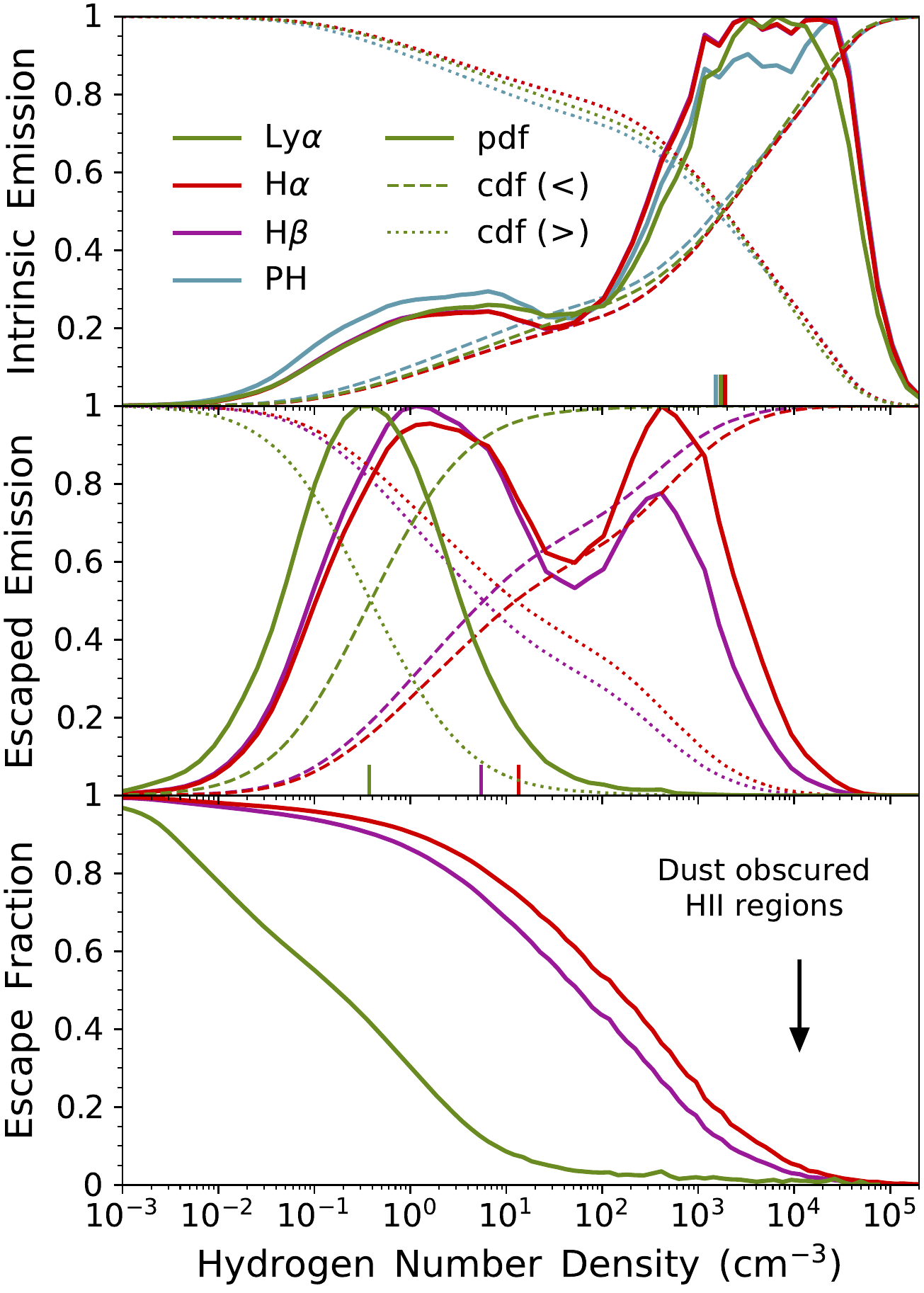}
  \caption{Same as Fig.~\ref{fig:T_hist} but for the hydrogen number density. These distributions are also bimodal with diffuse and compact components. The escape fractions depend strongly on gas and dust density, significantly altering the environments probed by these lines, especially for Ly$\alpha$ photons.}
  \label{fig:nH_hist}
\end{figure}

\begin{figure*}
  \centering
  \includegraphics[width=.33\textwidth]{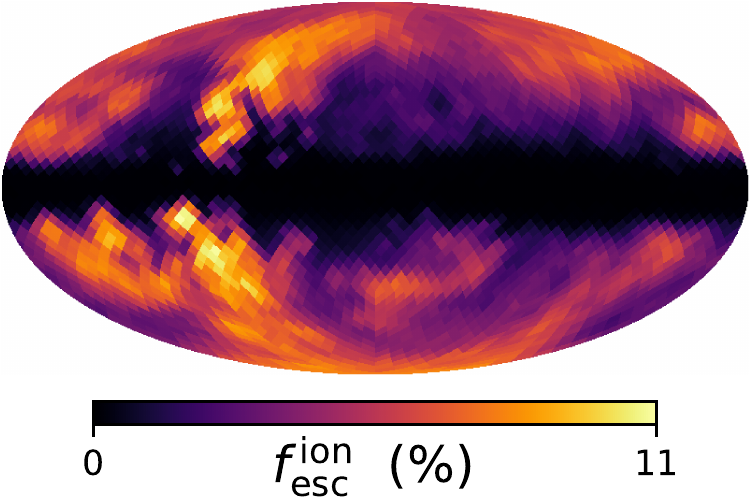}
  \includegraphics[width=.33\textwidth]{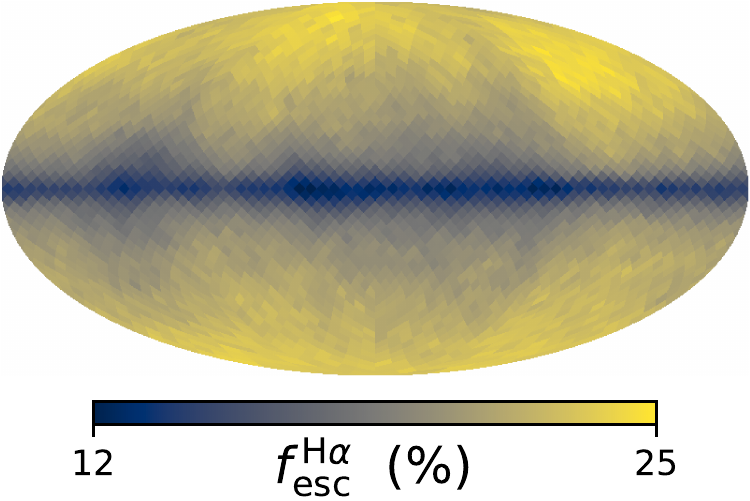}
  \includegraphics[width=.33\textwidth]{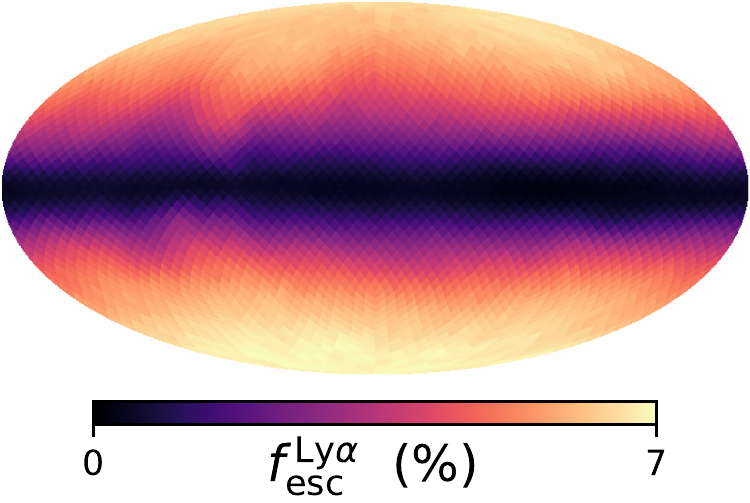}
  \caption{The angular distributions of the line-of-sight escape fractions ($f_\text{esc}$) for ionizing (left), H$\alpha$ (middle), and Ly$\alpha$ (right) photons at $1$\,Gyr. The images are based on Gaussian binning of the emergent photons onto 3072 healpix directions of equal solid angle. The disc in projection is the most prominent feature followed by various fluctuations reflecting the structures probed. Ionizing radiation is the most anisotropic, with Ly$\alpha$ scattered to high latitudes, and H$\alpha$ being relatively smooth aside from the thin disc of enhanced dust absorption.}
  \label{fig:healpix_fesc}
\end{figure*}

\subsection{Hydrogen lines}
Due to the close connection between photoionization and recombination, we find similar trends and fluctuations for the hydrogen emission lines, as revealed in the evolutionary diagnostics shown in Fig.~\ref{fig:f_esc_lines}. In particular, for lines listed as $X \in \{\text{Ly}\alpha, \text{H}\alpha, \text{H}\beta\}$ we calculate intrinsic luminosities of
$\langle L_X \rangle \approx \{42.7, 3.68, 1.24\} \times 10^{41}\,\text{erg\,s}^{-1}$, or equivalent photon rates of $\langle \dot{N}_X \rangle \approx \{26.1, 12.1, 3.02\} \times 10^{52}\,\text{s}^{-1}$. Interestingly, this results in intrinsic line ratios of $\langle L_{\text{Ly}\alpha} / L_{\text{H}\alpha} \rangle \approx 11.5$ and $\langle L_{\text{H}\alpha} / L_{\text{H}\beta} \rangle \approx 2.97$, which are higher than the commonly assumed values of $8.16$ and $2.86$ as a result of including collisional excitation and realistic temperature distributions. Specifically, the average fraction of the intrinsic emission originating from collisional excitation is $\langle f_{\text{col},X} \rangle \approx \{27.2, 5.3, 3.1\}$ for each line, respectively. In the second panel we show the escaping luminosity, including asymmetric $1\sigma$ confidence levels considering different viewing angles. The Ly$\alpha$ escape fractions are much lower due to resonant scattering, leading to observed luminosities of $\langle L_{\text{esc},X} \rangle \approx \{23.7, 10.7, 2.98\} \times 10^{40}\,\text{erg\,s}^{-1}$, corresponding to escape fractions of $\langle f_{\text{esc},X} \rangle \approx \{5.8, 31.7, 26.1\}$ as shown in the third panel. We note that the escape fractions for collisional excitation are noticeably higher with values of $\langle f_{\text{esc},X}^\text{col} \rangle \approx \{8.3, 48.3, 41.7\}$. This leads to a boost increasing the fraction of observed luminosity due to collisional excitation to $\langle f_{\text{col},X}^\text{esc} \rangle \approx \{40.3, 7.3, 4.5\}$, as shown in the bottom panel. This emphasizes the importance of the line cooling channel for Ly$\alpha$ observations of disc-like galaxies, and provides a warning as the strong temperature dependence introduces a large amount of uncertainty in theoretical modelling.

In the top panel of Fig.~\ref{fig:f_esc_cdf}, we present the probability that a given sightline has an escape fraction larger than a certain value across all snapshots. Ionizing escape fractions exhibit significant variation, with roughly $23\%$ ($28\%$) from relatively low-(high-)escaping sightlines of $f_\text{esc} < 1\%$ ($f_\text{esc} > 10\%$). The absence of realistic cosmological accretion and self-shielded clumpy structures throughout the CGM in our idealized simulation may explain our slightly higher LyC escape fractions compared to local observational estimates \citep[e.g.][]{Leitherer1995,Deharveng2001,Heckman2001,Grimes2009,Bergvall2013,Rutkowski2016}. The Ly$\alpha$ line has slightly less variation, while the H$\alpha$ and H$\beta$ lines are much more concentrated around the median values. In the bottom panel we show the probability distribution of the directional LyC and Ly$\alpha$ escape fractions over all snapshots. We note that most sightlines have higher Ly$\alpha$ escape fractions than LyC ones, specifically $f_{\text{esc,Ly}\alpha} / f_{\text{esc,LyC}} = 1.3^{+4.5}_{-0.8}$. For reference we include observational data from low-mass local starburst galaxies \citep[as reported by][]{Izotov2021}, although LyC leaking galaxies are typically much more gas rich and have significantly higher specific SFRs than the Milky Way. Given the above discussion, our results are consistent with general observational and theoretical expectations that Ly$\alpha$ resonant scattering facilitates escape through low-density channels \citep[e.g.][]{Dijkstra2016,Smith2019}.

\begin{figure*}
  \centering
  \includegraphics[width=\textwidth]{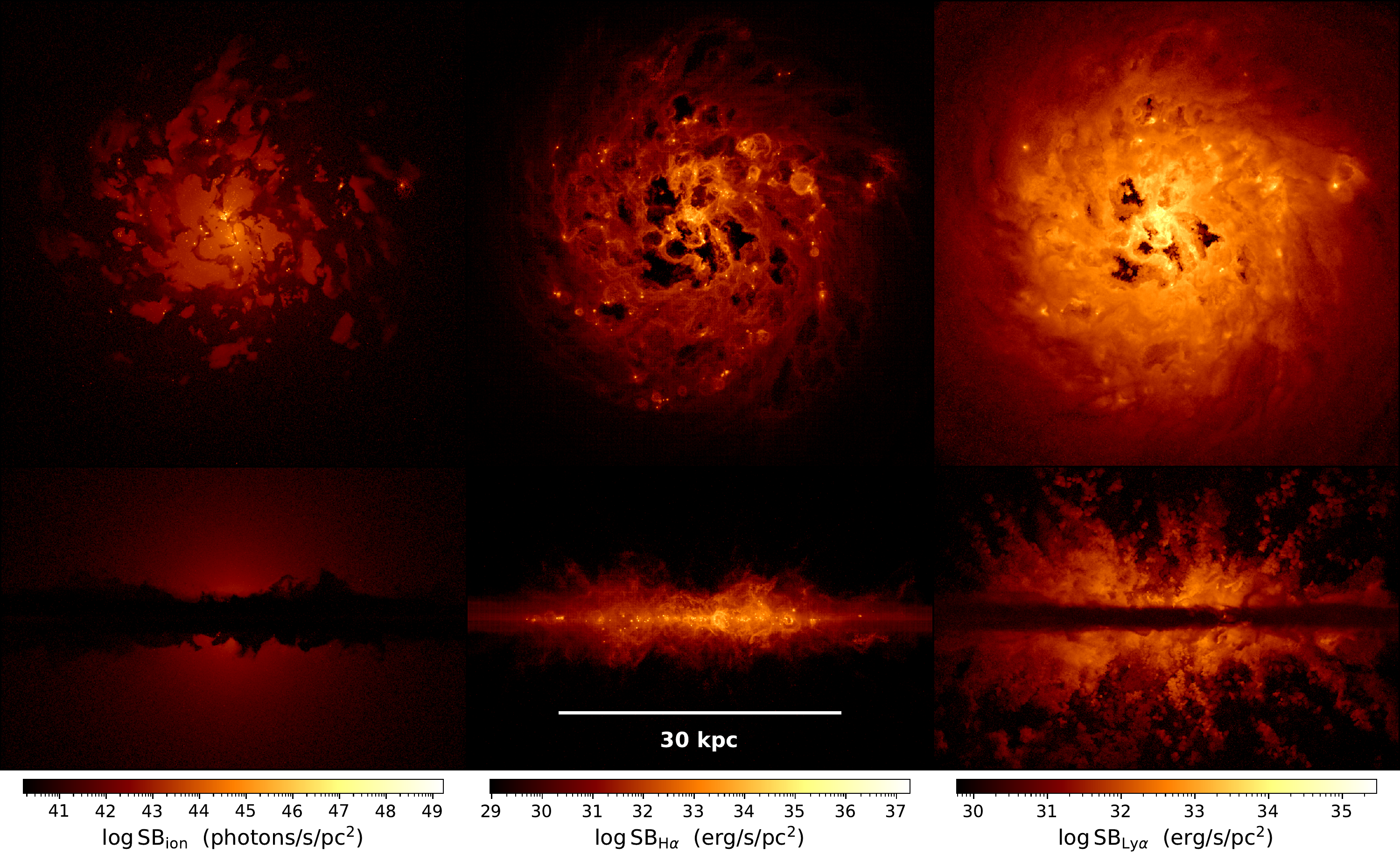}
  \caption{The escaping surface brightness for ionizing, H$\alpha$, and Ly$\alpha$ radiation, which illustrates the physical connections between each stage in the radiative transfer process. For example, ionizing photons are primarily emitted from highly clustered young sources but are only visible if the surrounding \HII regions successfully break out of the disc. H$\alpha$ photons show the internal structure of the reprocessed ionizing radiation after accounting for dust effects, while Ly$\alpha$ also includes resonant scattering resulting in a much more diffuse images.}
  \label{fig:ion_Lya_Ha}
\end{figure*}

\begin{figure*}
  \centering
  \includegraphics[width=\textwidth]{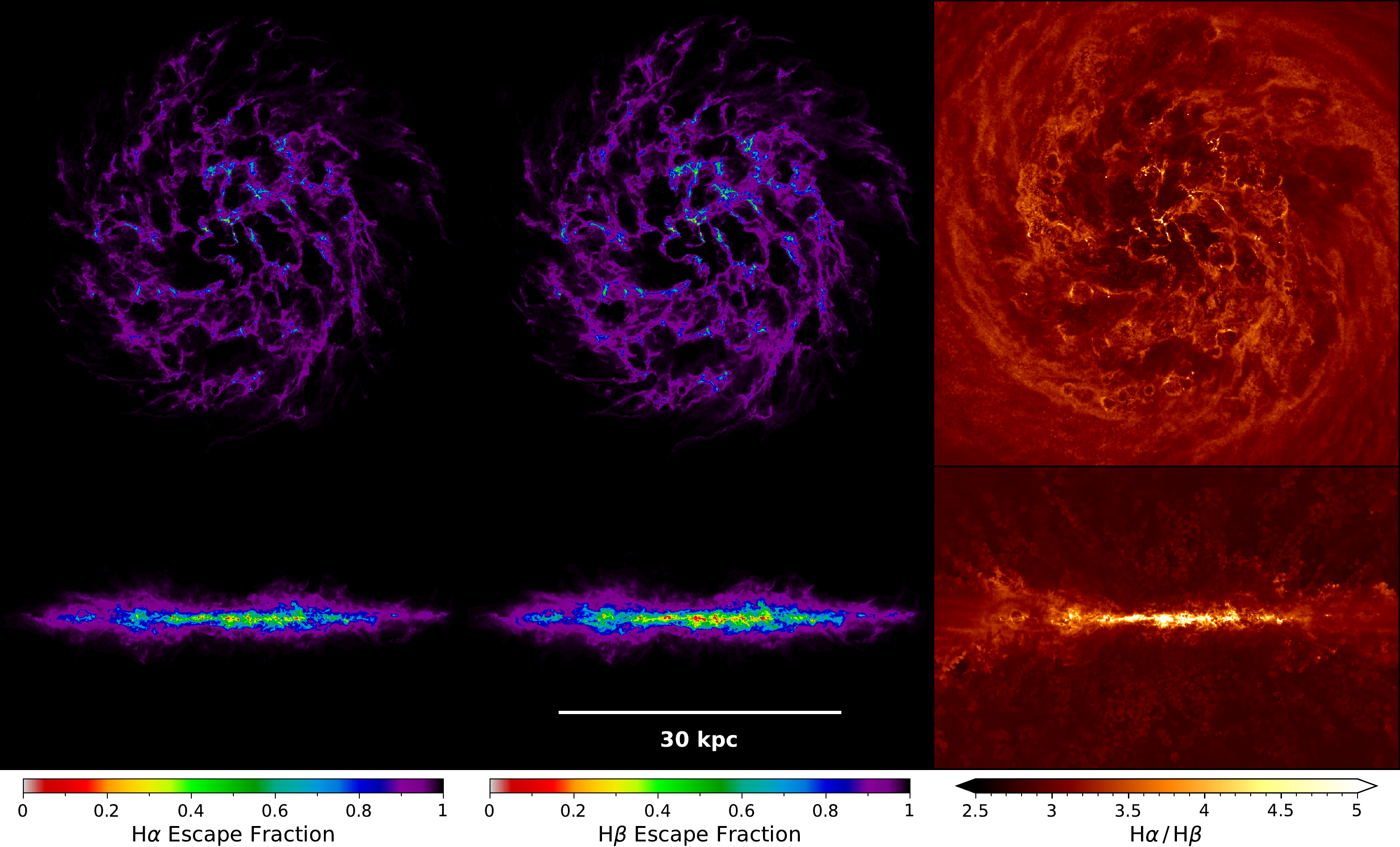}
  \caption{The escape fractions for the non-resonant H$\alpha$ and H$\beta$ lines as well as the Balmer decrement $\text{H}\alpha / \text{H}\beta$ in the rightmost panel, which illustrates the impact of differential absorption through high density regions ($n_\text{H} \gtrsim 10^3\,\text{cm}^{-3}$). The higher dust opacity of H$\beta$ compared to H$\alpha$ ($\kappa_{\text{d,H}\beta} / \kappa_{\text{d,H}\alpha} \approx 1.54$) gives rise to lower escape fractions and an effective map of dust intervening star-forming regions.}
  \label{fig:fesc_ratio}
\end{figure*}

To further investigate the sources of these photons in Fig.~\ref{fig:phase_nT}, we illustrate the relative line luminosities and photoheating across the density--temperature phase space. Most of the emission comes from dense \HII regions with a characteristic temperature of $T \sim 10^4$\,K. On the other hand, strong emission features at $T \lesssim 10^3$\,K are associated with underresolved and consequently underheated \HII regions, although the presence of molecular clouds, partial ionization, and transient phenomena complicate a purely numerical explanation. The mass contours highlight the presence of molecular ($\sim 10$\,K) as well as underresolved and underheated \HII regions ($\sim 10^3$\,K) cooling tracks or extended fingers coincident with the formation of collapsing structures prior to heating and disruption via feedback channels. In fact, there seems to be an insignificant fraction of gas by mass in the warm ($\sim 10^4$\,K) and high-density ($\gtrsim 10^2\,\text{cm}^{-3}$) phase, even though this is one of the most efficient regions for line production. Still, it is clear from the lower right panel that photoionization shapes the distributions for intrinsic line emission with respect to density and temperature. Interestingly, although the intrinsic Ly$\alpha$ and H$\alpha$ logarithmic emission maps are nearly indistinguishable in appearance, the line ratio immediately reveals the presence of the additional Ly$\alpha$ cooling channel at $T \gtrsim 10^4\,\text{K}$ (see the top panels). We also observe significant phase space reprocessing for Ly$\alpha$ photons after accounting for radiative transfer effects, which is not as drastic for the escaped H$\alpha$ radiation (middle panels). Ly$\alpha$ is efficiently trapped and destroyed due to the additional resonant scattering with neutral hydrogen gas reservoirs, while at the same time suffering from the almost order of magnitude increase in dust opacity compared to H$\alpha$. The difference between resonant scattering and direct dust absorption is underscored by contrasting the escape fractions of these lines (bottom panels). Ly$\alpha$ and H$\alpha$ respectively exhibit order unity escape fractions ($f_\text{esc} \approx 1$) only for diffuse gas at hydrogen number densities of $n_\text{H} \lesssim 10^{-3}$ and $1\,\text{cm}^{-3}$, negligible escape fractions ($f_\text{esc} \ll 1$) for gas at $n_\text{H} \gtrsim 1$ and $10^3\,\text{cm}^{-3}$, with a continuous transition for photons originating from intermediate densities. Finally, the escaped $\text{H}\alpha / \text{H}\beta$ line ratio, i.e. the Balmer decrement in the central right-hand panel, clearly illustrates the differential emission and attenuation of these lines at $n_\text{H} \gtrsim 10^3\,\text{cm}^{-3}$, as they have different dust opacities ($\kappa_{\text{d,H}\beta} / \kappa_{\text{d,H}\alpha} \approx 1.54$) and to a lesser extent density and temperature dependence in their recombination conversion probabilities ($P_\text{B}$).

To quantify these findings in greater detail, we marginalize over the phase space to investigate the radiative transfer outcomes as functions of temperature and density alone. In Fig.~\ref{fig:T_hist}, we show the relative distribution of gas temperatures contributing to the intrinsic and escaped reprocessed line emission (upper panels) and the escape fraction (lower panel) for each line. For comparison we also include (reverse) cumulative distribution functions shown as (dotted) dashed curves. The intrinsic emission is bimodal as underresolved and underheated \HII regions ($\sim 10^3\,\text{K}$) comprise almost half of the recombination budget. We note that due to the strong $\rho^2$ dependence for emission this corresponds to a much smaller mass-weighted fraction for unresolved \HII regions. However, the escape fractions from these same dense dusty regions are much lower in comparison to resolved \HII regions ($\sim 10^4\,\text{K}$). This results in drastically different escaped distributions that are unimodal with long tails to lower temperatures. Also, while H$\alpha$ and H$\beta$ have similar behaviours, Ly$\alpha$ has a stronger peak at $10^4\,\text{K}$ due to the additional collisional excitation and also has much lower escape fractions at all temperatures. For quantitative comparison, we calculate median and 1$\sigma$ \{Ly$\alpha$, H$\alpha$, H$\beta$\} statistics of $T_\text{int} \approx \{6.42^{+6.65}_{-5.58}, 1.67^{+10.89}_{-0.926}, 1.57^{+11.05}_{-0.831}\} \times 10^3\,\text{K}$ and $T_\text{esc} \approx \{11.3^{+1.73}_{-1.17}, 9.96^{+3.22}_{-1.20}, 10.5^{+2.61}_{-9.10}\} \times 10^3\,\text{K}$ for the intrinsic and escaped emission, respectively. In Fig.~\ref{fig:nH_hist}, we show the same quantities for the hydrogen number density. Interestingly, the intrinsic emission is also somewhat bimodal with a smaller contribution from diffuse ionized gas at $n_\text{H} \lesssim 100\,\text{cm}^{-3}$ and a larger fraction from compact \HII regions at densities up to $\sim 10^5\,\text{cm}^{-3}$. The escape fraction depends strongly on density as these environments are enshrouded by dust. In fact, the truncation for escape varies smoothly from unity at the lowest densities to zero at the highest. As expected, H$\alpha$ and H$\beta$ are least affected while the impact on Ly$\alpha$ photons is enhanced by scattering with neutral hydrogen gas. For concreteness, the median and 1$\sigma$ \{Ly$\alpha$, H$\alpha$, H$\beta$\} statistics are $\log(n_\text{H,int} / 1\,\text{cm}^{-3}) \approx \{3.24^{+1.00}_{-2.35}, 3.28^{+1.01}_{-2.26}, 3.28^{+1.00}_{-2.25}\}$ and $\log(n_\text{H,esc} / 1\,\text{cm}^{-3}) \approx \{-0.430^{+0.854}_{-0.766}, 1.13^{+1.76}_{-1.56}, 0.74^{+1.88}_{-1.28}\}$ for the intrinsic and escaped emission, respectively.

\subsection{Viewing angle dependence}
In Fig.~\ref{fig:healpix_fesc}, we illustrate the angular distributions of viewing angle dependent escape fractions ($f_\text{esc}$) for ionizing (left), H$\alpha$ (middle), and Ly$\alpha$ (right) photons at $1$\,Gyr, based on 3072 healpix directions of equal solid angle. Including multiple viewing angles also allows us to mimic observations of different galaxies with similar properties. Overall, the most prominent feature is the disc in projection followed by various fluctuations reflecting the structures probed. As seen in Fig.~\ref{fig:f_esc_lines}, although the maps are similar in appearance the variation across sightlines is at least five times larger for Ly$\alpha$ than H$\alpha$. This is because Ly$\alpha$ is scattered to high latitudes, while H$\alpha$ primarily suffers from enhanced dust absorption along the mid-plane of the thin disc. To quantify this, we consider the degree of isotropy $\langle F_\text{LOS} / F_\Omega \rangle \approx \{1^{+0.84}_{-0.95}, 1^{+0.62}_{-0.70}, 1^{+0.11}_{-0.13}, 1^{+0.13}_{-0.15}\}$ for ionizing photons, Ly$\alpha$, H$\alpha$, and H$\beta$, respectively. Specifically, this normalizes out the median at a given snapshot, so the asymmetric time-weighted confidence levels can roughly be thought of as a Poisson dispersion in sightlines (i.e. $\sigma / \mu$). Ionizing radiation is the most anisotropic as it is highly sensitive to the multiphase distribution of \HI without being smoothed out by scattering. In fact, the exponential $e^{-\tau}$ absorption leads to bimodal transmission either for (i) different directions for the same source and (ii) different source positions and integration paths for the same direction. The healpix maps show small and large-scale structures in projection such that neighboring sightlines are correlated despite the spatial integration over the entire image aperture.

In Fig.~\ref{fig:ion_Lya_Ha}, we demonstrate that there is also significant variation in the spatial distributions of the escaped surface brightness images. As expected, ionizing photons are primarily emitted from highly clustered young sources but are only visible if the surrounding super bubble regions successfully break out of the disc. Large holes dominate the observed morphology but beyond the obscured layers is a network of invisible compact \HII regions. The internal structure of the reprocessed ionizing radiation is revealed by H$\alpha$ photons after accounting for dust effects mostly associated with the highest density gas. Finally, resonant scattering of Ly$\alpha$ photons results in significant differences compared to the intrinsic emission, most notably in much more diffuse face-on images and scattering away from the mid-plane to higher latitudes for edge-on views. To illustrate further connections between radiative transfer processes in Fig.~\ref{fig:fesc_ratio}, we show spatially-resolved escape fractions for the non-resonant H$\alpha$ and H$\beta$ lines as well as the Balmer decrement $\text{H}\alpha / \text{H}\beta$. To simplify the interpretation and eliminate Monte Carlo noise, we employ the vanishing albedo ($A = 0$) ray-tracing calculations, which shows the impact of differential absorption through high-density regions ($n_\text{H} \gtrsim 10^3\,\text{cm}^{-3}$). The higher dust opacity of H$\beta$ compared to H$\alpha$ ($\kappa_{\text{d,H}\beta} / \kappa_{\text{d,H}\alpha} \approx 1.54$) gives rise to lower escape fractions and a map of dust absorption.

\begin{figure*}
  \centering
  \includegraphics[width=.33\textwidth]{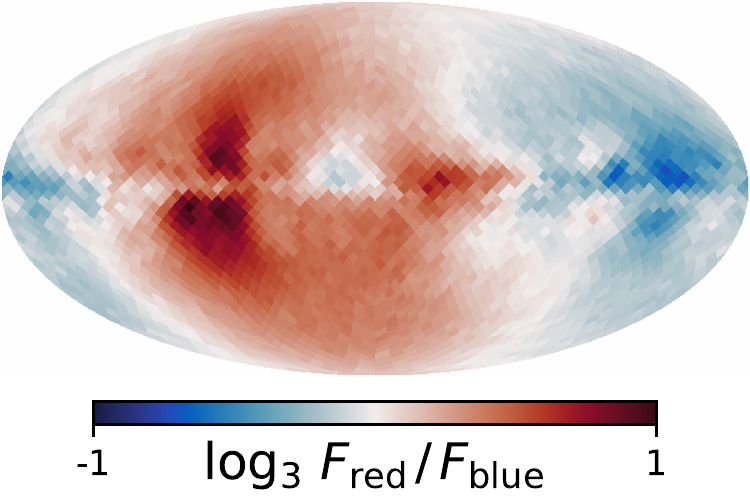}
  \includegraphics[width=.33\textwidth]{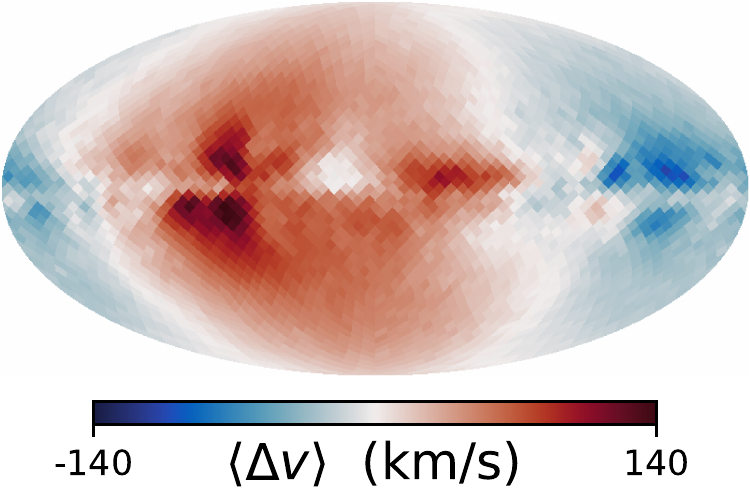}
  \includegraphics[width=.33\textwidth]{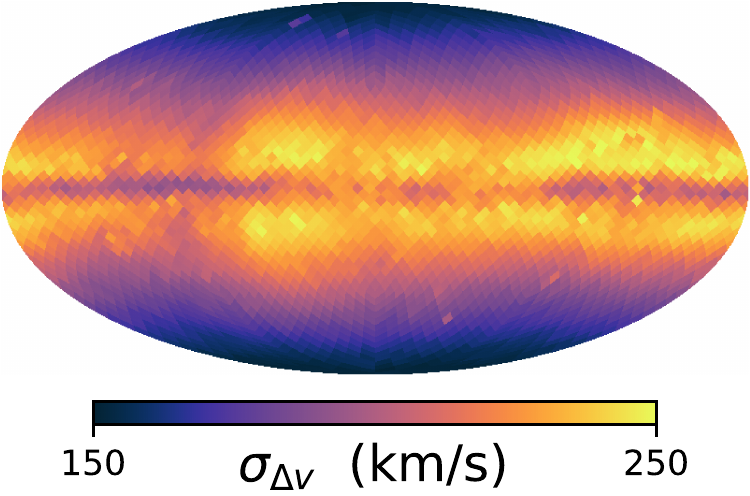}
  \caption{The angular distributions of the line-of-sight red-to-blue flux ratio $F_\text{red} / F_\text{blue}$ (left), flux-weighted frequency centroid $\langle \Delta v \rangle$ (middle), and flux-weighted standard deviation $\sigma_{\Delta v}$ (right) for Ly$\alpha$ photons at $1$\,Gyr. A moderate dipole moment is present in the average velocity offset corresponding to the direction of highest LyC escape shown in Fig.~\ref{fig:healpix_fesc}. For this snapshot the red enhancement also seems to correlate with reduced resonant line broadening.}
  \label{fig:healpix_freq}
\end{figure*}

\begin{figure*}
  \centering
  \includegraphics[width=\textwidth]{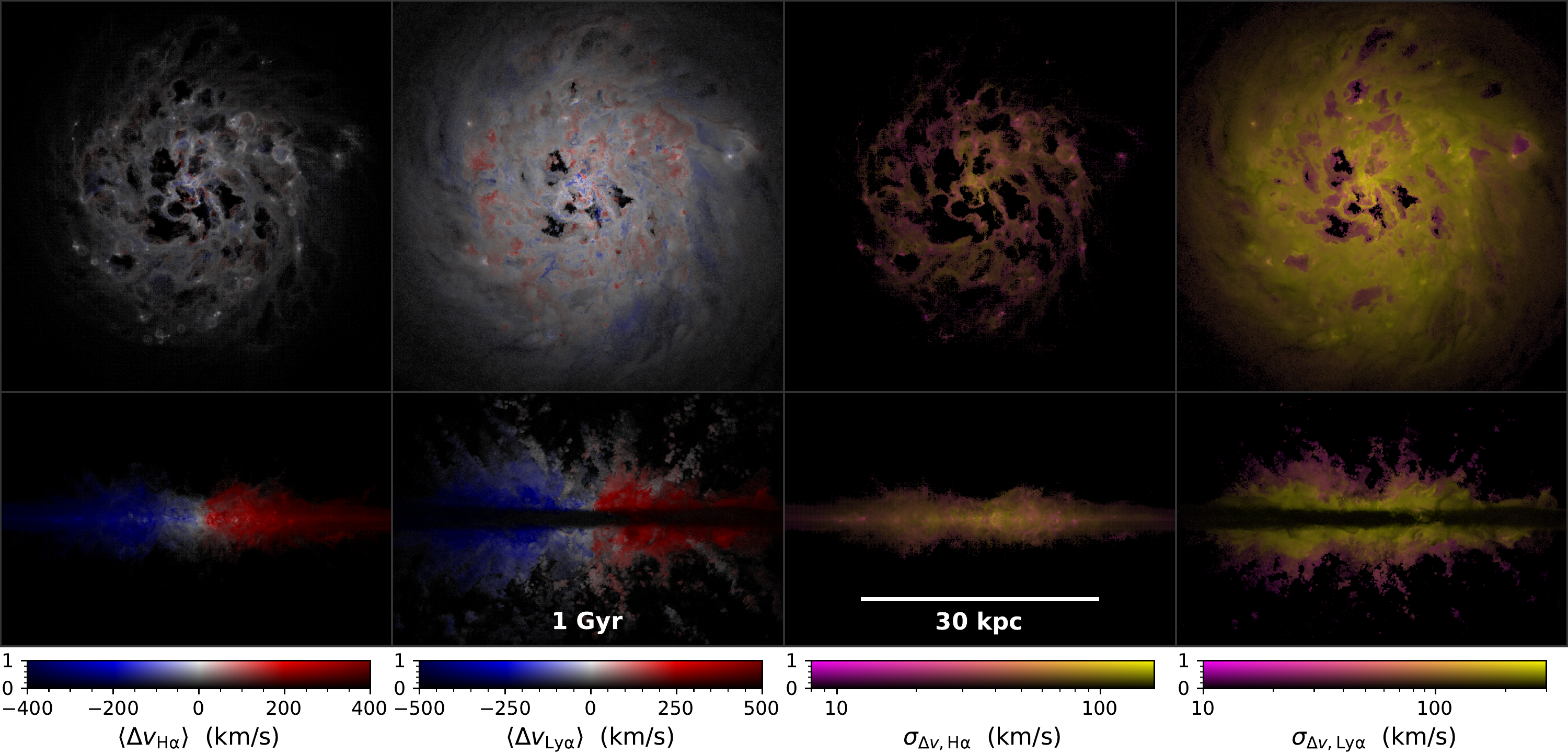}
  \caption{Spatially-resolved frequency moment maps for H$\alpha$ and Ly$\alpha$ lines for the same face-on and edge-on views as previous figures. The colour transparency of the flux-weighted frequency centroid $\langle \Delta v \rangle$ and standard deviation $\sigma_{\Delta v}$ is determined by the logarithmic surface brightness according to $\alpha = 1 - \log(\text{SB}/\text{SB}_\text{max}) / \log(f_\text{cut}) \in [0,1]$, where $\text{SB}_\text{max}$ denotes the maximum image intensity and $f_\text{cut}$ is an arbitrary relative lower limit set to $5 \times 10^{-9}$ and $2 \times 10^{-6}$ for H$\alpha$ and Ly$\alpha$, respectively. Edge-on views reveal the expected blue--red rotation signatures while face-on views exhibit a variety of local red or blue dominated regions. The standard deviations for H$\alpha$ and Ly$\alpha$ are drastically different due to resonant broadening within optically thick gas for Ly$\alpha$ photons.}
  \label{fig:Dv_avg_std}
\end{figure*}

\subsection{Frequency dependence}
Spectral profiles provide an additional dimension for understanding line emission and encode information affecting the transport physics, e.g. among other things the neutral hydrogen, dust, and velocity distributions. For simplicity we visualize the frequency information in terms of the Doppler velocity $\Delta v = c \Delta \lambda / \lambda$ and include the flux-weighted frequency centroid $\langle \Delta v \rangle \equiv \int \Delta v f_\lambda\,\text{d}\lambda / \int f_\lambda\,\text{d}\lambda$ and standard deviation $\sigma_{\Delta v} \equiv (\langle \Delta v^2 \rangle - \langle \Delta v \rangle^2)^{1/2}$ of the escaped radiation. Specifically, in Fig.~\ref{fig:healpix_freq}, we illustrate the angular distributions of the line-of-sight red-to-blue flux ratio $F_\text{red} / F_\text{blue}$ and flux-weighted frequency centroid $\langle \Delta v \rangle$ and standard deviation $\sigma_{\Delta v}$ for Ly$\alpha$ photons at $1$\,Gyr. Interestingly, there is a moderate dipole moment (i.e. higher red-to-blue flux ratio and redward centroid shift reflecting the outflow substructure) and reduced resonant line broadening in the average velocity offset corresponding to the direction of highest LyC escape shown in Fig.~\ref{fig:healpix_fesc}. To further summarize the spatial-spectral morphology in Fig.~\ref{fig:Dv_avg_std}, we display for both H$\alpha$ and Ly$\alpha$ lines the frequency moment maps for the same face-on and edge-on views as previous figures. While there are certainly strong connections to the gas velocity images in Fig.~\ref{fig:v_4}, detailed comparisons are not straightforward and deserve further study beyond the scope of this paper. In particular, edge-on views reveal the expected blue--red gradient rotation signatures while face-on views show the patchwork of lower velocity red and blue dominated regions. For standard deviations the H$\alpha$ line widths are narrower in bright dense regions as these originate from individual cloud regions while the more diffuse ionized gas exhibits broader spectral features. On the other hand, Ly$\alpha$ photons undergo resonant scattering that significantly broadens lines, so this is actually highest near bright dense optically thick regions and lowest in ultra-diffuse ionized pockets.

\begin{figure}
  \centering
  \includegraphics[width=\columnwidth]{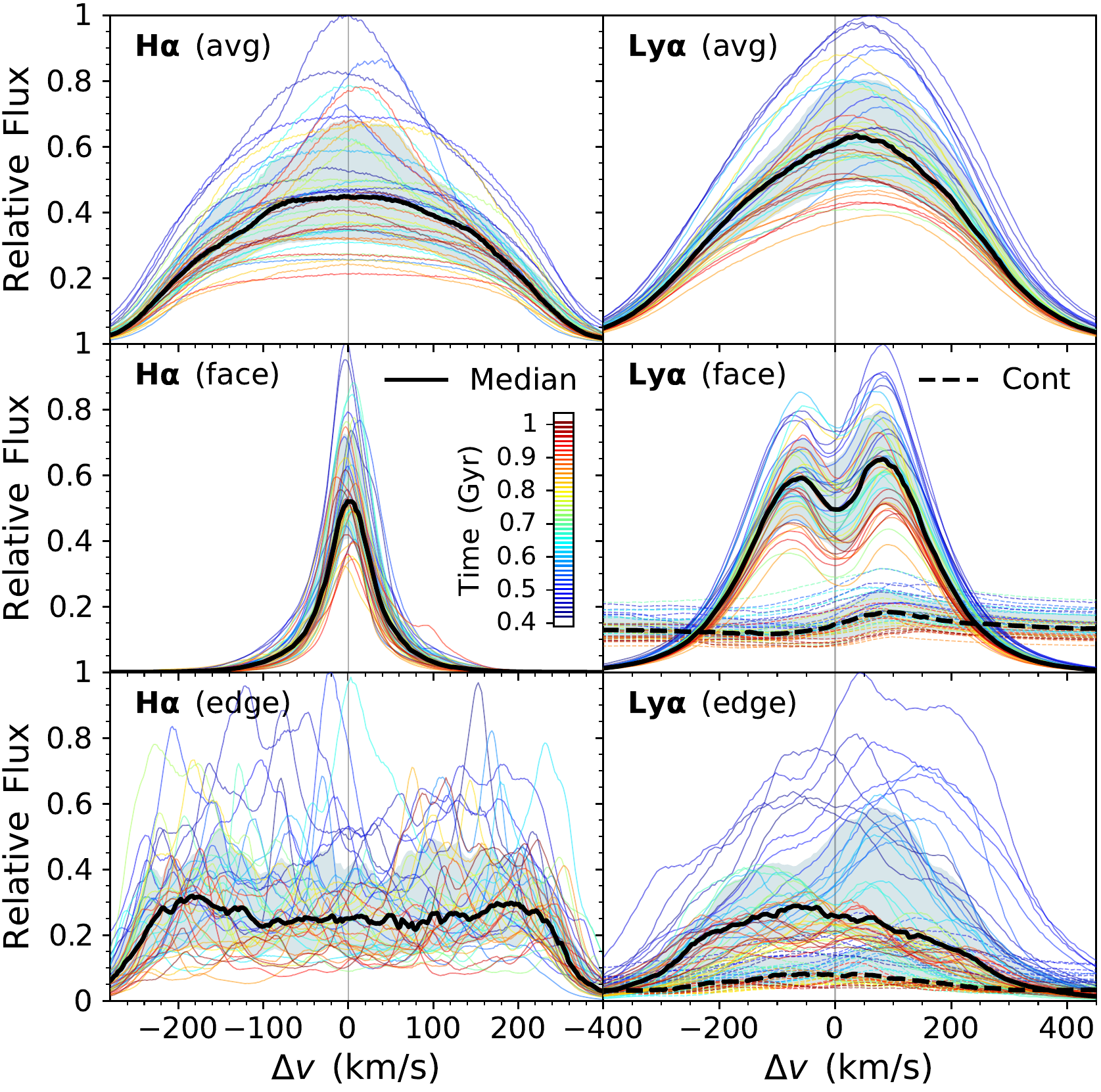}
  \caption{Spatially integrated emergent H$\alpha$ and Ly$\alpha$ line flux densities as a function of Doppler velocity $\Delta v = c \Delta \lambda / \lambda$ at each time (coloured curves) and with medians (black curves). The panel normalizations are $3.06 \times 10^{38}\,(2.57 \times 10^{39})\,[4.50 \times 10^{38}]\,\text{erg/s/(km/s)}$ for H$\alpha$ and $3.77 \times 10^{38}\,(1.79 \times 10^{39})\,[1.88 \times 10^{38}]\,\text{erg/s/(km/s)}$ for Ly$\alpha$ for angular-averaged (face-on) [edge-on] directions. We also show the stellar continuum as dashed curves around the Ly$\alpha$ line, which provides additional asymmetry favouring the red peak. See the text for further discussion about broadening due to viewing angle and resonant-scattering.}
  \label{fig:flux}
\end{figure}

\begin{figure}
  \centering
  \includegraphics[width=\columnwidth]{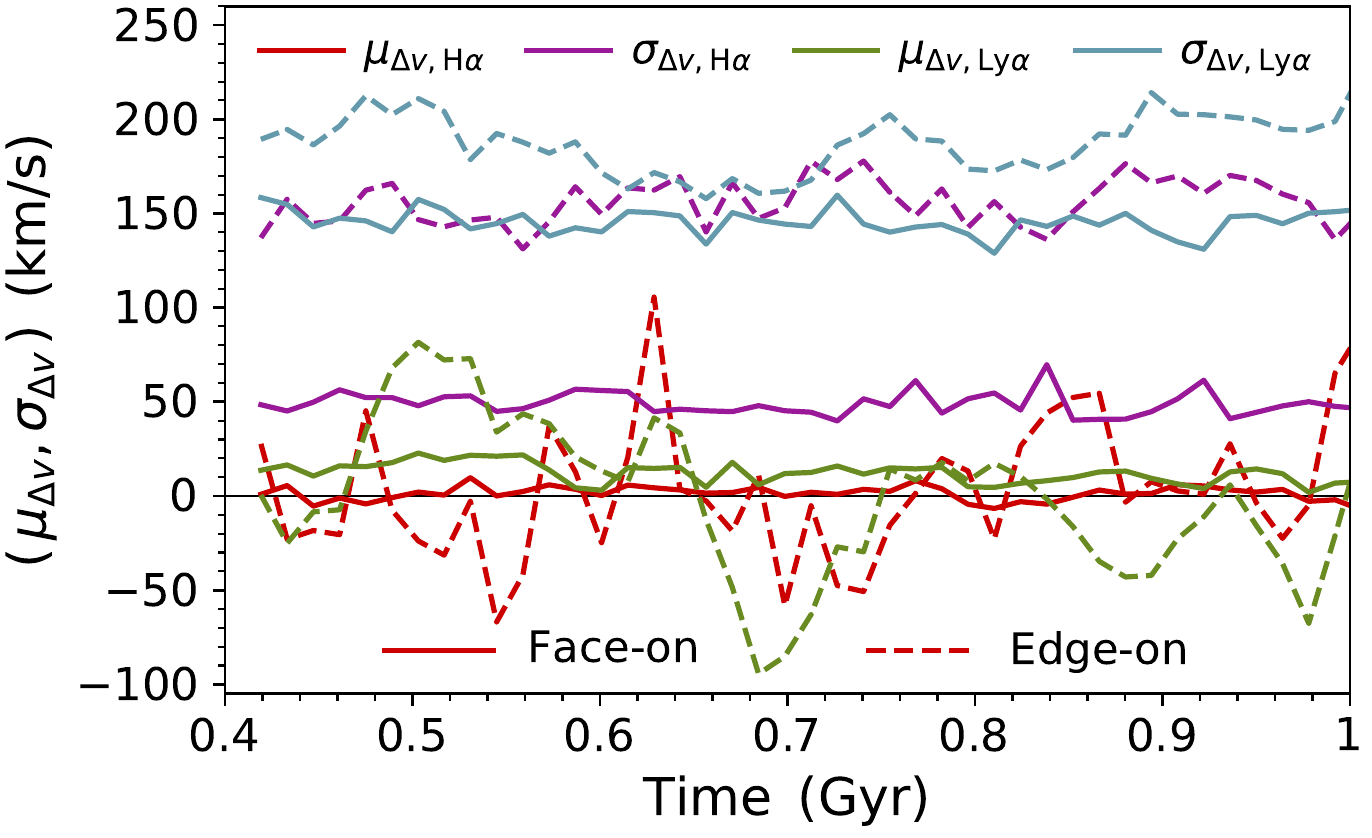}
  \caption{Time evolution of the flux-weighted frequency centroid $\langle \Delta v \rangle$ and standard deviation $\sigma_{\Delta v}$ for face-on and edge-on views for the H$\alpha$ and Ly$\alpha$ lines. The values remain fairly consistent throughout the simulation.}
  \label{fig:time_avg_std}
\end{figure}

We further investigate the spectral information in Fig.~\ref{fig:flux} by showing the spatially integrated emergent line flux under arbitrary normalizations. These curves capture the time evolution via line colouring and extreme viewing dependence contrasting face-on (top panels) and edge-on (bottom panels) views. The black curves give the median over all times for a representative shape, which for H$\alpha$ is a narrow peak at line centre for face-on views and a variety of multiple blended peaks for an average flat plateau appearance for edge-on views. Similarly, Ly$\alpha$ profiles are generally double peaked for face-on views and highly broadened for edge-on views. We note that Ly$\alpha$ profiles have substantial emission at line centre, which is due to the breakout of large \HII regions perpendicular to the disc facilitating more direct escape channels, combinations of turbulent and structured velocity shifts blending broad lines from a range of positions and directions, and the lack of a proper cosmological CGM environment around isolated disc simulations. Furthermore, it is also interesting that the Ly$\alpha$ red peak is dominant overall but to an extent that is insufficient to explain high-redshift and local observations. This is likely due to the more violent starburst environments of typical LAEs and the interplay of disruptive feedback to CGM scales. However, it could also indicate that further resolution is necessary to capture local outflows on ISM and molecular cloud scales. Such sub-grid spectral reddening on scales below a few parsecs (corresponding to both the gas softening length and the effective cell radii at the star-formation density threshold) would be a transient phenomenon in current state-of-the-art galaxy formation simulations. Thus, while we believe it is most fruitful for LAE modelling to move to cosmological zoom-in simulations, where our preliminary results reveal red peak dominated spectra, improved understanding of Ly$\alpha$ radiative transfer on turbulent molecular cloud scales certainly provides valuable insights as well \citep{Kimm2019,Kimm2022,Kakiichi2021}.

\begin{figure}
  \centering
  \includegraphics[width=\columnwidth]{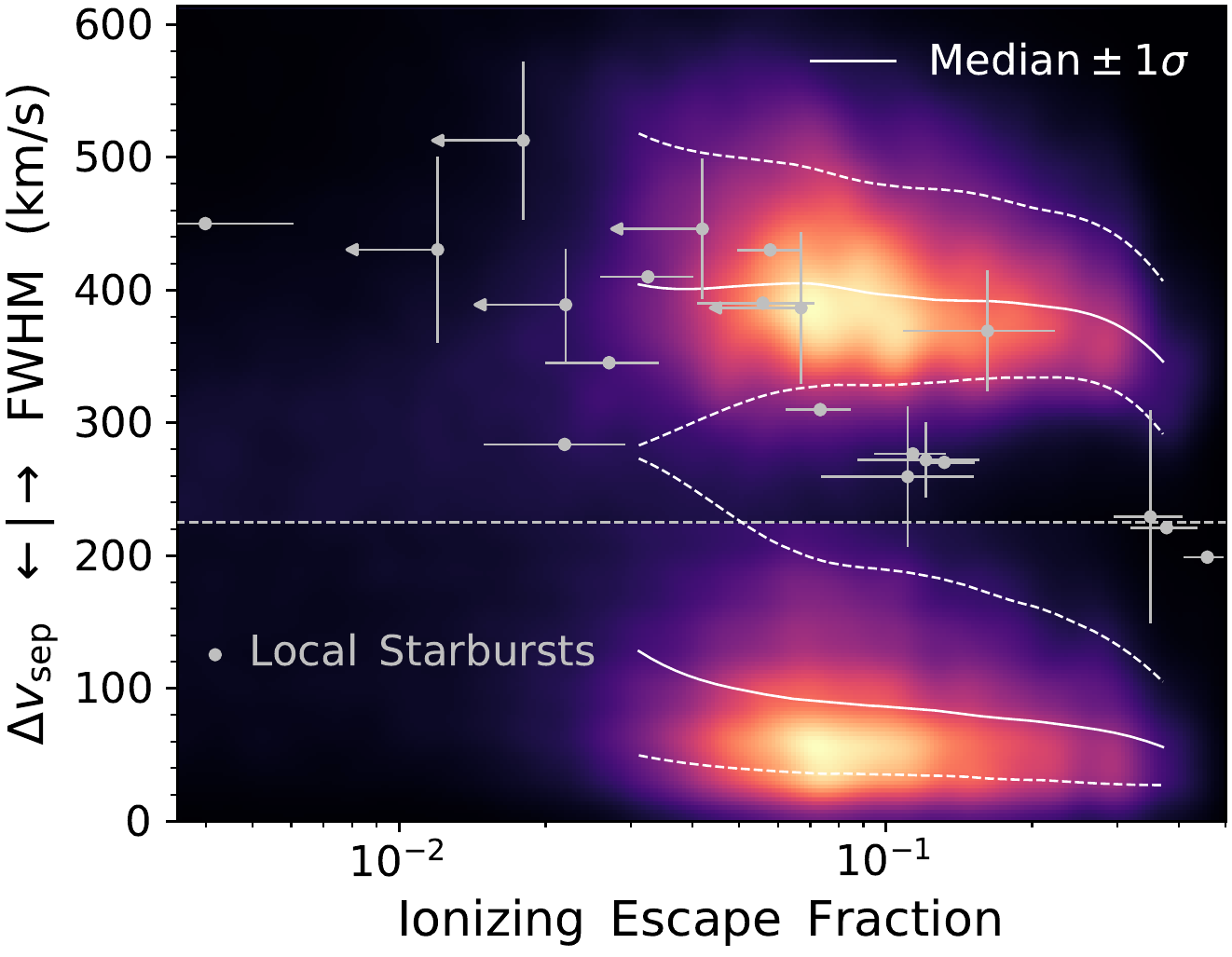}
  \caption{Probability distributions of the directional Ly$\alpha$ full-width at half-maxima (FWHM) and peak separation ($\Delta v_\text{sep}$) with the LyC escape fractions over all snapshots. For reference we show observational data from low-mass local starburst galaxies \citep[as reported by][]{Izotov2021}, which is meant to guide intuition that both quantities are expected to decrease at higher LyC escape fractions. We note that the distributions have been smoothed to avoid aliasing artefacts from the $10\,\text{km\,s}^{-1}$ spectral binning, while the median and $1\sigma$ curves are taken from the underlying unsmoothed data.}
  \label{fig:Lya_FWHM}
\end{figure}

To further condense the flux information in Fig.~\ref{fig:time_avg_std}, we show the time evolution of the frequency moments for face-on and edge-on views for the H$\alpha$ and Ly$\alpha$ lines. In general, the flux-weighted frequency centroids $\langle \Delta v \rangle$ are consistent with zero, although there are consistent modest red offsets for Ly$\alpha$ above the disc. The standard deviations $\sigma_{\Delta v}$ encapsulating line widths and resonant broadening are consistently $\sim 150$--$200\,\text{km}\,\text{s}^{-1}$ throughout the simulation. Edge-on views have higher LOS velocities and dust/\HI column densities, which means frequency centroids can be influenced by the motions of relatively fewer bright nearby sources. We note that the time variations are still much smaller than the line widths, i.e. $\langle \Delta v \rangle < \sigma_{\Delta v}$. While each of the frequency dependent perspectives considered in this section are informative on their own, we note that spatially-integrated spectra and frequency-integrated moment maps reduce the dimensionality and can therefore mask complex phenomena that is otherwise preserved in the full spatial-spectral data. For completeness, for $X \in \{\text{Ly}\alpha, \text{H}\alpha, \text{H}\beta\}$ we calculate the time- and sightline-averaged red-to-blue flux ratios to be $\langle F_{\text{red},X} / F_{\text{blue},X} \rangle \approx \{1.2, 1, 1\}$, flux-weighted frequency centroids are $\langle \langle \Delta v \rangle_X \rangle \approx \{19.3, 2.0, 2.1\}\,\text{km\,s}^{-1}$, flux-weighted standard deviations are $\langle \sigma_{\Delta v,X} \rangle \approx \{199, 132, 132\}\,\text{km\,s}^{-1}$, characteristic peak frequency offsets are $\langle \Delta v_{\text{peak},X} \rangle \approx \{52, 0, -2\}\,\text{km\,s}^{-1}$, and full-width at half-maxima are $\langle \text{FWHM}_X \rangle \approx \{370, 107, 124\}\,\text{km\,s}^{-1}$, collectively highlighting the role of Ly$\alpha$ resonant scattering to broaden line profiles and imprint red peak asymmetry. Finally in Fig.~\ref{fig:Lya_FWHM}, we show probability distributions of the directional Ly$\alpha$ FWHM and peak separation ($\Delta v_\text{sep}$) with the LyC escape fractions over all snapshots. For reference we include observational data from low-mass local starburst galaxies \citep[as reported by][]{Izotov2021}, although LyC leaking galaxies are typically much more gas rich and have significantly higher specific SFRs than the Milky Way. Therefore, we caution against over comparison of these samples, especially given the spectral shortcomings discussed in this section. Still, the intuition holds that both quantities are expected to decrease at higher LyC escape fractions \citep{Verhamme2015}, though the range in observed $\Delta v_\text{sep}$ is likely more easily explained by the wide range of halo masses, SFR variability, and galaxy ISM/CGM environments.

\begin{figure}
  \centering
  \includegraphics[width=\columnwidth]{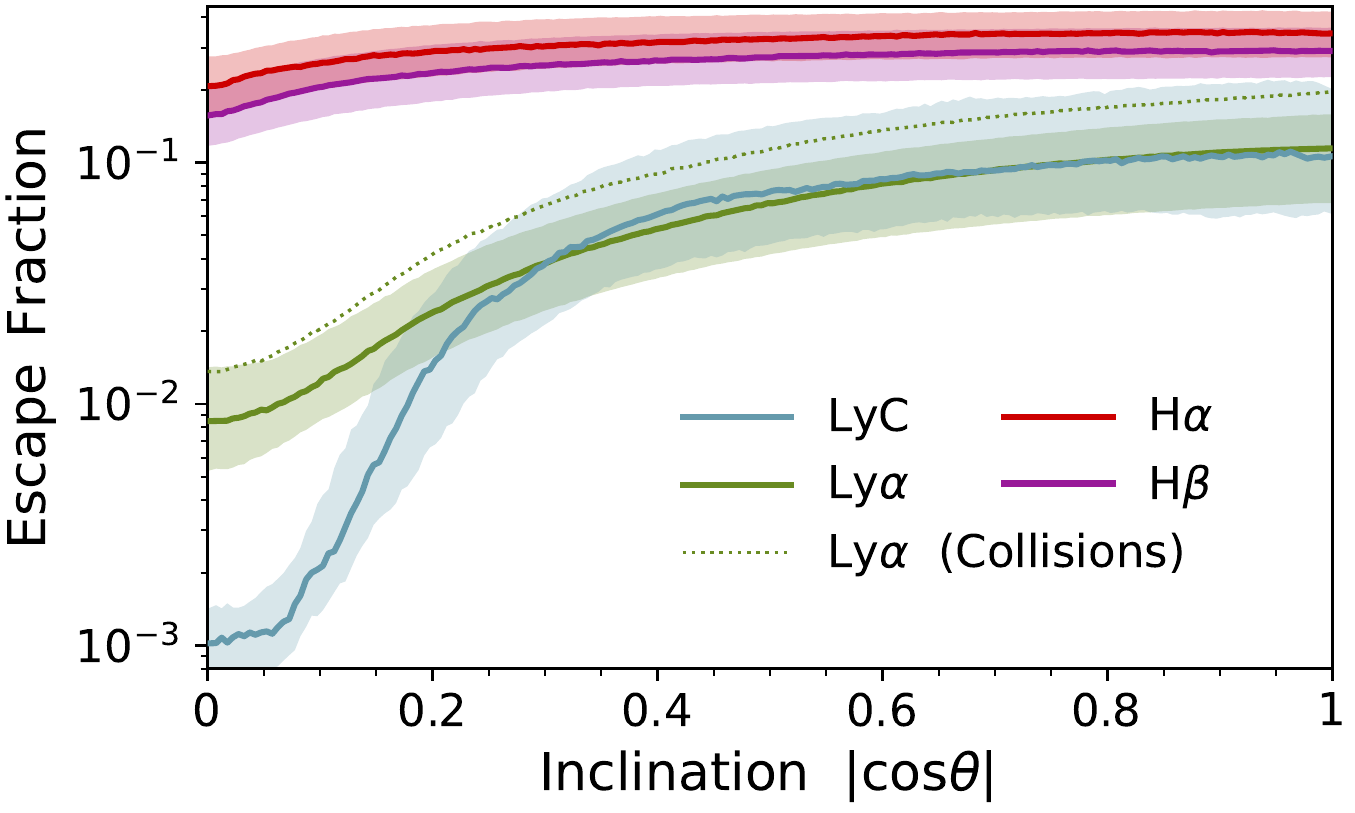}
  \caption{The escape fractions for ionizing and line photons as a function of inclination angle $|\cos\theta|$ with respect to the disc rotation axis. The shaded regions show the 1$\sigma$ confidence levels considering the time evolution. The viewing angle dependence is strongest for LyC (blue) and Ly$\alpha$ (green) photons whereas H$\alpha$ (red) and H$\beta$ (purple) are typically suppressed by less than a factor of two. For comparison we also show Ly$\alpha$ from collisional excitation only, which retains the same dependence but has higher escape fractions.}
  \label{fig:f_esc_mu}
\end{figure}

\begin{figure}
  \centering
  \includegraphics[width=\columnwidth]{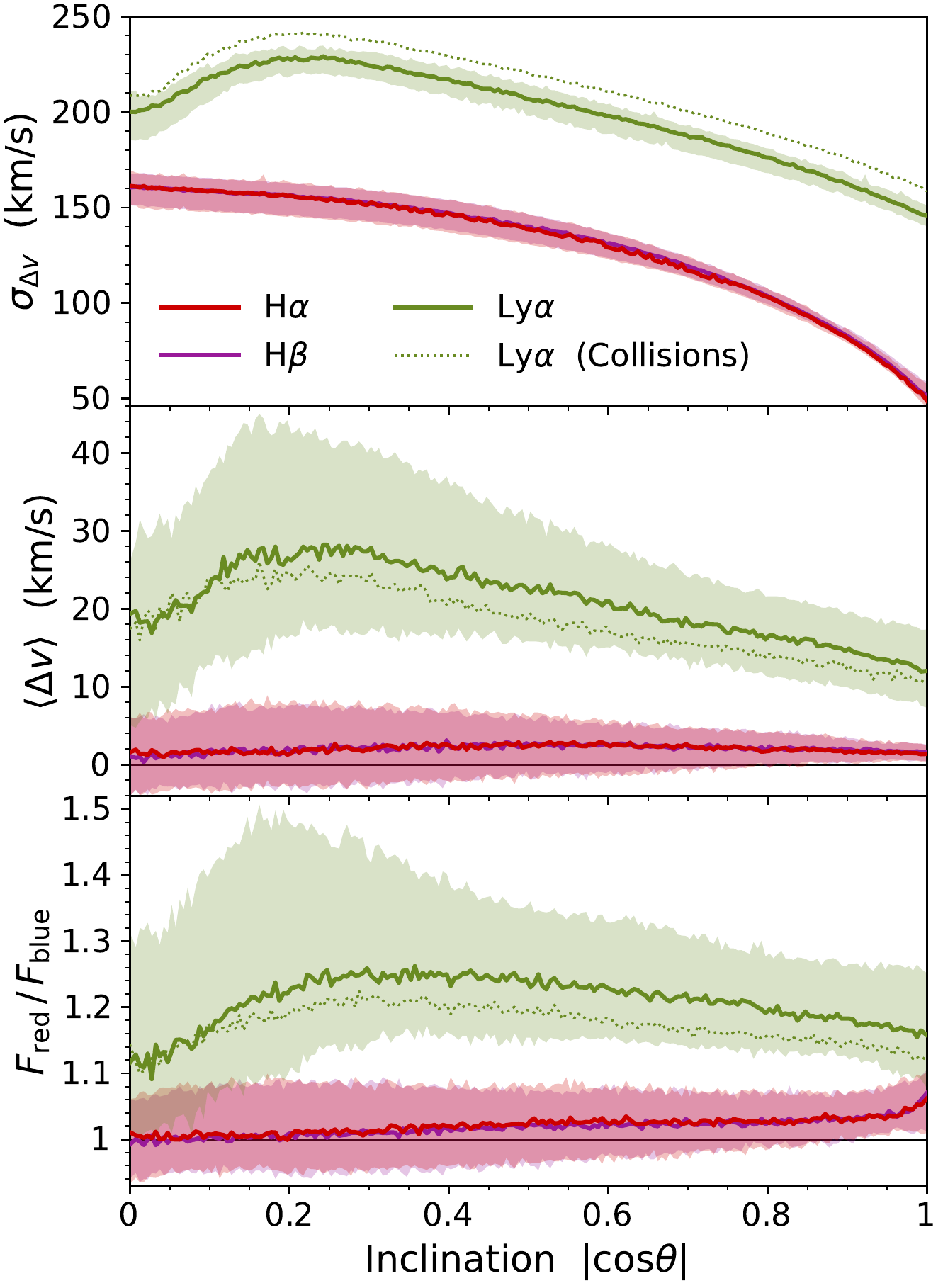}
  \caption{\emph{From top to bottom:} Inclination angle $|\cos\theta|$ dependence of the flux-weighted frequency standard deviation $\sigma_{\Delta v}$ and centroid $\langle \Delta v \rangle$ along with the red-to-blue flux ratio $F_\text{red} / F_\text{blue}$ for the H$\alpha$ (red), H$\beta$ (purple), and Ly$\alpha$ (green) lines. The non-resonant lines are approximately symmetric and broadened based on the disc rotation, while Ly$\alpha$ is slightly red peak dominated with resonant scattering significantly increasing the line widths.}
  \label{fig:Dv_avg_std_mu}
\end{figure}

\begin{figure}
  \centering
  \includegraphics[width=\columnwidth]{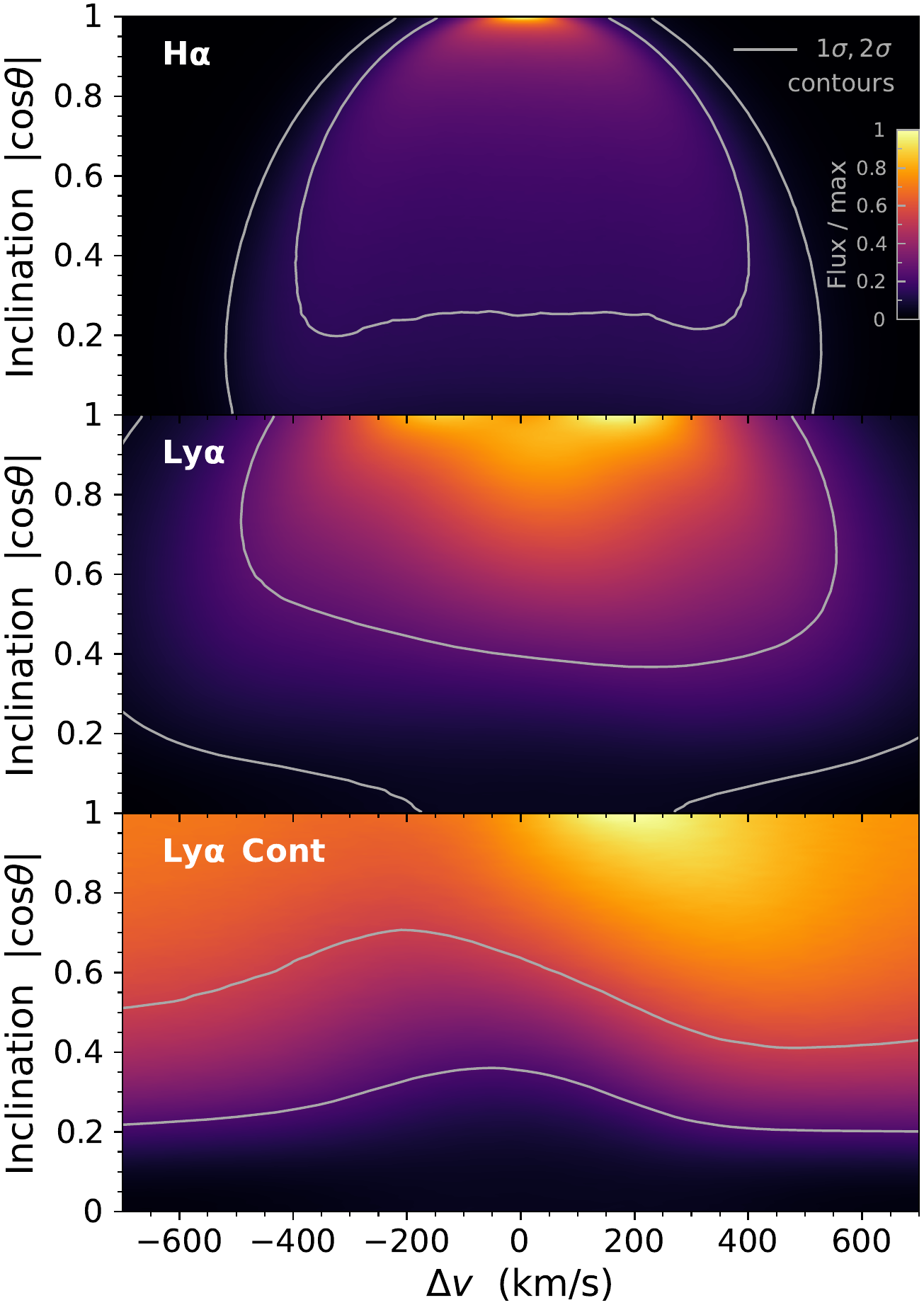}
  \caption{Spectral flux relative to the maximum value as functions of velocity offset $\Delta v$ and inclination angle $\cos \theta$ for the H$\alpha$ (top panel) and Ly$\alpha$ (middle panel) lines, and stellar continuum around the Ly$\alpha$ line (bottom panel). These continuous illustrations compactly summarize the time-averaged viewing dependent escaping flux density, contrasting the different line behaviours. We include grey $1\sigma$ and $2\sigma$ contours to help guide the eye.}
  \label{fig:Flux_mu}
\end{figure}

\subsection{Inclination dependence}
We now directly explore the role of inclination viewing angle $|\cos\theta|$ with respect to the rotation axis on the escape properties. First of all, in Fig.~\ref{fig:f_esc_mu}, we show that the escape fractions of LyC (blue), Ly$\alpha$ (green), H$\alpha$ (red), and H$\beta$ (purple) photons are all affected by the disc geometry due to different extinction properties near the mid-plane. This is most dramatic for LyC and Ly$\alpha$ radiation that is respectively absorbed and scattered by the hydrogen gas that remains neutral in the outskirts of the disc. For most directions $f_\text{esc}^{\text{Ly}\alpha} > f_\text{esc}^\text{ion}$, except for a particular range around $\theta \approx 20$--$30^{\circ}$ where the advantage of Ly$\alpha$ escape through LyC leaking holes is negated by diffusive trapping and isotropization from the disc boundary. The shaded regions show the 1$\sigma$ variation considering the time evolution. For comparison we also show Ly$\alpha$ from collisional excitation only, which retains the same inclination dependence but has higher escape fractions.

We next investigate the relationship between spectral line properties and inclination angle. In Fig.~\ref{fig:Dv_avg_std_mu}, we show the flux-weighted frequency centroid $\langle \Delta v \rangle$ and standard deviation $\sigma_{\Delta v}$ along with the red-to-blue flux ratio $F_\text{red} / F_\text{blue}$ for the H$\alpha$ (red), H$\beta$ (purple), and Ly$\alpha$ (green) lines. The non-resonant lines are approximately symmetric and broadening is determined by the LOS orientation with respect to the disc rotation. We interpret the small systematic redward velocity centroid shift in Ha/Hb ($0 < \langle \Delta v \rangle \ll \sigma_{\Delta v}$) as a subtle signature that star-forming gas is on average slightly infalling and less dust obscured on the near side. On the other hand, Ly$\alpha$ profiles are slightly red peak dominated and have significantly broadened line widths as resonant scattering redistributes photons into the wings. We note that escaping Ly$\alpha$ photons from collisional excitation generally have broader and more symmetric distributions. Ly$\alpha$ spectral properties are also strongly affected by the disc boundary when $|\cos\theta| \lesssim 0.2$. In Fig.~\ref{fig:Flux_mu}, we illustrate these properties further with the (relative) spectral flux for H$\alpha$ and Ly$\alpha$ as functions of velocity offset and inclination angle. The colour map and grey $1\sigma$ and $2\sigma$ contours clearly show that face-on views resemble what is expected from highly porous static or turbulent clouds, while most other views are strongly impacted by the rotation, notably including additional Ly$\alpha$ reddening in both the line and stellar continuum spectral flux.

\subsection{Stellar continuum and equivalent widths}
\begin{figure}
  \centering
  \includegraphics[width=\columnwidth]{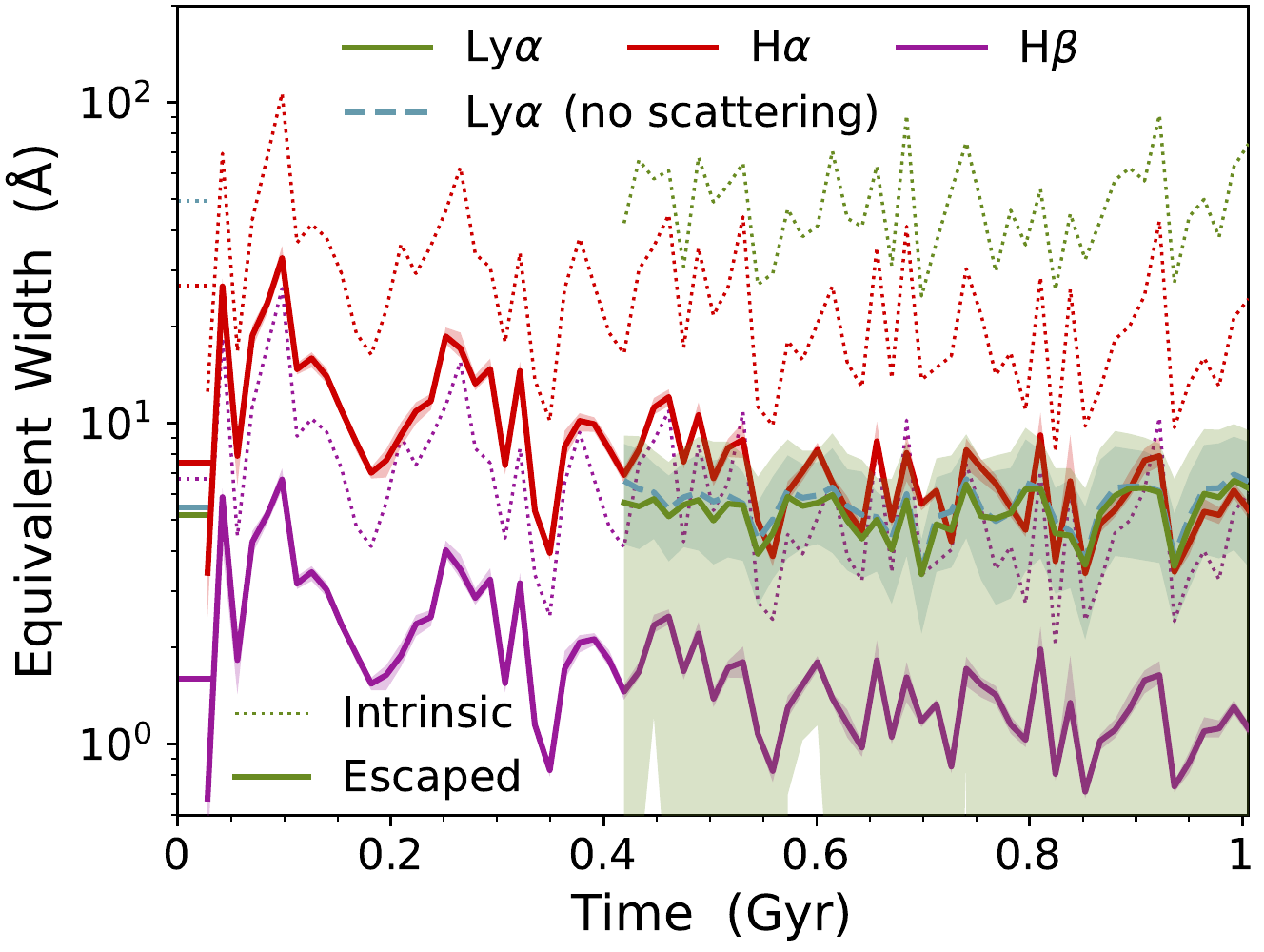}
  \caption{The evolution of the equivalent width for the Ly$\alpha$ (green includes resonant scattering while blue does not), H$\alpha$ (red), and H$\beta$ (purple) emission lines. The intrinsic (dotted curves) and escaped (solid curves) values mirror the time variability of the star-formation activity. The shaded regions show the $1\sigma$ deviations considering different viewing angles, which is most significant for the Ly$\alpha$ line even though the median value does not change much.}
  \label{fig:EW_esc}
\end{figure}

The equivalent width (EW) characterizes the strength of the observed line relative to the continuum flux, and is defined as:
\begin{equation}
  \text{EW} \equiv \int \frac{f_\lambda - f_\text{cont}}{f_\text{cont}}\,\text{d}\lambda \, .
\end{equation}
For non-resonant lines like H$\alpha$ and H$\beta$ the continuum flux around the line is independent of frequency and mainly reflects the different spectral emission and dust escape fractions from star particles compared to the spatially reprocessed ionization. Thus, the EW simplifies to the ratio of the emergent bolometric line luminosity and the frequency-averaged continuum luminosity density, i.e. $\text{EW} = L_\text{esc} / L_{\lambda,\text{esc}}^\text{cont}$. In the case of Ly$\alpha$ resonant scattering leads to preferential absorption near line centre as these photons tend to have larger traversal distances. We self-consistently account for these effects by running isolated continuum radiative transfer simulations both with and without resonant scattering. In Fig.~\ref{fig:EW_esc}, we show the time evolution of the EW for the Ly$\alpha$ (green with resonant scattering and blue without), H$\alpha$ (red), and H$\beta$ (purple) emission lines. We observe intrinsic and escaping fluctuations driven by the star-formation activity as the line emission probes shorter time-scales than the more stable continuum. We emphasize that these EWs are taken from the SEDs and that old stellar populations pre-existing from the initial conditions have a significant contribution. This imprints a bias towards lower EWs but also results in different spatial and escape morphologies compared to the line emission. In comparison, the studies of \citet{Verhamme2012} and \citet{Behrens2014} only emit line and continuum photons from young stars ($<10\,\text{Myr}$) and assume intrinsic Ly$\alpha$ EWs of 200\,\AA\ from these populations. High-redshift galaxies often have larger EWs than seen here due to their lower dust absorption and more realistic distribution of stellar ages.

\begin{figure}
  \centering
  \includegraphics[width=\columnwidth]{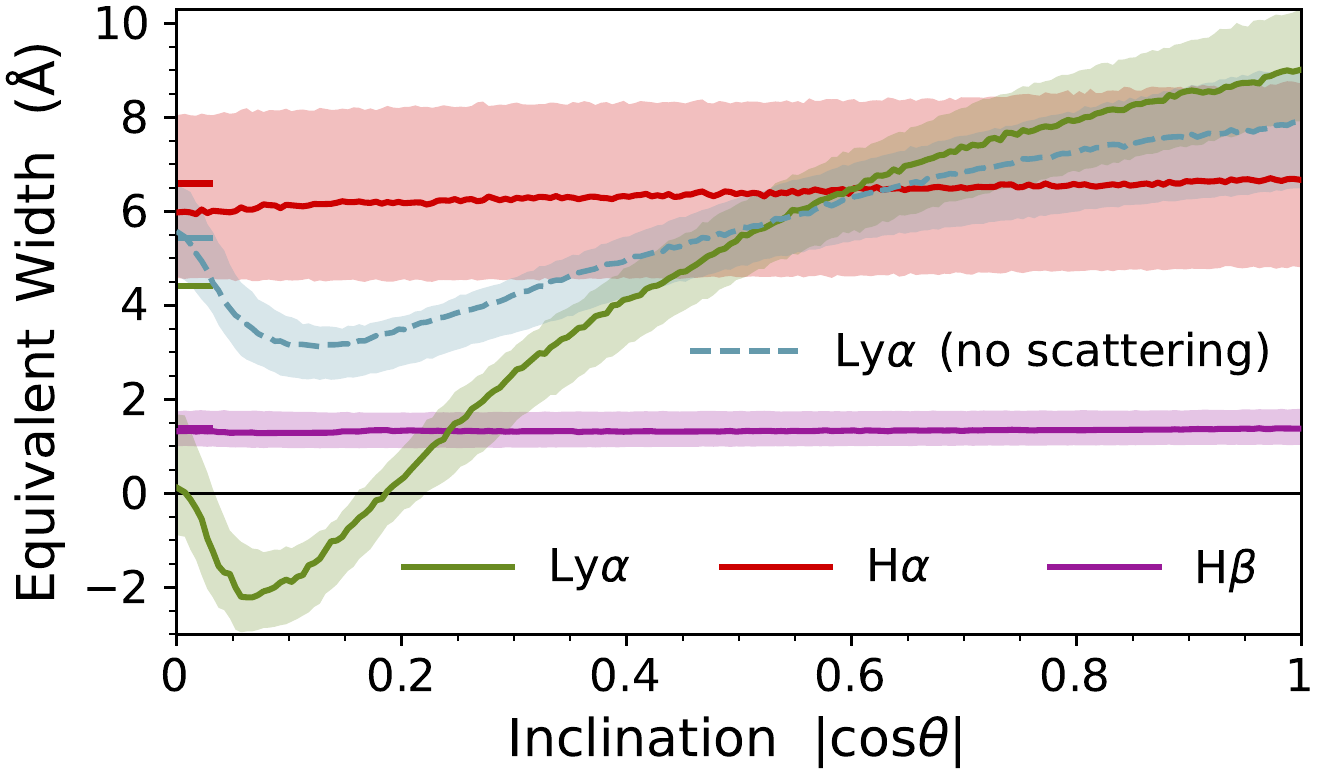}
  \caption{Inclination angle $|\cos\theta|$ dependence of the equivalent width for the Ly$\alpha$ (green includes resonant scattering while blue does not), H$\alpha$ (red), and H$\beta$ (purple) emission lines. The escape fraction dependence seen in Fig.~\ref{fig:f_esc_mu} nearly cancels out for the Balmer lines as the line and continuum escape are both similarly attenuated along the mid-plane, i.e. $f_\text{esc}(\theta) \propto f_\text{esc}^\text{cont}(\theta)$. However, the larger dust opacity and different UV continuum sourcing behaviour results in interesting viewing angle dependence for the Ly$\alpha$ line even without resonant scattering included. This is accentuated when accounting for scattering to higher latitudes by neutral hydrogen reservoirs in the disc, leading to sightlines for which Ly$\alpha$ is a net absorber against the continuum.}
  \label{fig:EW_mu}
\end{figure}

Specifically, we calculate escape fractions for line continuum photons of $\langle f_{\text{esc},X}^\text{cont} \rangle \approx \{53.5, 98.7, 97.3\}$ for $X \in \{\text{Ly}\alpha, \text{H}\alpha, \text{H}\beta\}$, which reveals that stellar continuum flux is considerably less affected by dust than line emission due to the less dramatic age dependence that reduces the high-density environmental bias inherent to recombining \HII regions. However, it is still clear that the continuum around Ly$\alpha$ can be significantly affected by dust even without resonant scattering, which is ignored in this calculation for direct comparison to the Balmer line results. When we include resonant scattering for Ly$\alpha$ photons and integrate over a frequency window of $2500\,\text{km\,s}^{-1}$ around line centre the EW decreases to $\langle \text{EW}_{\text{Ly}\alpha}^\text{scat} \rangle \approx 5.2\,\text{\AA}$. Although this is a relatively minor change in Fig.~\ref{fig:EW_mu}, we demonstrate that the inclination dependence can be significantly impacted by accounting for scattering to higher latitudes by neutral hydrogen reservoirs in the disc. In other contexts Ly$\alpha$ scattering may also not significantly change the median statistics but can still strongly impact certain sightlines associated with filamentary and disc-like structures. Therefore, it is important to model the stellar continuum as self-consistently as the Ly$\alpha$ radiation in terms of sourcing and transport. Overall, we find the EWs averaged over all times and viewing angles to be $\langle \text{EW}_X \rangle \approx \{5.5, 6.3, 1.3\}\,\text{\AA}$. We note the $\text{EW}_{\text{Ly}\alpha}(\theta)$ is lowest at $|\cos\theta| \approx 0.1$ rather than $\approx 0$ because of the overall reduction in stellar continuum away from the resonance line due to enhanced dust absorption (similar to H$\alpha$ in Fig.~\ref{fig:healpix_fesc}).

\begin{figure}
  \centering
  \includegraphics[width=\columnwidth]{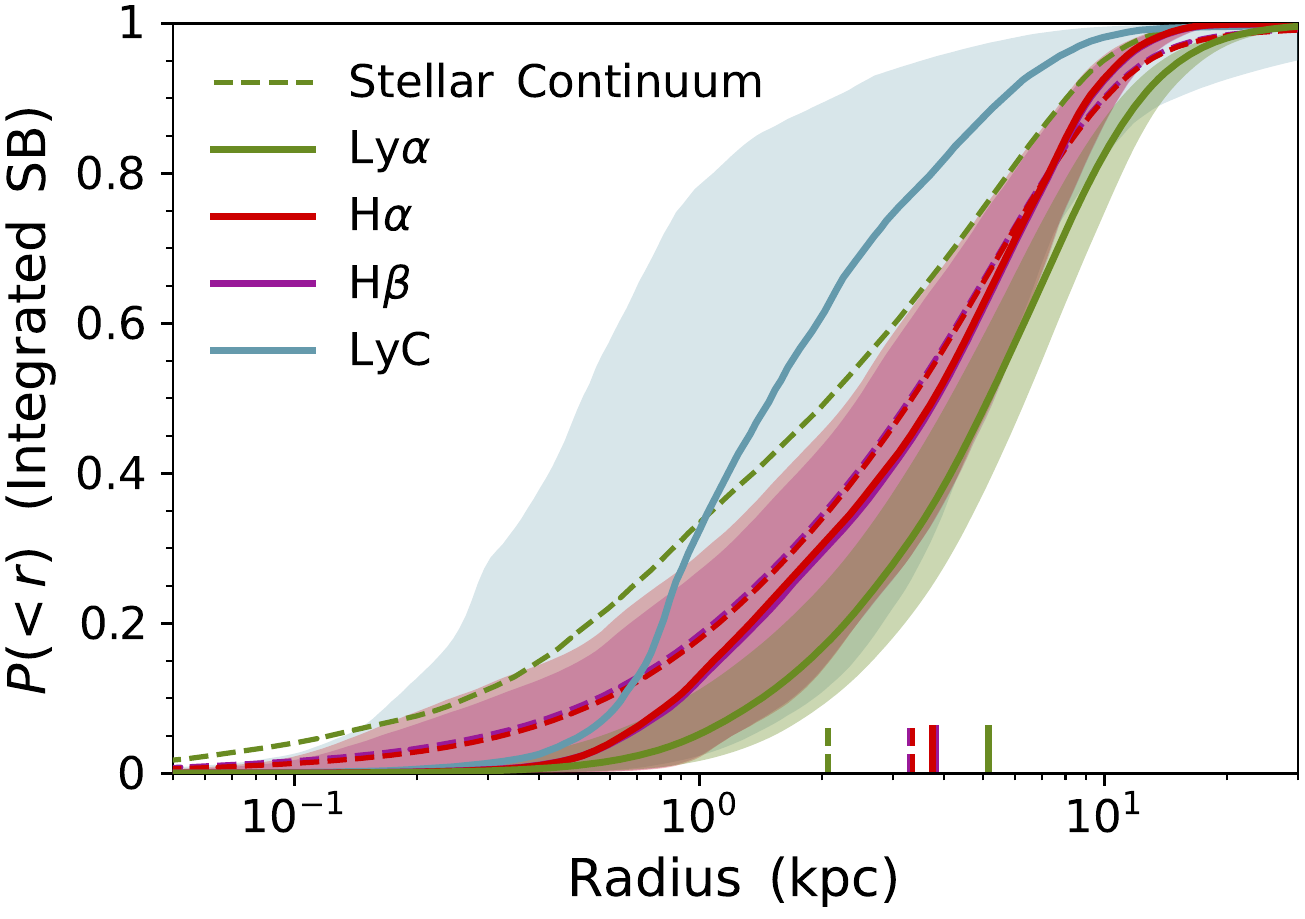}
  \caption{The cumulative distribution functions of integrated observed light within a given projected 2D radius, $P(<r) \propto \int_0^r \text{SB}(r') r' \text{d}r'$, showing the median and $1\sigma$ confidence regions for Ly$\alpha$ (green), H$\alpha$ (red), H$\beta$ (purple), and LyC (blue) photons. The escaped ionizing radiation is the most compact while resonant scattering makes Ly$\alpha$ more extended than the Balmer lines. The stellar continuum (dashed curves) for each line is also more centrally concentrated than the reprocessed ionizing emission.}
  \label{fig:ISB}
\end{figure}

As an additional metric we consider the half-light radius, defined as $P(<R_{1/2}) = 1/2$ normalized such that $P(<R_\text{box}) = 1$, of each line and report the time- and angular-averaged values to be $\langle R_{1/2,X} \rangle \approx \{5.2, 3.8, 3.8\}\,\text{kpc}$. It is common to fit the radial surface brightness data with exponential profiles, which we do for each sightline after removing the central kpc so that the annular radii are at least a few times larger than the bin widths. Specifically, we calculate the radial exponential scale lengths as $\langle R_{\text{h},X} \rangle \approx \{5.6, 11.1, 11.2\}\,\text{kpc}$. Interestingly, in comparison the average continuum half-light radii are smaller than the line emission sizes with $\langle R_{1/2,X}^\text{cont} / R_{1/2,X} \rangle \approx \{0.40, 0.87, 0.84\}$ and the associated radial exponential scale lengths are larger with $\langle R_{\text{h},X}^\text{cont} / R_{\text{h},X} \rangle \approx \{2.3, 1.6, 1.5\}$. In Fig.~\ref{fig:ISB}, we show the cumulative luminosity distribution of the escaping Ly$\alpha$ (green), H$\alpha$ (red), H$\beta$ (purple), and LyC (blue) emission within a given projected 2D radius, $P(<r) \propto \int_0^r \text{SB}(r') r' \text{d}r'$. The ionizing escape has significant variation but is biased toward the central kpc where large \HII cavities open up due to the concentrated star formation and supernovae feedback. The Balmer lines illustrate a baseline for understanding source differences for line and stellar continuum emission (dashed curves), while the escaping Ly$\alpha$ emission is significantly more extended due to resonant scattering.

\begin{figure}
  \centering
  \includegraphics[width=\columnwidth]{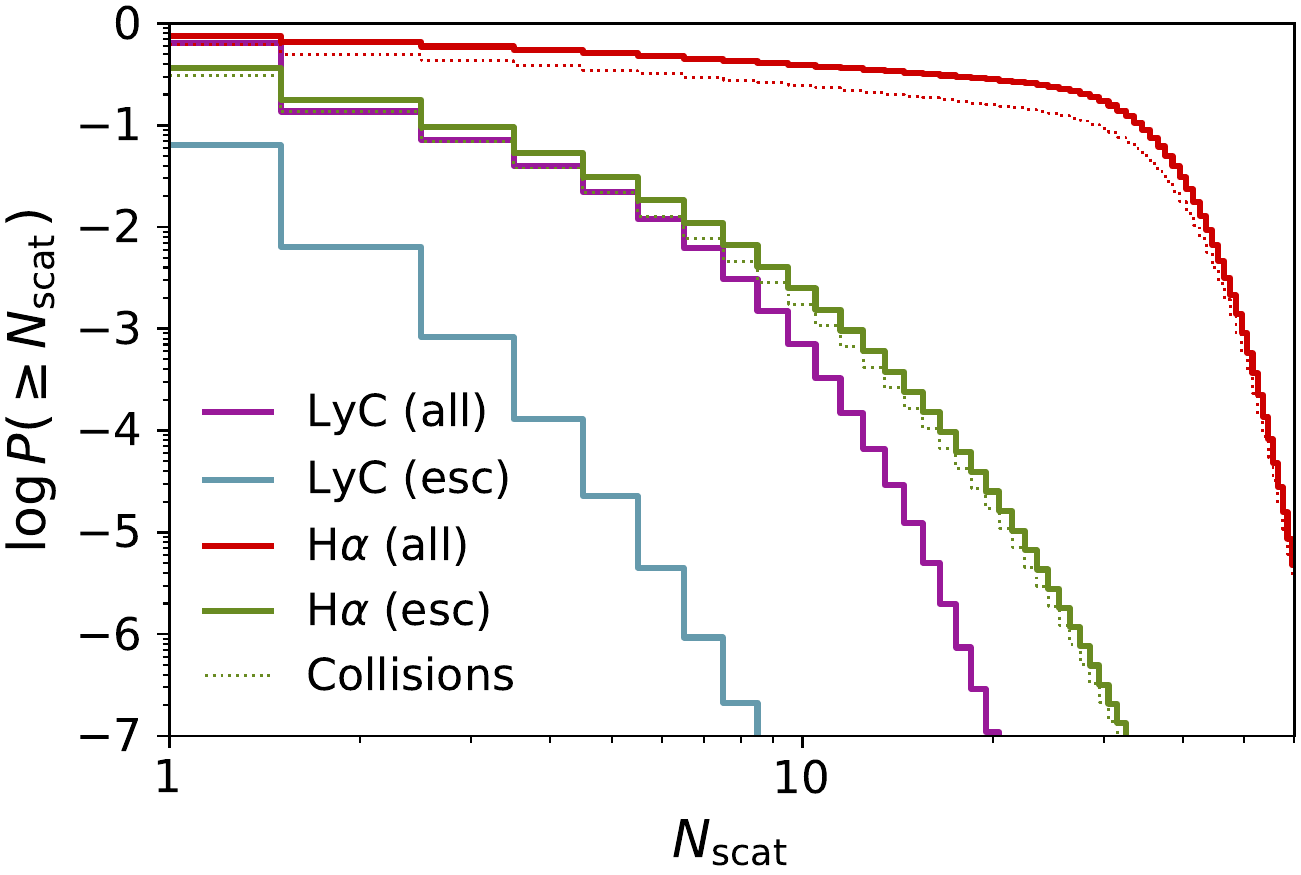}
  \caption{The probability that H$\alpha$ and LyC photons scatter at least $N_\text{scat}$ times with dust when considering all photons (red and purple curves) or only escaped photons (green and blue curves), illustrating the intuitive observed bias for lower scattering counts. These results are averaged over all photon trajectories from all snapshots and sightlines. In addition, we show that photons arising from collisional excitation emission (dotted curves) tend to scatter a fewer number of times than recombination emission.}
  \label{fig:n_scat_cdf}
\end{figure}

\subsection{Photon statistics of scattering and absorption}
We now investigate a selection of theoretical questions related to the absorption and scattering of photons including: what are the photon trajectory distributions of absorption optical depths, ionization outcomes, and interaction distances? These help us distinguish between smooth or multimodal scenarios and identify physical signatures of photon transport processes. First, in Fig.~\ref{fig:n_scat_cdf}, we show the probability that H$\alpha$ and LyC photons scatter at least $N_\text{scat}$ times with dust accounting for interactions of all photons (red and purple curves) or only escaped photons (green and blue curves). In the context of H$\alpha$ the number of dust scatterings acts as a proxy for the absorption optical depth but also directly encodes information about the diffusion of these photons and the potential for direct observation versus deflection away from the mid-plane. For LyC the number of scattering events is significantly lower due the reduced effective scattering albedo from the additional absorption from hydrogen and helium photoionization. Specifically, the $e^{-\tau}$ bias lowers the expectation of having at least $\{1, 2, 3, 4, 5, 7, 10\}$ scattering events from $\{64.73, 13.62, 7.13, 3.98, 2.21, 0.62, 0.07\}$ to $\{6.36, 0.64, 0.083, 0.013, 2.3 \times 10^{-3}, 9.3 \times 10^{-5}, 1.1 \times 10^{-6}\}$ per cent for LyC and from $\{75.74, 65.72, 59.30, 54.54, 50.78, 45.10, 39.18\}$ to $\{36.47, 17.77, 9.56, 5.38, 3.11, 1.10, 0.25\}$ per cent for H$\alpha$ photons. In fact, the expected number of scattering events, $\bar{N}_\text{scat} = \sum N_\text{scat} p(N_\text{scat})$, considering all (escaped) photons is $0.941$ ($0.071$) times for LyC and $11.71$ ($0.77$) times for H$\alpha$. These results are averaged over all photon trajectories from all snapshots and sightlines. In addition, we show that photons arising from collisional excitation emission tend to scatter fewer times than recombination emission. Finally, we note that the scattering counts for H$\beta$ are nearly identical to H$\alpha$, and that we did not explore $N_\text{scat}$ for Ly$\alpha$ photons as there are subtle complications here, e.g. discriminating between dust, core, and wing scattering events.

\begin{figure}
  \centering
  \includegraphics[width=\columnwidth]{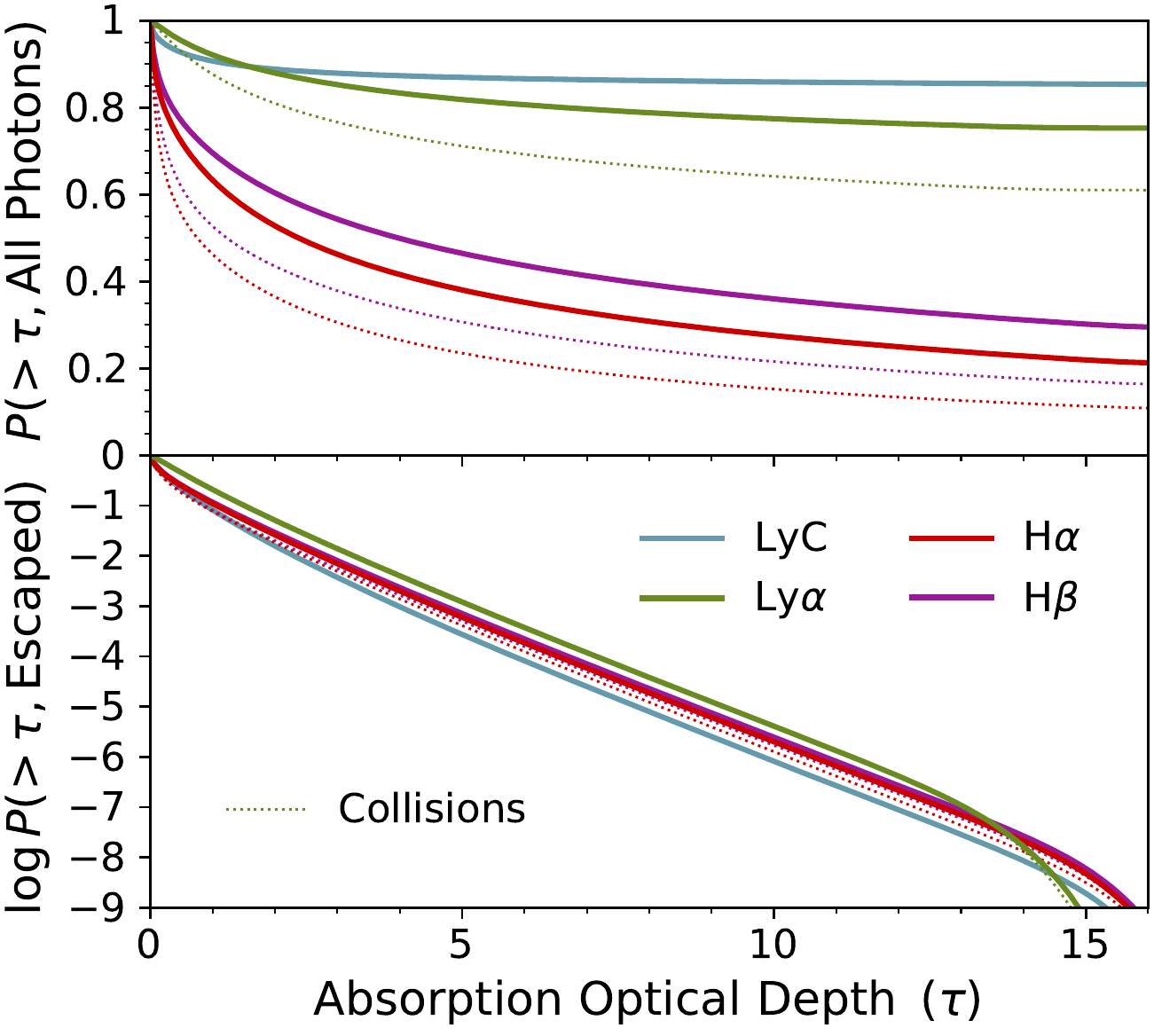}
  \caption{The probability that individual LyC (blue), Ly$\alpha$ (green), H$\alpha$ (red), and H$\beta$ (purple) photons accumulate an absorption optical depth of at least a given value. The top panel gives the result for all photons and illustrates that photons are typically highly absorbed ($\tau > 10$) or minimally absorbed ($\tau \lesssim 1$). The bottom panel shows the same cumulative distribution for escaping photons and reveals the strong impact of the $e^{-\tau}$ cut-off. Collisional excitation emission (dotted curves) systematically undergoes less absorption than recombination emission.}
  \label{fig:tau}
\end{figure}

In Fig.~\ref{fig:tau}, we show the cumulative probability distribution function for the absorption optical depths of individual Ly$\alpha$ (green), H$\alpha$ (red), H$\beta$ (purple), and LyC (blue) photons. The top panel gives the result for all photons and illustrates that photons are typically highly absorbed ($\tau > 10$) or minimally absorbed ($\tau \lesssim 1$), which is more true for LyC (flat) than H$\alpha$ (inclined). The bottom panel is for escaping photons only and reveals the strong impact of the $e^{-\tau}$ cut-off while also showing that it is less likely for LyC photons to escape with significant optical depth ($\tau \gtrsim 1$) than Ly$\alpha$ photons. We again note that collisional excitation emission systematically undergoes less absorption than recombination emission.

\begin{figure}
  \centering
  \includegraphics[width=\columnwidth]{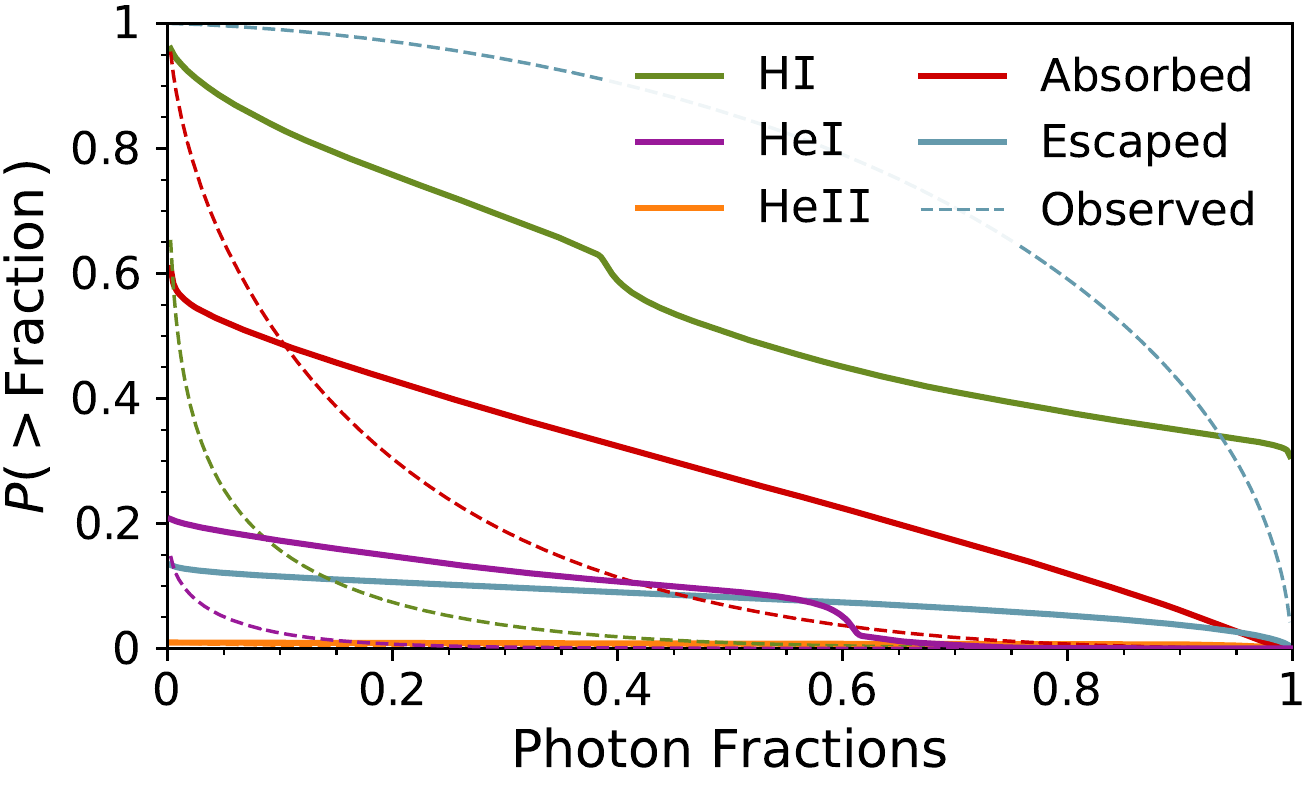}
  \caption{The probability that individual photon trajectories give rise to fractional outcomes of at least a given value, including channels for \{\HI, \HeI, \HeII\} photoionization (green, purple, and orange curves), dust absorption (red curves), and escaped emission (blue curves). For reference we also include the biased observed emission for each outcome (dashed curves). Overall, most channels exhibit a broad range of values and can result in any value between zero and one. See the text for further discussion, e.g. it is noteworthy that roughly one-third of photons are entirely absorbed by hydrogen.}
  \label{fig:weight}
\end{figure}

Ionizing radiation has the additional complexity of having several absorption channels, so we explore these different outcomes in Fig.~\ref{fig:weight} by showing the cumulative probability distribution functions of the fraction of LyC photons consumed by \{\HI, \HeI, \HeII\} photoionization (green, purple, and orange curves), dust absorption (red curves), and escaped emission (blue curves). For reference we also include the biased observed emission for each outcome as dashed curves, which reveals high-photon escape fractions with minor absorption from dust followed by ionization, indicative of unobscured LyC leakage. Overall, it is clear that most channels exhibit a broad range of values, e.g. the amount of ionization compared to dust absorption can be any value between zero and one. The hydrogen ionization is particularly interesting as there is a significant fraction (one-third) of photon sources that are entirely absorbed by hydrogen, likely corresponding to emission from unresolved \HII regions in the \HI band. Conversely, large fractions of the ionizing emission are not impacted by the dust absorption (40\%), helium ionization (80\%), or escape channels (10--15\%). In addition to the smooth probability distributions, we also see a sharp feature mirrored in \HI and \HeI at 0.4 and 0.6, respectively. This is due to the higher \HeI cross-section in this energy band when simultaneously ionizing hydrogen and helium, highlighting the interconnected nature of photon outcomes.

\begin{figure}
  \centering
  \includegraphics[width=\columnwidth]{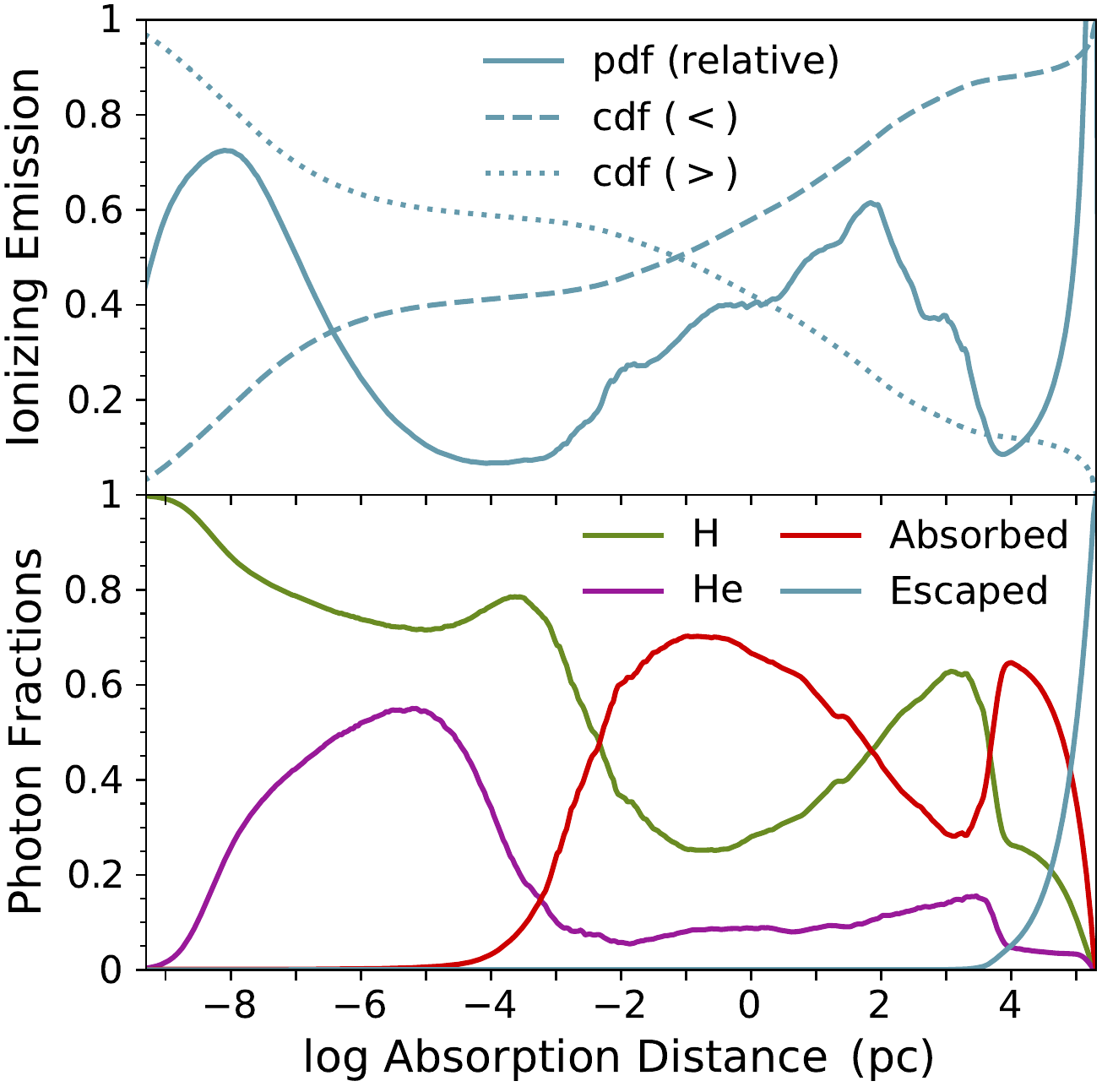}
  \caption{\textit{Top:} The relative distribution of the average distance to absorption $\langle \ell \rangle$ as defined in equation~(\ref{eq:mean_dist}), with (reverse) cumulative distribution functions shown as (dotted) dashed curves. \textit{Bottom:} The corresponding H and He ionization (green and purple), absorption (red), and escape (blue) fractions, which add up to unity. Together the outcomes reveal a complex multimodal landscape of rich photon transport physics from sub-grid to galactic scales.}
  \label{fig:dist}
\end{figure}

Finally, to probe the photon transport physics in Fig.~\ref{fig:dist}, we explore the average distance to absorption $\langle \ell \rangle$ as defined in equation~(\ref{eq:mean_dist}). In the top panel we show the (cumulative) probability distribution function as a function of absorption distance and in the bottom panel we show the corresponding ionization, absorption, and escape fractions, which add up to unity. The distances cover nearly 15 orders of magnitude and reveal a complex picture in which sub-grid to galactic scales all play unique roles. Most importantly, the distribution is multimodal with peaks representing unresolved, resolved, and galaxy escape scales. The numbers that follow are intended to guide the reader rather than provide exact distance ranges. The first region with $\langle \ell \rangle \lesssim 10^{-3}\,\text{pc}$ is dominated by H and He ionization corresponding to stellar sources that are unable to ionize their host neutral cells. The next region spans $\langle \ell \rangle \sim 10^{-3}\text{--}10^2\,\text{pc}$ characterized by a transition to being dominated by dust absorption once individual cells become ionized in compact but resolved dusty \HII regions. At this point $\langle \ell \rangle \sim 10^2\text{--}10^4\,\text{pc}$, the more diffuse and well-resolved bubble structures result in ionization dominating over dust absorption once again. On even larger scales $\langle \ell \rangle \gtrsim 10\,\text{kpc}$, the photon counts are relatively low but their trajectories are either terminated by dust absorption or allowed to successfully escape after long flights through ionized pathways. While some of this is numerical and far from converged, the main physics is captured and the analysis in this section can help inform us about how ionization and absorption behave on local and long-range scales in the context of state-of-the-art galaxy formation simulations.

\section{Summary and Discussion}
\label{sec:summary}
In this paper we have presented a comprehensive LyC, Ly$\alpha$, and Balmer line (H$\alpha$ and H$\beta$) radiative transfer study of a high-resolution isolated Milky Way galaxy simulation employing the state-of-the-art \textsc{arepo-rt} radiation hydrodynamics code \citep{Kannan2020}. The realistic framework includes self-consistent photoheating and radiation pressure, non-equilibrium thermochemistry accounting for molecular hydrogen, and dust grain evolution in the ISM within the SMUGGLE galaxy formation model for supernova feedback and stellar winds \citep{Marinacci2019}. The main motivation is to elucidate the nature of star formation processes though the reprocessing of ionizing radiation by a multiphase ISM within local and high-redshift galaxies. To accomplish this we extended our publicly available \textsc{colt} code with the necessary algorithms for self-consistent end-to-end (non-)resonant line predictions from next-generation cosmological zoom-in simulations. At the same time we obtain numerous insights about spatially, temporally, and spectrally resolved line emission, photon transport physics, and observed signatures wile exploring the numerical and physical properties of line emission from disc-like galaxies. Our main conclusions are as follows:

\begin{enumerate}
    \item Most of the characteristic ionizing and emission line radiative transfer properties vary on time-scales similar to the star-formation activity. However, for quasi-equilibrium disc configurations this mainly leads to modulations around relatively stable averages for the intrinsic and escaped quantities (see Figs.~\ref{fig:f_esc_ion} and \ref{fig:f_esc_lines}).
    \item Hydrogen emission line studies based on reprocessed ionizing radiation are sensitive to the precise ionization states of high-density gas. Therefore, radiation hydrodynamics simulations must be able to either fully resolve all \HII regions or incorporate on-the-fly prescriptions for unresolved sub-grid physics. In any case, MCRT can correct for problematic transient numerical artifacts in post-processing, and we recommend careful validation to ensure the robustness of line radiative transfer predictions (see Fig.~\ref{fig:delta_HI}). We note that accurate conversion of photoionization to recombination emission in massive galaxies also requires modelling helium ionization ($f_\text{He} \approx 8.7\pm1.1\%$), pre-absorption of LyC photons by dust ($f_\text{abs} \approx 27.5\pm6.0\%$), and anisotropic escape fractions ($f_\text{esc} \approx 5.6^{+9.6}_{-5.3}\%$), each of which collectively reduces the available budget for hydrogen line emission ($f_\text{H} \approx 55.9\pm6.5\%$).
    \item We provide convenient fitting formulae for collisional excitation emission accounting for all transitions up to the $n \leq 5$ levels (see Appendix~\ref{appendix:col_rates}). This contribution can be sensitive to the thermal state of marginally resolved \HII regions so we develop a limiter based on the photoheating rate to control spurious cooling emission (see equation~\ref{eq:L_col_max}). We calculate time-averaged fractional contributions of $\{27.2, 5.3, 3.1\}$ per cent for the \{Ly$\alpha$, H$\alpha$, H$\beta$\} lines.
    \item Monte Carlo photon packet realizations enable the precise study of the origins and outcomes of the simulated radiation sources and trajectories. For example, the LyC output from the youngest stars ($\lesssim 1\,\text{Myr}$) goes almost entirely into ionization, while dust plays an important role in moderate aged \HII regions until older stars ($\gtrsim 10\,\text{Myr}$) are cleared from their birth clouds with enhanced escape fractions (see Fig.~\ref{fig:age_hist}). This also allows us to probe density--temperature phase space luminosity diagrams to demonstrate how various regions are visibly modified by the transport physics of resonant scattering, dust absorption, and so forth (see Figs.~\ref{fig:phase_nT}--\ref{fig:nH_hist}). We similarly explore the distributions for the number of scattering events $P(\geq N_\text{scat})$, absorption optical depths $\tau$, photon outcome fractions, and average distance to absorption $\langle \ell \rangle$ (see Figs.~\ref{fig:n_scat_cdf}--\ref{fig:dist}).
    \item We portray the spatial morphology of the galaxy with face-on and edge-on views of the gas properties and velocity structure (see Figs.~\ref{fig:rho_T_D_HI} and \ref{fig:v_4}). Likewise, spatially-resolved radiative transfer observables provide a powerful mechanism for LyC, Ly$\alpha$, H$\alpha$, and H$\beta$ connections, including the emergent surface brightness to reveal sources, Balmer decrement to capture dust absorption, and frequency moment maps for spectral characterizations (see Figs.~\ref{fig:ion_Lya_Ha}--\ref{fig:Dv_avg_std}).
    \item Inclination angle dependence is strong for LyC and Ly$\alpha$ but seems to mainly affect the H$\alpha$ and H$\beta$ lines along nearly edge-on sightlines (see Figs.~\ref{fig:healpix_fesc} and \ref{fig:f_esc_mu}). In addition, it is important to model resonant scattering of the stellar continuum around the Ly$\alpha$ line as this reduces the equivalent width near the mid-plane while producing an enhancement at higher latitudes (see Fig.~\ref{fig:EW_mu}).
    \item The simulated Ly$\alpha$ line profiles retain a significant amount of flux near line centre, likely due to the absence of a proper cosmological circumgalactic medium (see Figs.~\ref{fig:flux}--\ref{fig:time_avg_std} and \ref{fig:Flux_mu}). The low-specific SFRs and high-dust content especially in bright \HII regions also lead to relatively symmetric line profiles $F_\text{red}/F_\text{blue} \approx 1.2$ (see Fig.~\ref{fig:Dv_avg_std_mu}). Destroying dust in \HII regions beyond what is obtained from the empirical dust model can boost the Ly$\alpha$ escape fraction by a factor of two and further assist in shaping the line profile to resemble observations at $z \lesssim 3$ (see Fig.~\ref{fig:fion_test}).
\end{enumerate}

In a companion study \citet{Tacchella2022}, we investigate the observational implications for Balmer line emission, including its utility as a SFR tracer on spatially-resolved scales. Overall, we find that our isolated disc simulations are well-suited for comprehensive observational comparisons with local H$\alpha$ surveys. However, these idealized set-ups are not expected to serve as analogs for high-redshift, starbursting, Ly$\alpha$ emitting galaxies. Upcoming applications of our framework to next-generation cosmological radiation-hydrodynamic zoom-in simulations will provide a major step forward in interpreting hydrogen spectroscopy of high-redshift galaxies, e.g. with the \textit{James Webb Space Telescope}. Additional radiative transfer studies and dedicated observations are needed to better understand the connections between Ly$\alpha$ and H$\alpha$ emission, e.g. the Lyman Alpha Reference Sample (LARS) contains about 42 galaxies at $z \approx 0.1$ \citep{Rivera-Thorsen2015,Herenz2016,Runnholm2020,Rasekh2022}. Although the galaxies are significantly different from those modelled here, it is intriguing to draw insights from nearby spatially-resolved samples with Ly$\alpha$ counterparts as these inform us about escape mechanisms. 

For example, so-called picket fence models adopt a covering fraction framework to explain profile diversity as some sightlines escape directly while others undergo multiple scattering \citep{Tenorio-Tagle1999,Heckman2011,Reddy2022}. Of course, kinematics also play an important role to enhance Ly$\alpha$ \citep{Wofford2013}, which can also be viewed in terms of the evolutionary progression of birth clouds when these dominate the galactic emission \citep{Naidu2022,Matthee2022}. Similar conclusions have been drawn about the dust geometry and neutral hydrogen column density, e.g. based on observed relations between Ly$\alpha$/H$\alpha$ and H$\alpha$/H$\beta$ line ratios for 31 $z \approx 0.3$ LAEs \citep{Scarlata2009} and spectroscopy of 43 Green Pea galaxies at $z \approx 0.1$--$0.3$ \citep{Yang2017}. It is also instructive to compare our simulated profiles to multiple-peaked LAE samples at $z \sim 2$--$3$ \citep{Kulas2012,Trainor2015}, which exhibit a wide variety of Ly$\alpha$ line profiles that are consistent with our expectations for more clustered and bursty star formation in cosmological contexts.

Recently, \citet{Camps2021} argued that further research is needed to determine the required spatial resolution of hydrodynamical simulations to enable meaningful Ly$\alpha$ radiative transfer. Unfortunately, the answer may depend on the underlying galaxy formation model, even among different groups achieving state-of-the-art resolutions. The simulations used in this work employ significantly more advanced ISM modelling than the test simulation explored by \citet{Camps2021}, which crucially includes accurate ionization states and spatially resolved Ly$\alpha$ sourcing. Still, their demonstrations provide a timely warning to avoid a false sense of security, especially when employing moderate resolution cosmological simulations.

Eventually, we envision the development of subresolution radiative transfer procedures that are still strongly connected to and guided by the simulated physics. For example, we plan to incorporate novel higher-order ray-tracing techniques into \textsc{colt} and \textsc{arepo-mcrt} that explicitly account for density and velocity gradient information. Some of these have already been explored in recent studies \citep{LaoSmith2020,Smith2020,SmithThesan2022}. Beyond this, sub-grid radiation transport schemes may be motivated by idealized models such as clumpy multiphase geometries, especially in combination with empirical calibration to match observational constraints \citep[e.g.][]{Li2020}. However, the advantages of sub-grid assumptions must be weighed carefully as they also introduce additional uncertainties that impact the resulting theoretical and observational interpretations. Recent progress on ISM modelling in terms of physical realism and resolution is quite encouraging, yet adding RHD and other complex physics broadens the landscape of possible outcomes and reveals additional pitfalls to address in future model iterations.

In the future, we plan to expand our spectral coverage of synthetic observations to include dust continuum and nebular line radiative transfer based on state-of-the-art pipelines \citep[e.g.][]{CampsBaes2020,Narayanan2021}. Also, this could mirror the approach taken by many simulations to stitch local ionization region models with galaxy scale radiation transport for dust-reprocessed spectral energy distributions optimized for emission line properties \citep{Katz2019,Shen2020,Wilkins2020,KannanLIM2022}. However, there is an opportunity to extend our spatially-resolved MCRT framework to include metal ionization for self-consistent nebular line studies. These methodologies are mainly limited by the physics and resolution obtained by simulations, but would naturally fit within the scope of zoom-in simulations within the \textsc{thesan} project \citep{KannanThesan2022,GaraldiThesan2022,SmithThesan2022}. In this context, we will also pursue improving the spatial and time resolution around newly formed stars with super-Lagrangian schemes in order to resolve higher density \HII regions. Otherwise, we will incorporate sub-grid models of photoheating and radiation pressure feedback for a more accurate treatment of unresolved processes \citep[e.g.][]{Jeffreson2021}. With this outlook, we hope to improve our understanding of the physics of emission lines throughout the local and high-redshift Universe.

\section*{Acknowledgements}
We thank the referee for constructive comments and suggestions which have improved the quality of this work.
We thank Ewald Puchwein, Bryan Terrazas, Ben Johnson, Enrico Garaldi, Volker Bromm, Desika Narayanan, and Ben Kimock for insightful discussions related to this work.
AS and HL acknowledge support for Program numbers \textit{HST}-HF2-51421.001-A and \textit{HST}-HF2-51438.001-A provided by NASA through a grant from the Space Telescope Science Institute, which is operated by the Association of Universities for Research in Astronomy, Incorporated, under NASA contract NAS5-26555.
MV acknowledges support through NASA ATP 19-ATP19-0019, 19-ATP19-0020, 19-ATP19-0167, and NSF grants AST-1814053, AST-1814259, AST-1909831, AST-2007355 and AST-2107724.
FM acknowledges support through the program ``Rita Levi Montalcini'' of the Italian MUR.
LVS acknowledges support from the NSF AST-1817233 and NSF CAREER 1945310 grants.
PT acknowledges support from NSF AST-1909933, NSF AST-2008490, and NASA ATP Grant 80NSSC20K0502.
Computing resources supporting this work were provided by the Extreme Science and Engineering Discovery Environment (XSEDE), at Comet, Expanse, and Stampede2 through allocation TG-AST200007 and by the NASA High-End Computing (HEC) Program through the NASA Advanced Supercomputing (NAS) Division at Ames Research Center. Additional computing resources were provided by the MIT Engaging cluster.
% The Acknowledgements section is not numbered. Here you can thank helpful colleagues, acknowledge funding agencies, telescopes and facilities used etc. Try to keep it short.
% Remember to thank the referee.

%%%%%%%%%%%%%%%%%%%%%%%%%%%%%%%%%%%%%%%%%%%%%%%%%%
\section*{Data Availability}
The data underlying this article will be shared on reasonable request to the corresponding author.

\bibliographystyle{mnras}
\bibliography{biblio}

%%%%%%%%%%%%%%%%% APPENDICES %%%%%%%%%%%%%%%%%%%%%
\appendix

\section{New fitting formulae for collisional excitation rates}
\label{appendix:col_rates}

\begin{table}
  \centering
  \caption{\textit{Top:} Einstein A radiative rates in units of Hertz for allowed transitions in hydrogen needed to calculate the branching ratios $P(n,l \rightarrow n',l')$ in equation~(\ref{eq:branching-ratio}). \textit{Bottom:} Probability $P(n, l \rightarrow X)$ that a cascade originating from a given quantum state $(n,l)$ results in the specified line photon $X$ when considering the $n \leq 5$ levels in equation~(\ref{eq:prob-line}).}
  \label{table:EinsteinA}
  \begin{tabular}{cc cc cc}
    \hline
    $nl \rightarrow n'l'$ & $A_{ij}$ & $nl \rightarrow n'l'$ & $A_{ij}$ & $nl \rightarrow n'l'$ & $A_{ij}$ \\
    \hline
    $3p \rightarrow 2s$ & 2.25e7 & $4p \rightarrow 3s$ & 3.07e6 & $5f \rightarrow 3d$ & 4.54e6 \\
    $4p \rightarrow 2s$ & 9.67e6 & $5p \rightarrow 3s$ & 1.64e6 & $5p \rightarrow 4s$ & 7.38e5 \\
    $5p \rightarrow 2s$ & 4.95e6 & $4s \rightarrow 3p$ & 1.84e6 & $5s \rightarrow 4p$ & 6.45e5 \\
    $3s \rightarrow 2p$ & 6.32e6 & $4d \rightarrow 3p$ & 7.04e6 & $5d \rightarrow 4p$ & 1.49e6 \\
    $3d \rightarrow 2p$ & 6.47e7 & $5s \rightarrow 3p$ & 9.05e5 & $5p \rightarrow 4d$ & 1.89e5 \\
    $4s \rightarrow 2p$ & 2.58e6 & $5d \rightarrow 3p$ & 3.39e6 & $5f \rightarrow 4d$ & 2.59e6 \\
    $4d \rightarrow 2p$ & 2.06e7 & $4p \rightarrow 3d$ & 3.48e5 & $5d \rightarrow 4f$ & 5.05e4 \\
    $5s \rightarrow 2p$ & 1.29e6 & $4f \rightarrow 3d$ & 1.38e7 & $5g \rightarrow 4f$ & 4.26e6 \\
    $5d \rightarrow 2p$ & 9.43e6 & $5p \rightarrow 3d$ & 1.50e5 & & \\
    \hline
    %\caption{}
  \end{tabular}
  \addtolength{\tabcolsep}{2.75pt}
  \begin{tabular}{cc ccccc}
    \hline
    Line & Level & $s$ & $p$ & $d$ & $f$ & $g$ \\
    \hline
    $2\gamma$ & $n = 2$ & $1$ & $0$ & -- & -- & -- \\
    & $n = 3$ & $0$ & $1$ & $0$ & -- & -- \\
    & $n = 4$ & $0.416$ & $0.739$ & $0.255$ & $0$ & -- \\
    & $n = 5$ & $0.486$ & $0.692$ & $0.313$ & $0.093$ & $0$ \\
    \hline
    Ly$\alpha$ & $n = 2$ & $0$ & $1$ & -- & -- & -- \\
    & $n = 3$ & $1$ & $0$ & $1$ & -- & -- \\
    & $n = 4$ & $0.584$ & $0.261$ & $0.745$ & $1$ & -- \\
    & $n = 5$ & $0.514$ & $0.308$ & $0.687$ & $0.907$ & $1$ \\
    \hline
    H$\alpha$ & $n = 3$ & $1$ & $1$ & $1$ & -- & -- \\
    & $n = 4$ & $0.416$ & $0.261$ & $0.255$ & $1$ & -- \\
    & $n = 5$ & $0.378$ & $0.280$ & $0.267$ & $0.729$ & $1$ \\
    \hline
    H$\beta$ & $n = 4$ & $0.584$ & $0.739$ & $0.745$ & $0$ & -- \\
    & $n = 5$ & $0.168$ & $0.075$ & $0.077$ & $0.271$ & $0$ \\
    \hline
    H$\gamma$ & $n = 5$ & $0.454$ & $0.646$ & $0.657$ & $0$ & $0$ \\
    \hline
    Pa$\alpha$ & $n = 4$ & $0.416$ & $0.261$ & $0.255$ & $1$ & -- \\
    & $n = 5$ & $0.059$ & $0.046$ & $0.031$ & $0.093$ & $1$ \\
    \hline
    Pa$\beta$ & $n = 5$ & $0.319$ & $0.233$ & $0.236$ & $0.637$ & $0$ \\
    \hline
    Br$\alpha$ & $n = 5$ & $0.227$ & $0.121$ & $0.107$ & $0.363$ & $1$ \\
    \hline
    %\caption{}
  \end{tabular}
  \addtolength{\tabcolsep}{-2.75pt}
\end{table}

As discussed in Section~\ref{subsec:rt}, we include line emission due to radiative de-excitation of collisional excitation of neutral hydrogen by free electrons. The general strategy is to calculate weighted effective collision strengths for each line and level \citep[see also the discussion in][]{Dijkstra2019}. We adopt the Case B on-the-spot approximation in which higher Lyman series radiative transitions are excluded as these typically excite nearby atoms to the same level in realistic astrophysical environments where collisional excitation is important. Once an atom is in a quantum state $(n,l)$ the probability that a radiative cascade results in a given line $X$ is
\begin{equation} \label{eq:prob-line}
  P(n, l \rightarrow X) = \sum_{n',l'} P(n,l \rightarrow n',l')\,P(n',l' \rightarrow X) \, .
\end{equation}
In practice, we start from lower levels and work upwards to derive the probabilities for higher states. When multiple transitions are allowed then the probability is taken from the branching ratio given by the Einstein A coefficients for permitted transitions, i.e.
\begin{equation} \label{eq:branching-ratio}
  P(n,l \rightarrow n',l') = \frac{A_{n,l \rightarrow n',l'}}{\sum_{n'',l''} A_{n,l \rightarrow n'',l''}} \, .
\end{equation}
For convenience and transparency in Table~\ref{table:EinsteinA}, we provide the Einstein A coefficients employed in our calculations entering equation~(\ref{eq:branching-ratio}), as well as the resulting probabilities from equation~(\ref{eq:prob-line}) for all lines originating from the $n \leq 5$ levels.

\begin{table}
  \centering
  \caption{Coefficients for the polynomial fits to cascade-weighted effective collision strengths for each line for the $n \leq 5$ levels. Specifically, the functional form is $\Upsilon_{1n,X}(T) = a_0 + a_1 T_6 + a_2 T_6^2 + a_3 T_6^3$, where the temperature is normalized as $T_6 \equiv T / (10^6\,\text{K})$ to allow order unity coefficients.}
  \label{table:Upsilon}
  \addtolength{\tabcolsep}{-1.4pt}
  \begin{tabular}{cc cccc}
    \hline
    Line & Level & $a_0$ & $a_1$ & $a_2$ & $a_3$ \\
    \hline
    Cool & $n = 2$ & $0.616414$ & $16.8152$ & $-32.0571$ & $35.5428$ \\
    & $n = 3$ & $0.217382$ & $3.92604$ & $-10.6349$ & $13.7721$ \\
    & $n = 4$ & $0.0959324$ & $1.89951$ & $-6.96467$ & $10.6362$ \\
    & $n = 5$ & $0.0747075$ & $0.670939$ & $-2.28512$ & $3.4796$ \\
    \hline
    $2\gamma$ & $n = 2$ & $0.267486$ & $1.57257$ & $-6.44026$ & $11.5401$ \\
    & $n = 3$ & $0.0940821$ & $2.90748$ & $-6.69528$ & $7.86419$ \\
    & $n = 4$ & $0.0415028$ & $1.13673$ & $-3.97454$ & $6.05941$ \\
    & $n = 5$ & $0.032277$ & $0.40856$ & $-1.34604$ & $2.07558$ \\
    \hline
    Ly$\alpha$ & $n = 2$ & $0.348928$ & $15.2426$ & $-25.6168$ & $24.0027$ \\
    & $n = 3$ & $0.1233$ & $1.01857$ & $-3.93958$ & $5.90791$ \\
    & $n = 4$ & $0.0544296$ & $0.762786$ & $-2.99013$ & $4.57684$ \\
    & $n = 5$ & $0.0424305$ & $0.262379$ & $-0.939078$ & $1.40402$ \\
    \hline
    H$\alpha$ & $n = 3$ & $0.217382$ & $3.92604$ & $-10.6349$ & $13.7721$ \\
    & $n = 4$ & $0.0372036$ & $0.508162$ & $-1.86495$ & $2.82946$ \\
    & $n = 5$ & $0.029568$ & $0.165593$ & $-0.52784$ & $0.772594$ \\
    \hline
    H$\beta$ & $n = 4$ & $0.0587288$ & $1.39135$ & $-5.09972$ & $7.80679$ \\
    & $n = 5$ & $0.00844856$ & $0.0682112$ & $-0.277137$ & $0.450694$ \\
    \hline
    H$\gamma$ & $n = 5$ & $0.0366909$ & $0.437134$ & $-1.48014$ & $2.25631$ \\
    \hline
    Pa$\alpha$ & $n = 4$ & $0.0372036$ & $0.508162$ & $-1.86495$ & $2.82946$ \\
    & $n = 5$ &  $0.00825154$ & $-0.0217447$ & $0.177446$ & $-0.334415$ \\
    \hline
    Pa$\beta$ & $n = 5$ &  $0.0213164$ & $0.187338$ & $-0.705286$ & $1.10701$ \\
    \hline
    Br$\alpha$ & $n = 5$ &  $0.0167001$ & $0.0464665$ & $-0.0996903$ & $0.116279$ \\
    \hline
    %\caption{}
  \end{tabular}
  \addtolength{\tabcolsep}{1.4pt}
\end{table}

\begin{figure}
  \centering
  \includegraphics[width=\columnwidth]{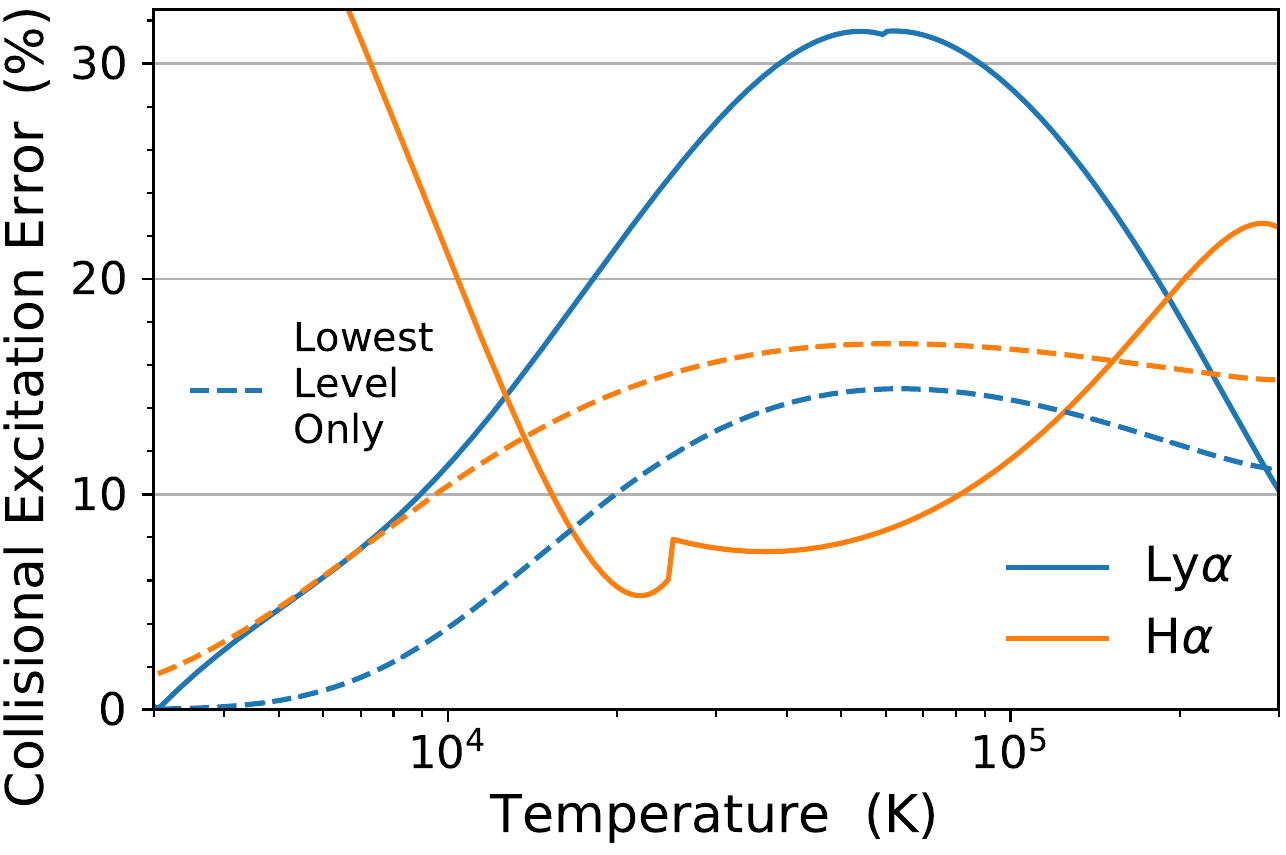}
  \caption{Relative error in collisional excitation rates $q_\text{col}$ as a function of temperature, comparing $q_{1s2p}$ for Ly$\alpha$ from \citet{Scholz1991} and $q_{13}$ for H$\alpha$ from \citet{Aggarwal1983} to the approach of including all transitions in the $N \leq 5$ levels. For reference we also include dashed curves showing the error within our framework when only considering the lowest level.}
  \label{fig:q_col_error}
\end{figure}

\begin{figure}
  \centering
  \includegraphics[width=\columnwidth]{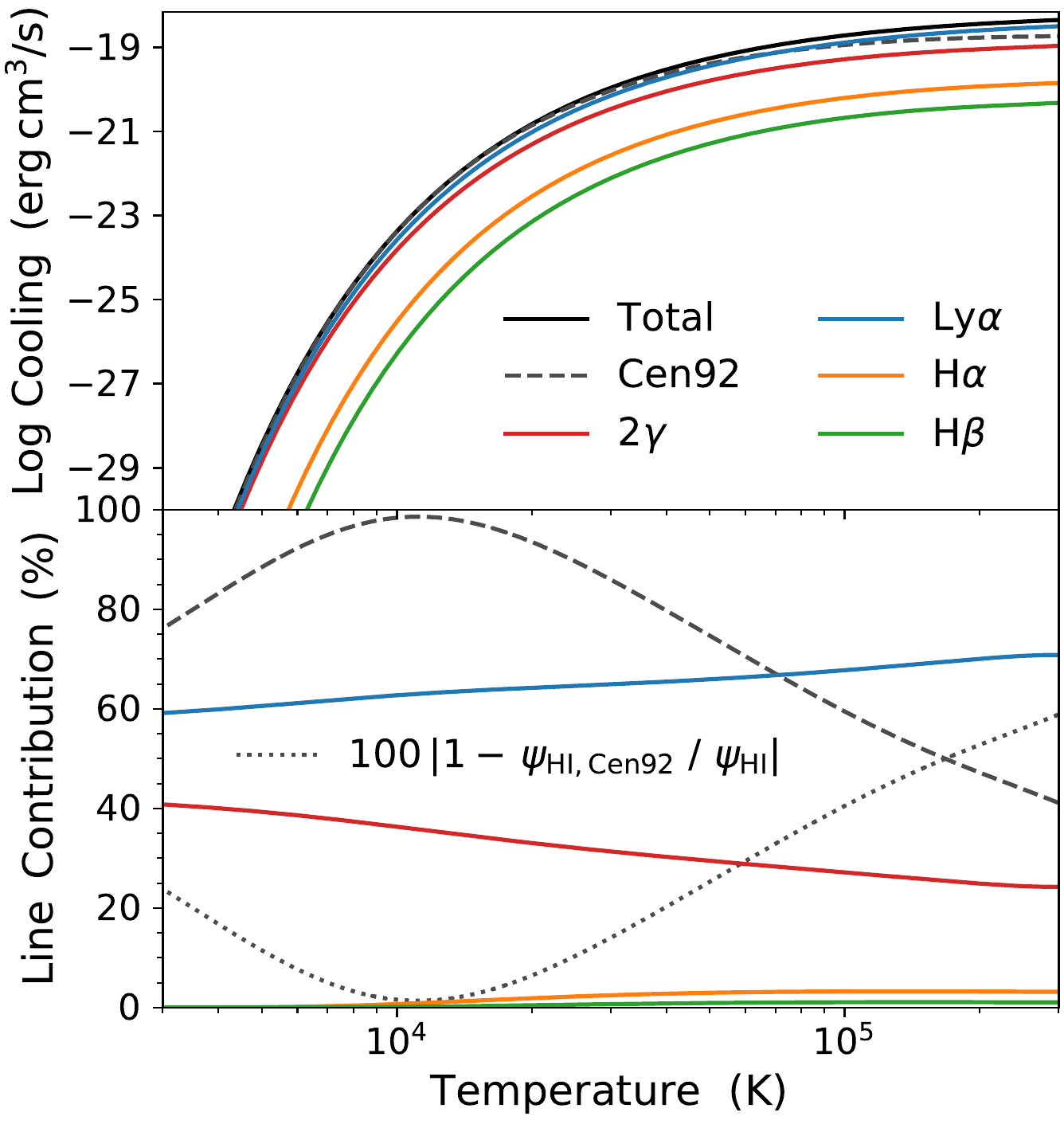}
  \caption{Absolute and fractional contributions of each line to the total hydrogen collisional excitation cooling rate $\psi_\text{\HI}$ as functions of temperature. For comparison we include the commonly adopted fitting formula of \citet{Black1981} modified following \citet{Cen1992}, which agrees near $T = 10^4\,\text{K}$ but underpredicts cooling by factors $\lesssim 2$ at other temperatures. Ly$\alpha$ emission accounts for over $60$ per cent of the overall cooling, with two photon emission dominating the remainder, and lines like H$\alpha$ and H$\beta$ at the percent level.}
  \label{fig:line_cool}
\end{figure}

At this point we obtain the cascade weighted effective collision strengths from the individual Maxwellian-averaged values $\Upsilon_{ij}$ by adding all contributions of the type $(1s \rightarrow n,l \rightarrow X)$ via
\begin{equation} \label{eq:line-collision-strength}
  \Upsilon_{1n,X} = \sum_l \Upsilon_{1s \rightarrow n,l} P(n, l \rightarrow X) \, .
\end{equation}
Thus, the collisional excitation rate coefficients are
\begin{equation}
  q_{\text{col},X} = \frac{8.63 \times 10^{-6}}{\omega_1 \sqrt{T}} \sum_n \Upsilon_{1n,X}\,\exp\left(-\frac{T_{1n}}{T}\right) \, ,
\end{equation}
where $\omega_1 = 2$ is the statistical weight of the ground state, the constant prefactor results from $(2\pi \hbar^4 / k_\text{B} m_e^3)^{1/2}$, and $T_{1n} = h \nu_{1n} / k_\text{B}$ is the temperature corresponding to the energy of the transition $h \nu_{1n} = (1 - n^{-2})\,\text{Ryb}$. Similarly, the total cooling rate coefficient due to collisional excitation is
\begin{equation} \label{eq:cooling}
  \psi_\text{\HI} = \frac{8.63 \times 10^{-6}}{\omega_1 \sqrt{T}} \sum_n h \nu_{1n} \Upsilon_{1n,\text{Cool}}\,\exp\left(-\frac{T_{1n}}{T}\right) \, ,
\end{equation}
where $\Upsilon_{1n,\text{Cool}} = \sum_l \Upsilon_{1s \rightarrow n,l}$. For reference in Fig.~\ref{fig:q_col_error}, we plot the relative error in collisional excitation rates $q_\text{col}$ as a function of temperature. Specifically, we find that the commonly used $q_{1s2p}$ for Ly$\alpha$ from \citet{Scholz1991} is $\approx10$--$30\%$ low while $q_{13}$ for H$\alpha$ from \citet{Aggarwal1983} is $\approx 10\%$ high compared to the approach of including all transitions in the $N \leq 5$ levels. Similarly, in Fig.~\ref{fig:line_cool}, we plot the absolute and fractional contributions of each line to the total collisional excitation cooling rate $\psi_\text{\HI}$. We also compare against the commonly adopted fitting formula of \citet{Black1981} modified following \citet{Cen1992}, which agrees well near $T = 10^4\,\text{K}$ but underpredicts cooling by factors $\lesssim 2$ at other temperatures. We note that Ly$\alpha$ emission accounts for over $60$ per cent of the overall cooling, with two photon emission dominating the remainder.

We find the data for $\Upsilon_{1n,X}$ based on \citet{Anderson2002} are well fit by third-order polynomials in temperature, i.e. in most cases the accuracy remains below one per cent over the provided range of $0.5$--$25$\,eV. In practice, we set the effective collision strengths $\Upsilon_{1n,X}$ to the values obtained at $T = 3 \times 10^5\,\text{K}$ for higher temperatures. We note that the line rate coefficients $q_{\text{col},X}$ retain the $T^{-1/2}$ scaling and that such hot gas is highly ionized so this is typically an unphysical range for collisional excitation. The polynomial form is quite stable so we employ these $\Upsilon_{1n,X}$ formulae at all lower temperatures. In fact, the zero-order coefficient corresponds to the low-temperature limit. We provide our derived coefficients in Table~\ref{table:Upsilon}, which are convenient to incorporated into codes that calculate line emission due to collisional excitation based on equation~(\ref{eq:L_col}).

We note that the effective collision strengths themselves are still uncertain at the 10--20\% level for calculations between different groups \citep[see the discussion in][]{Dijkstra2019}. However, the strong temperature dependence mainly enters through the exponential functions $\exp(-T_{1n}/T)$ so the accuracy of the thermochemistry in hydrodynamical simulations is the main source of uncertainty for collisional excitation line emission. Finally, we note that at high densities ($n_e \gtrsim 10^4\,\text{cm}^{-3}$) we must also account for collisional de-excitation processes, which have subtle implications for hydrogen line radiative transfer. Firstly, at high densities collisions can give rise to a non-negligible fraction of electrons in excited states, thereby affecting certain emission and transport properties. Specifically, optical depths and collisional rates (equation~\ref{eq:line-collision-strength}) assume electrons overwhelmingly originate in the ground state. Secondly, collisions mix different $l$-levels at a fixed $n$, with the limiting case setting populations based on the statistical weights $\propto (2 l - 1)$, modifying various excitation and cascade probabilities, although this is already accounted for in the \citet{StoreyHummer1995} recombination tables. Finally, frequent collisions open up the forbidden $2s \rightarrow 1s$ transition providing an additional destruction mechanism for Ly$\alpha$ photons.

\section{Comparison of ionization states}
\label{appendix:ion_accuracy}

\begin{figure*}
  \centering
  \includegraphics[width=.85\textwidth]{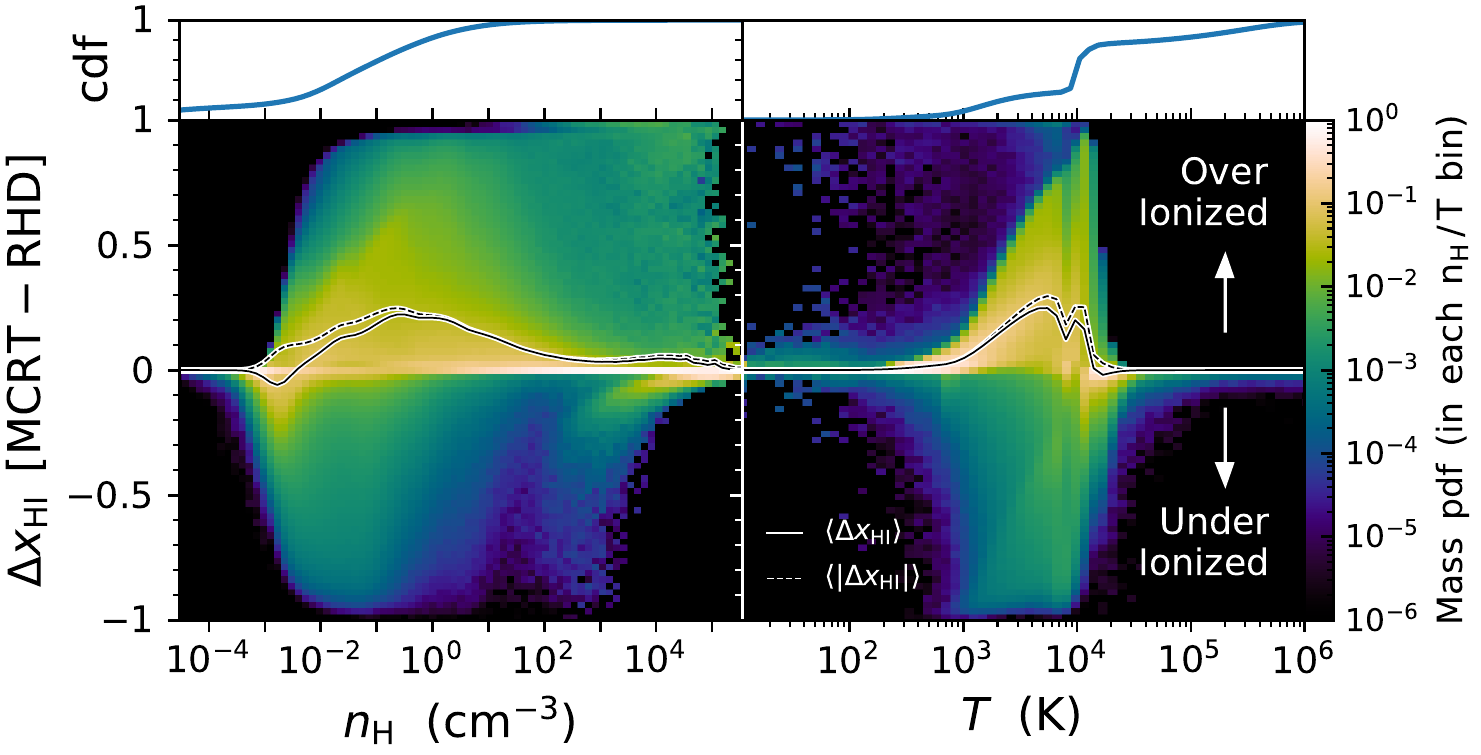}
  \caption{Mass-weighted histograms (normalized to the total mass in each bin) comparing the time-averaged change in ionization state $\Delta x_\text{\HI}$ between the post-processing MCRT photoionization equilibrium and on-the-fly radiation hydrodynamics results as functions of hydrogen number density $n_\text{H}$ and temperature $T$. To guide the eye, we include sign and absolute value averages as solid and dashed curves, respectively. Overall, the agreement between the two methods is quite impressive, but a non-negligible fraction of gas mass can be over ($\Delta x_\text{\HI} > 0$) or under ($\Delta x_\text{\HI} < 0$) ionized and significantly affect resolved line emissivities. For reference in the top panels we show the normalized cumulative distribution functions of gas mass in the simulation.}
  \label{fig:delta_HI}
\end{figure*}

As discussed in Section~\ref{subsec:rt-ion}, transient numerical phenomena can give rise to ionization states that are not sufficiently robust for the detailed hydrogen line emission study presented in this work. However, to better understand where differences arise, in Fig.~\ref{fig:delta_HI}, we show mass-weighted histograms comparing the post-processing MCRT photoionization equilibrium and on-the-fly radiation hydrodynamics results as functions of hydrogen number density $n_\text{H}$ and temperature $T$. As the MCRT solver does not explicitly track molecular hydrogen we combine both the H and H$_2$ gas phases, i.e. we define $\Delta x_\text{\HI} \equiv x_\text{\HI,MCRT} - x_\text{\HI,RHD} - 2 x_{\text{H}_2,\text{RHD}}$. Overall, the agreement between the two methods is quite impressive and provides an independent and highly non-trivial test confirming the general consistency of the treatment of radiation in the simulation. However, it is clear that certain densities and temperatures are susceptible to over or under ionization. This is most likely due to not fully resolving the temperature and density substructure of a fraction of the young \HII regions, which is a challenging problem for radiation hydrodynamics simulations in general. On the other hand, some variations also arise from the less accurate flux directionality of the M1 closure scheme \citep[e.g. see Section~4.7 of][]{Kannan2019}. Specifically, photons encounter different optical depth distributions along the slightly divergent transport paths tracked in the MCRT and M1 methods. While the resolution of \textsc{arepo-rt} is fully determined by the underlying mesh geometry, sub-grid ray-tracing methods are able to model radiation processes below the grid scale which allows us to explicitly correct for on-the-fly resolution artifacts. Beyond this, our MCRT approach has other fundamental algorithmic differences including subresolution sourcing from star particles, dust absorption and scattering that changes effective path lengths, and stochastic photon packet sampling of the radiation field compared to the smoother moment-based fluid representation (although we have gone to great lengths to ensure convergence). More quantitatively, we calculate time-averaged mass (volume) weighted global neutral fractions of $\langle x_\text{\HI,MCRT} \rangle \approx 0.57\,(3.38 \times 10^{-5})$ for post-processed ionization states and $\langle x_\text{\HI,RHD} \rangle \approx 0.46\,(3.33 \times 10^{-5})$ for on-the-fly ones. This implies a relative difference between the two methods of $1 - \langle x_\text{\HI,RHD} \rangle / \langle x_\text{\HI,MCRT} \rangle \approx 19\,(1.4)\%$ (see additional statistics in Table~\ref{tab:time_avg_RHD}). The change in recombination emission is significant enough to require the MCRT calculations, i.e. an order of magnitude. This is because resolved line luminosities depend on the square of the density and are therefore highly sensitive to errors associated with unresolved \HII regions that mainly affect the higher density gas. In summary, we recommend cross validating line emissivities in the context of radiation hydrodynamic simulations, which may reveal interesting behaviour depending on the simulation resolution and model implementation, including distinguishing between physical and numerical phenomena.

\begin{table}
  \centering
  \caption{Time-averaged comparison of angle-averaged radiative transfer outcomes calculated from the RHD and MCRT ionization states, along with the relative difference. For simplicity we only show the most relevant quantities for LyC photons, including the mass-weighted neutral fraction ($\langle x_\text{\HI} \rangle$), fractions of photons ionizing hydrogen and helium ($f_\text{H},f_\text{He}$), and dust absorption and escape fractions ($f_\text{abs},f_\text{esc}$).}
  %, as well as for H$\alpha$ recombination photons, including the intrinsic luminosity ($L_\text{rec}$) and escape fraction ($f_\text{esc,rec}$) showing the strong sensitivity of line emission to ionization states.}}
  \label{tab:time_avg_RHD}
  \addtolength{\tabcolsep}{-3pt}
  \renewcommand{\arraystretch}{1.1}
  \begin{tabular}{@{} l ccc @{}}
    \hline
    Quantity & RHD & MCRT & $1 - \text{RHD}/\text{MCRT}$\ [\%] \\
    \hline
    $\langle x_\text{\HI} \rangle\ [\%]$ & $46.2$ & $57.0$ & $19.0$ \\
    $f_\text{H}\ [\%]$ & $76.6$ & $55.9$ & $37.1$ \\
    $f_\text{He}\ [\%]$ & $4.66$ & $8.73$ & $46.7$ \\
    $f_\text{abs}\ [\%]$ & $15.5$ & $27.5$ & $43.8$ \\
    $f_\text{esc}\ [\%]$ & $3.31$ & $7.91$ & $58.2$ \\
    \hline
    % $\log L_\text{rec}\ [\text{erg\,s}^{-1}]$ & $42.7$ & $41.5$ & $1650$ \\
    % $f_\text{esc,rec}\ [\%]$ & $20.2$ & $31.1$ & $35.1$ \\
    % \hline
  \end{tabular}
  \addtolength{\tabcolsep}{3pt}
  \renewcommand{\arraystretch}{0.9090909090909090909}
\end{table}

\section{Survival of dust in ionized gas}
\label{appendix:fion_test}
Our simulations incorporate spatially-dependent dust-to-gas ratios based on the self-consistent dust formation and destruction model presented in \citet{McKinnon2017} and \citet{Kannan2020}. When such models are not available it is common to assume a constant dust-to-metal ratio such that the dust cross-section scales linearly with the local metallicity, although this is a biased proxy \citep[e.g. see the discussions in][]{McKinnon2016,McKinnon2018,Aoyama2018,LiNarayanan2019}. In addition, it has become standard for Ly$\alpha$ radiative transfer simulations to incorporate an additional parameter $f_\text{ion}$ denoting the survival fraction of dust within ionized regions. Dust abundances may be lowered by radiation pressure on dust \citep{Draine2011} or rotational disruption of dust grains by radiative torques in strong radiation fields \citep{Hoang2019}. However, we emphasize that $f_\text{ion}$ is not very well constrained and likely varies across and within galaxies, so should be viewed as a practical but approximate way of modelling the destruction of dust in physically hostile environments \citep[see the discussion in][]{Laursen2009}. While this motivation certainly makes sense in the absence of dust modelling, it is not clear that including $f_\text{ion} \ll 1$ on top of our current scheme is any more realistic. In fact, the commonly adopted value of $f_\text{ion} = 0.01$ may overcompensate beyond what is expected from detailed state-of-the-art simulations where the resolved environmental variation of $\mathcal{D}$ is generally only an order of magnitude for gas in the ISM without efficient thermal sputtering $T \lesssim 10^6\,\text{K}$ \citep{Hu2019,Kannan2020}.

\begin{figure}
  \centering
  \includegraphics[width=\columnwidth]{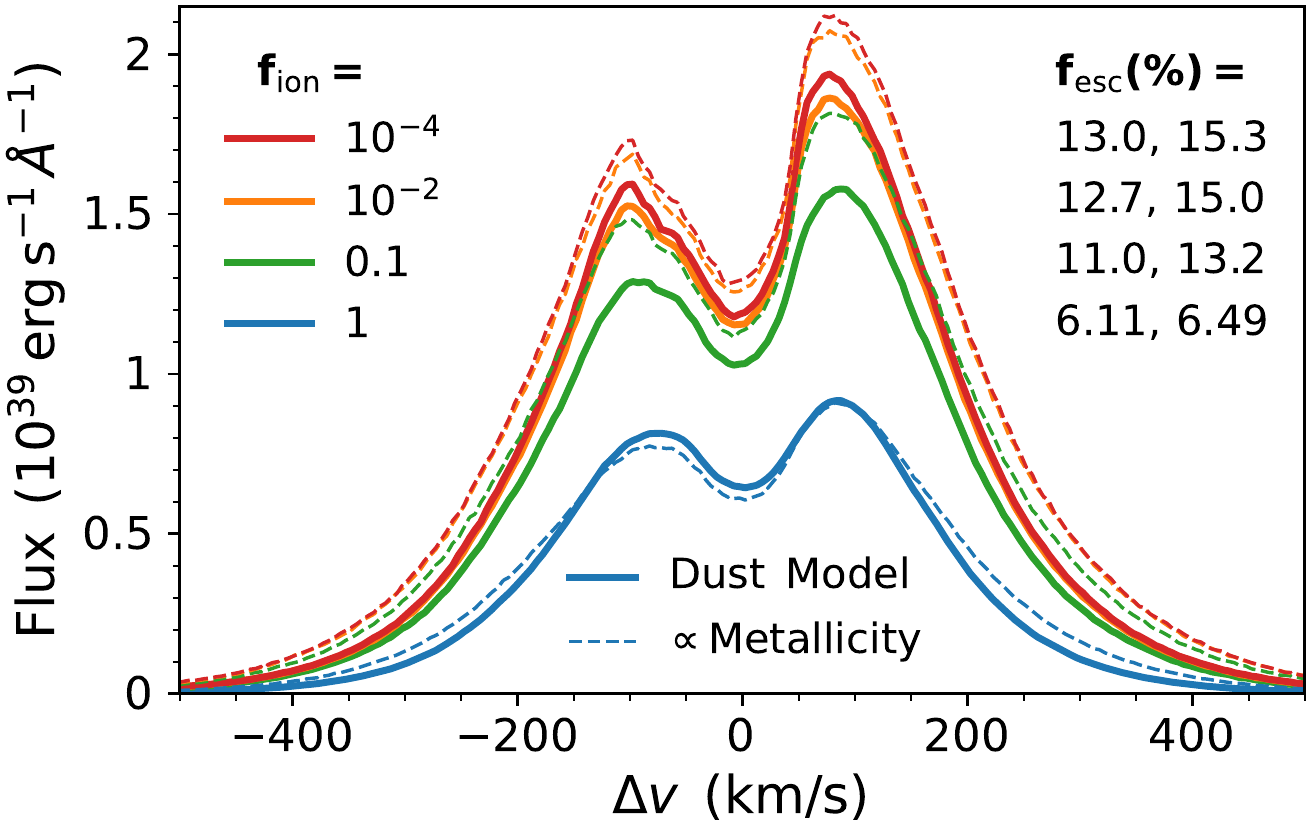}
  \caption{A numerical experiment comparing the face-on emergent spectral line profiles from simulations with different survival fractions of dust in ionized regions $f_\text{ion} = \{1, 10^{-1}, 10^{-2}, 10^{-4}\}$ for a snapshot at 1\,Gyr. The solid and dashed curves respectively show results based on the fiducial dust model and scaling with the metallicity (with a dust-to-metal ratio of $\text{DTM} = 1.5$ chosen to produce similar escape fractions). Although the profile shapes are not significantly affected, the escape fractions increase by up to a factor of $\approx 2$ when dust is removed from ionized gas with high column densities.}
  \label{fig:fion_test}
\end{figure}

As dust modelling is an active area of research, we simply show the results of a numerical experiment in Fig.~\ref{fig:fion_test} in which we compare the face-on emergent flux assuming different survival fractions $f_\text{ion} = \{1, 10^{-1}, 10^{-2}, 10^{-4}\}$ for a snapshot at 1\,Gyr. Interestingly, the spectral line profiles are not significantly affected, which we interpret as a minimal change in the overall escaping trajectories. However, the escape fractions increase by up to a factor of $\approx 2$ when dust is removed from ionized gas with substantial dust column densities. As most of the dust absorption occurs on relatively local scales with respect to the emission, we conclude that the main role of $f_\text{ion}$ is to preserve more photons from compact \HII regions, and consequently boost the escape fraction $f_\text{esc} \approx \{6.1, 11.0, 12.7, 13.0\}\%$ and red-to-blue flux ratio $F_\text{red} / F_\text{blue} \approx \{1.10, 1.19, 1.20, 1.20\}$ for the respective values of $f_\text{ion}$, noting that angular averages are qualitatively similar but with lower escape fractions. For comparison, we also provide control runs based on scaling with metallicity and find very similar results but with overall elevated escape fractions $f_\text{esc} \approx \{6.5, 13.2, 15.0, 15.3\}\%$ and corresponding red-to-blue flux ratios $F_\text{red} / F_\text{blue} \approx \{1.15, 1.22, 1.24, 1.23\}$, noting that in this case we adopt a constant dust-to-metal ratio of $\text{DTM} = 1.5$ chosen so the resulting escape fraction roughly matches our fiducial model. For reference if we adopt the commonly used SMC cross-section normalization from \citet{Laursen2009} (not plotted) then we obtain $f_\text{esc} \approx \{9.1, 15.7, 17.1, 17.3\}\%$ and $F_\text{red} / F_\text{blue} \approx \{1.17, 1.22, 1.22, 1.23\}$. Of course, lower metallicity, high-redshift Ly$\alpha$ emitters will exhibit a much smaller dependence on $f_\text{ion}$ as the dust content is already reduced enough to produce relatively high escape fractions, i.e. $f_\text{esc} \gtrsim 1/2$. In addition, preferential destruction of small dust grains could flatten the extinction curve to reduce the Ly$\alpha$ dust opacity while leaving H$\alpha$ mostly unaffected \citep{Aguirre2001}. Overall, employing $f_\text{ion} \ll 1$ provides more optimistic Ly$\alpha$ escape without significantly altering the spectral properties such as peak separations \citep[see Section~4.4 of][]{Kimm2022}. Many of these processes occur below the resolution scales of typical galaxy formation simulations, so understanding dust physics in high-resolution multiphase simulations also provides information about subresolution transport. It is important to continue improving dust models in the context of Ly$\alpha$ radiative transfer, especially as high dust absorption around compact \HII regions affects inferred Ly$\alpha$ fluxes.

%%%%%%%%%%%%%%%%%%%%%%%%%%
% Don't change these lines
\bsp % typesetting comment
\label{lastpage}
\end{document}